\pdfoutput=1
\documentclass[3p,times]{elsarticle}
\usepackage{ecrc}

\usepackage{graphicx}

\usepackage{amsmath,amssymb,amsfonts}
\usepackage{algorithmic}
\usepackage{graphicx}
\usepackage{textcomp}
\usepackage{xcolor}
\usepackage{xspace}
\usepackage{booktabs}
\usepackage{hyperref}

\usepackage{paralist}

\newcommand\crule[3][black]{\textcolor{#1}{\rule{#2}{#3}}}

\def\BibTeX{{\rm B\kern-.05em{\sc i\kern-.025em b}\kern-.08em
    T\kern-.1667em\lower.7ex\hbox{E}\kern-.125emX}}

\usepackage{rotating}   
  \usepackage{multirow}  
  \usepackage{rotating}
\usepackage[english]{babel}
\usepackage[utf8]{inputenc} %utf8x
\usepackage[T1]{fontenc}
\usepackage{amssymb}
\usepackage{tabularx}
\usepackage{pdfpages}
\usepackage{balance}
\usepackage{url}
\usepackage{adjustbox}
\newcolumntype{R}[2]{%
    >{\adjustbox{angle=#1,lap=\width-(#2)}\bgroup}%
    l%
    <{\egroup}%
}

\usepackage{lscape}

\usepackage{pdflscape}

\usepackage[most]{tcolorbox}

\usepackage{etoolbox}
\AtBeginEnvironment{tcolorbox}{\small}
\usepackage{listings}
\usepackage{color}
\usepackage[labelfont=bf]{caption} % optional
\usepackage{ragged2e}
\usepackage{tabularx}
\usepackage{lscape} 

\usepackage{subcaption}

\usepackage{pgfplots} %\pgfplotsset{compat=1.14}

\usepackage[title]{appendix}

\def\fig {Fig.~}

\usepackage{xcolor}
\usepackage{soul}

\usepackage{amsmath}
\usepackage{tcolorbox}
\definecolor{lightGrey}{rgb}{0.9, 0.9, 0.9}

\usepackage{xcolor}

\usepackage{enumitem}
\newcommand{\nd}{\vspace{1mm}\noindent}
\usetikzlibrary{shapes.misc,shadows}
\definecolor{airforceblue}{rgb}{0.36, 0.54, 0.66}
\definecolor{dkgreen}{rgb}{0,0.6,0}
\definecolor{gray}{rgb}{0.5,0.5,0.5}
\definecolor{mauve}{rgb}{0.58,0,0.82}
\definecolor{antiquefuchsia}{rgb}{0.57, 0.36, 0.51}
\definecolor{applegreen}{rgb}{0.55, 0.71, 0.0}
\definecolor{asparagus}{rgb}{0.53, 0.66, 0.42}
\usepackage[graphicx]{realboxes}
\usepackage{color, colortbl} 
 \definecolor{Gray}{gray}{0.9}
  \usepackage{listings,xcolor,tikz}

\usepackage[normalem]{ulem}

\usepackage{xspace}
\usepackage{epstopdf} 
\usepackage{amsmath}
\usepackage{amssymb}
\newcolumntype{Y}{>{\centering\arraybackslash}X}
\usepackage{multirow} 
\usepackage{adjustbox}
\definecolor{findOptimalPartition}{HTML}{D7191C}
\definecolor{storeClusterComponent}{HTML}{FDAE61}
\definecolor{dbscan}{HTML}{ABDDA4}
\definecolor{constructCluster}{HTML}{2B83BA}
\definecolor{mygray}{gray}{0.85}
\DeclareRobustCommand{\hlcyan}[1]{{\sethlcolor{mygray}\hl{#1}}}
\definecolor{definedGray}{HTML}{1C4966}

\newcommand{\basicalert}[2]{\fbox{\bfseries\sffamily\scriptsize\color{blue}#1}{\sf\small$\blacktriangleright$\textit{\color{red} #2}$\blacktriangleleft$} }

\newcommand{\RQone}{\textbf{RQ1}\xspace}

\newcommand{\RQthree}{\textbf{RQ3}\xspace}

\newcommand{\Mona}[1]{\basicalert{From Mona}{#1}}

\volume{00}

\firstpage{1}

\journalname{The Journal of Systems and Software}

\runauth{Moses Openja, Mohammad Mehdi Morovati, Le An, Foutse Khomh, Mouna Abidi}

\jid{jnltitle}

\CopyrightLine{2022}{Published by Elsevier Ltd.}

\usepackage{amssymb}
\begin{document}

\begin{frontmatter}



%% e.g. \dochead{17th International Conference on Dynamical Processes in Excited States of Solids}

\title{Technical Debts and Faults in Open-source Quantum Software Systems: An Empirical Study}

\author[swat]{Moses Openja\corref{cor1}} \ead{openja.moses@polymtl.ca}

\author[swat]{Mohammad Mehdi Morovati\corref{cor1}} \ead{mehdi.morovati@polymtl.ca}

%\author[swat]{Mouna Abidi\corref{cor1}} 
%\ead{mouna.abidi@polymtl.ca}

\author[swat]{Le An\corref{cor1}} 
\ead{le.an@polymtl.ca}

\author[swat]{Foutse Khomh\corref{cor1}} \ead{foutse.khomh@polymtl.ca}
%\address{River Valley Technologies, SJP Building, Cotton Hills, Trivandrum, Kerala, India 695014}

\author[swat]{Mouna Abidi\corref{cor1}} \ead{mouna.abidi@polymtl.ca}

%\address{River Valley Technologies, SJP Building, Cotton Hills, Trivandrum, Kerala, India 695014}
\address[swat]{SWAT Lab, \'{E}cole Polytechnique  de Montr\'{e}al}

%\address[fn1]{Department of Computer and Software Engineering, \'{E}cole Polytechnique  de Montr\'{e}al, Canada}

%\address[fn2]{1475 Ren\'{e}-L\'{e}vesque E, Montr\'{e}al, QC, H2L 2M4, Canada}
%\address[02]{SWAT Lab, \'{E}cole Polytechnique  de Montr\'{e}al, Canada}
%\address[03]{SWAT Lab, \'{E}cole Polytechnique  de Montr\'{e}al, Canada}
%\address[04]{SWAT Lab, \'{E}cole Polytechnique  de Montr\'{e}al, Canada}

\begin{abstract}
Quantum computing is a rapidly growing field attracting the interest of both researchers and software developers. Supported by its numerous open-source tools, developers can now build, test, or run their quantum algorithms.  Although the maintenance practices for traditional software systems have been extensively studied, the maintenance of quantum software is still a new field of study but a critical part to ensure the quality of a whole quantum computing system. In this work, we set out to investigate the distribution and evolution of technical debts in quantum software and their relationship with fault occurrences. Understanding these problems could guide future quantum development and provide maintenance recommendations for the key areas where quantum software developers and researchers should pay more attention. In this paper, we empirically studied 118 open-source quantum projects, which were selected from GitHub. The projects are categorized into 10 categories. We found that the studied quantum software suffers from the issues of code convention violation, error-handling, and code design. We also observed a statistically significant correlation between code design, redundant code or code convention, and the occurrences of faults in quantum software. 

\section*{Highlights}

\begin{itemize}
   \item Quantum software suffers from code convention, error handling, and code redundancy.
   
   \item There is a correlation between technical debts and fault in quantum software.
   
   \item Quantum developers should use existing static analysis tools to examine their code. 
   
   \item New tools should be introduced to support identifying quantum-specific problems. 
   
   \item Future works should study other aspects of maintenance and reliability of quantum.

\end{itemize}

%\begin{itemize}
%   \item The quantum software systems suffer from code convention violation, error-handling, and code redundancy, mostly in the initial versions.
   
%   \item There is a statistically significant correlation between technical debts (such as code convention, error-handling, redundant code, cognitive complexity of code) and fault occurrences in quantum software systems.

%   \item The quantum software developers should use the existing static analysis tools to examine their code. Code reviewers and quantum quality assurance team should pay attention to the code quality and code size, especially when new files are added. 
   
%   \item New tools should be introduced to support identifying quantum-specific problems, such as the technical debts and faults that only occur in a quantum software system. 
   
%   \item Future works are appealed to study other aspects of quantum software in terms of maintenance and reliability, such as code review, verification methods to ensure the correctness of a quantum program, and practical fault detection techniques for supporting quantum systematic testing and debugging.
%\end{itemize}

\end{abstract}

\begin{keyword}


Quantum Computing \sep Technical Debts\sep Software Bugs \sep Software Maintenance \sep Software Reliability
\end{keyword}

\end{frontmatter}

% \linenumbers

\section{Introduction}
Quantum Computing is a paradigm that intersects computer science, mathematics, and physics~\cite{spector:1999:quantum, kaye:2007:introduction}. Unlike other computing fields, quantum computing uses the law of quantum mechanics with the goal
% \Foutse{the goal?}  %concern 
to achieve high computation efficiency. In contrast, \emph{quantum software} can be software applications that run quantum algorithms, provide a platform for testing and simulating quantum algorithms, or govern quantum computers' operations. The example of operational quantum software include
applications developed for
% \Foutse{applications developed for?} 
checking and correcting errors, maintaining the stability of quantum computers or software for supporting complex and highly computational tools such as medical equipment.

% Software maintenance support is crucially an important aspect 
% of any computer 
% in software development. 
Software maintenance support is considered one of the most important aspects of the software development process. 
% The maintenance effort is well-known to take between $60$\% and $80$\% of the total cost of software development
It is a well-known fact that between $60$\% and $80$\% of the total software development cost is taken by the maintenance phase
~\cite{vandoren:1997:maintenance}. 
Previous studies have shown that 
modifying and revising previously released software's versions are the main activities of software maintenance~\cite{ieee:1990:ieee,kruger:2020:quantum,perez:2020:reengineering}.
%Software maintenance activities include modifying and revising the previously released software versions~\cite{ieee:1990:ieee,kruger:2020:quantum,perez:2020:reengineering}. 
Therefore, to ensure a %one key contributor to produce 
maintainable software, developers have % is 
%the design and the implementation of the software source code, for example, that can
to design and implement the source code, in a way that facilitates future modifications and evolution changes. %to be easily changed or modified.
%easily be changed or modified. 
However, maintenance activities have never been a straightforward task, it is considered as the most time/effort-consuming and complex task throughout the life cycle of software development~\cite{pigoski:1996:practical}.
The maintenance practices for traditional software systems have been extensively studied~\cite{Andrew:2006:Exploratory-study,Gyimothy:2005:Empiricalvalidation,Xiong:2009:open-source,raymond:1999:cathedral,pigoski:1996:practical,Lenarduzzi:2019:Empirical-TechnicalDebt,Zengyang:2015:MappingStudyonTechnicalDebt,cataldo:2008:socio,saika:2016:developers}. These studies focus on a broad range of areas, %related to maintenance of traditional software systems, 
including examining source-code complexity and software enhancement effort~\cite{banker:1998:software}, estimating the reliability in the maintenance phase~\cite{mateen:2016:estimating}, studying the distributions and evolution of technical debts~\cite{Fowler:Technical-debt:2019, Antonio:TechnicalDebt:2015, Reimanis:Technical-Debt:2016, kruchten:2012:technical}, and predicting the prevalence of bugs using anti-patterns~\cite{taba:2013:predicting,ubayawardana:2018:bug,cairo2018:impact} among others. 

To the best of our knowledge, there is no prior study exploring maintenance efforts for 
%the new types of software systems such as 
quantum software yet. %In particular, studies on quantum software maintenance are yet emerging. 
The recent emerging studies~\cite{perez:2020:reengineering,kruger:2020:quantum} related to quantum software systems maintenance mostly focused on re-engineering new quantum algorithms within traditional software systems. For example, Pérez-Castillo~\cite{perez:2020:reengineering} proposed a model-driven re-engineering~\cite{seacord:2003:modernizing} that allows the migration of classical or legacy systems together with quantum algorithms and the integration of new quantum software during the re-engineering of classical or legacy systems while preserving knowledge. While maintenance effort is a broader dimension, examining the maintenance in terms of the technical debt composition in a quantum software system is one direction to understand how this software is being maintained. Studies~\cite{Antonio:TechnicalDebt:2015,Cunningham:TheWyCash:1992,Lenarduzzi:2019:Empirical-TechnicalDebt,Zengyang:2015:MappingStudyonTechnicalDebt,Fowler:Technical-debt:IsHighquality:2019} have shown that technical debts provide relevant and actionable insight into the design and implementation deficiencies of software systems. For instance, according to Martin Fowler~\cite{Fowler:Technical-debt:2019}, technical debts indicate the internal quality issues of software that make the software more difficult to modify or develop (for example the source code conventions for readability~\cite{smit2011:maintainability}). Understanding the distribution and evolution of technical debts in quantum software could %help %Exploring the related study on quantum systems can 
guide future development and provide maintenance recommendations for the key areas that may require further attention to both the practitioners and researchers in quantum computing. In this paper, we examine the distribution and evolution of technical debts in quantum software and their relation with fault occurrences. In particular, we answered the following research questions:

\begin{enumerate}

    \item[\textbf{RQ1)}]  \textbf{What Are the Characteristics of Technical Debts in Quantum Software?} 
	
	We examined the distribution of technical debts in quantum software systems represented as code smells and coding errors and their severity (categorise as critical, major, minor, and blocker)~\cite{Sonarqube:issues:2020}. We summarized the technical debts based on the types of technical debts and highlighted the critical debts. Results show that about 80\% of the technical debts are related to the code smells and more than half of technical debts in all software types %(quantum and non-quantum) 
	belong to the major severity. The major severity are quality issues or flaw that can highly impact the productivity of developer, for example, an uncovered piece of code, unused parameters, or duplicated blocks. %\Foutse{what is 'major' category? is this a severity level? please clarify what it is!!!}. 
	In addition, we found that a few types of technical debts (such as `code convention' (problem with coding convention such as formatting, naming, white-space), `design issues' (\emph{e.g.,} duplicate string literals), `brain-overload' (related to cognitive complexity), and `error-handling') dominate the total number of technical debts.

    \item[\textbf{RQ2)}]  \textbf{How Do Technical Debts Evolve Over Time?} 
    
    We investigated how new technical debts are added into the code-base with respect to the total file size over time. We observe that technical debts tend to be added in the initial versions of a project (when most new codes and files are added). Besides, we found that LOC %and code complexity 
    can be considered as key indicators of the existence of technical debts in quantum computing software systems. 
%most of \Foutse{can we provide statistics? how much is 'most'?} the technical debts are associated with the software systems' initial versions (when most new codes or source files are added).
    This result is in line with the studies on traditional software~\cite{Arthur:Discovering:2017,Arthur-Jozsef:Longitudinal:2019}. We recommend quantum software developers pay more attention to the code quality and code size, especially when new files are added to the code base.
    
    \item[\textbf{RQ3)}] \textbf{What Is the Relationship Between Technical Debts and Faults?}
    
    In this research question, we used
    % Using 
    regression models to examine the correlation between technical debts (and their types~\cite{sqdocumentation}) 
    % \footnote{\url{https://docs.sonarqube.org/latest/user-guide/built-in-rule-tags/}}
 and fault-inducing commits in quantum software at the file level. Our results indicate a statistically significant correlation. Particularly, we found that the highest significance in all studied quantum software systems is related to `convention' and `unused' technical debts. 
    
\end{enumerate}

The rest of this paper is organized as follows. In Section \ref{sec:background}, we provide the background of this study. Then, we discuss the related works on quantum computing, technical debts, and fault analysis in the software systems in Section \ref{sec:related-works}. We describe the methodology that is followed in Section \ref{sec:study-design}. In Section \ref{sec:results}, we present the results of our analysis. In Section \ref{sec:discussion}, we discuss the implication of our findings. Section \ref{sec:threats} describes the threats to the validity of our study. Finally, we provide the conclusion in Section \ref{sec:conclusion}.

\section{Background}\label{sec:background}
The main focus of this study is to investigate technical debts and their evolution in quantum software and how the technical debts affect the reliability of quantum software. 
%explore the quantum software practices \Foutse{we should be more specific and state what we do...this is too broad and we don't really examine developers practices!!! we simply analyse technical debt and their evolution, and impact on reliability!!!} and understand the efforts related to software reliability and maintainability, how the practices differ from other types of software and identify any gap that may need improvement in order to develop a fully maintainable  quantum software. 
This section describes the background of quantum computing and quantum software development. %We then give the general background of technical debt and the sonar platform used in this study to identify the reliability and maintainability issues at the source code level. 

\subsection{Quantum Computing}

Quantum computers can adopt a superposition form of $|0\rangle$ and $|1\rangle$, which enables a qubit to exist in all of these states simultaneously. A qubit is a two-dimensional quantum-mechanical system, presenting sizeable information to process in quantum computing. This basic concept of superposition allows quantum computers to perform computations on an extensive scale in parallel (also known as parallel computation). 

Another property of a quantum computer is the entanglement: given a two-qubit system, it is possible that the state of each qubit cannot be described separately. In mathematical form, the state cannot be factorized as the tensor product of two separate states. Certain types of highly entangled systems are challenging to model for classical computers. 
Generally, it is also possible to efficiently describe systems that do not exhibit too much entanglement using classical computational methods such as tensor networks~\cite{ORUS:2014:117}. However, highly entangled systems are almost impossible to be modeled in classical computers. Consequently, quantum algorithms must exploit high amounts of entanglement to reach higher possible capabilities than classical algorithms. This entanglement property has applications in many aspects of quantum computing, such as cryptography~\cite{ekert:1991:quantum}, and quantum computation~\cite{nielsen:2002:quantum,shor:1999:polynomial}.

\subsection{Quantum Software Development}

Zhao, Jianjun.~\cite{zhao:2020:quantum} defined Quantum software engineering as \textit{``the application of sound engineering principles for the development, operation, and maintenance of quantum software and the associated document to achieve economically quantum
software that operates efficiently on quantum computers and is reliable.''} 

%\nd Like in classical computing, the role of quantum software development is to allow in creating quantum software using steps and procedures also know as quantum software life cycle(QSDLC)~\cite{ghezzi:1991:fundamentals} from the requirements analysis, design, and implementation to testing and maintenance\mehdi{I cannot understand the concept of this sentence completely.}. 
Like classical software, quantum software is also developed by following a series of steps: requirements analysis, design, and implementation to testing and maintenance. %These steps are known as quantum software life cycle (QSDLC)~\cite{ghezzi:1991:fundamentals}.
The quantum design stage provides the means for modeling and defining quantum software systems~\cite{Kiefl:2020,rez-Delgado:2020}. The testing phase aims to identify any flaws or defects in the quantum software and verify that the software's behavior reflects the documentation defined at the early phase of analysis. Finally, the maintenance as the last phase of quantum development process represents any changes or updates later when quantum software products are released. Indeed, to develop maintainable and reliable quantum software systems, there are numerous factors that developers must take into consideration and respect in every phase of software development, for example, defining system requirements and quantum design patterns, implementing test and fault detection models, implementing development environments, and selecting quantum programming mechanics.
%For example, defining the system requirements, the quantum design pattern, implementing the test and fault detection model or development environments, and quantum programming (mechanics)\mehdi{What are these items? are they the factors that mentioned in the previous sentence?}. 
For further readings on quantum software development, we refer our readers to~\cite{zhao:2020:quantum,rez-Delgado:2020, thompson:2018:quantum,nielsen:2002:quantum,piattini:2020:talavera} containing well documented set of guidelines to assist a well-engineered quantum software system development. 

\nd Quantum programming is the design and implementation of a program executable on a quantum computer to meet the computing need~\cite{ying:2016:foundations,miszczak:2012:high}. Every chunk of code in a typical quantum program consists of both quantum and classical instructions~\cite{cook:1997:finding}. A classical instruction uses classical bit registers to measure the qubit states, including conditional operation, whereas quantum instruction uses qubit registers to operate on the quantum computer.  Unlike in the early stage of Quantum Turing Machine (QTM)~\cite{deutsch:1985:quantum}, quantum programming majorly focuses on quantum circuit models. This is followed by new quantum programming models such as the Quantum Random-Access Machine (QRAM) model~\cite{knill:2000:encyclopedia} and pseudocode~\cite{knill1:996:conventions}. Instead of just designing the quantum circuit, a quantum program is built to run on a classical computer to control the quantum system. Numerous other quantum programming languages such as 
% `Qiskit' \footnote{\url{https://qiskit.org}}, 
%\href{https://qiskit.org}{`Qiskit'},
`Quipper'~\cite{green:2013:quipper}, 
% `ProjectQ'\footnote{\url{https://projectq.ch/}}, 
%\href{https://projectq.ch/}{`ProjectQ'},
Q\#~\cite{svore:2018:q} and `Scaffold'~\cite{javadiabhari:2015:scaffcc,abhari:2012:scaffold} are now available. These programming languages are built on top of the traditional programming languages such as C\#, Python, Java, C/C++, and Julia. For the complete list of the programming languages and their history, we refer the readers to the studies~\cite{cook:1997:finding,zhao:2020:quantum,ying:2016:foundations}.

\section{Related Works}\label{sec:related-works}

%\Le{I just found that Hausi Muller has quite some publications about SE for quantum computing. \url{https://scholar.google.com/citations?hl=en&user=8hyNFkYAAAAJ&view_op=list_works&sortby=pubdate}. If his research topic is very close to ours, he may possibly selected to review this paper. So perhaps, we check if any of his papers can be discussed.}
\nd In this section, we discusses the related works on quantum computing, technical debts, and analysis of fault characteristics.

\subsection{Quantum Computing}
Piattini et al. \cite{piattini2020quantum} studied the emergence of quantum computing software systems and quantum software engineering. They also explained that based on evidences, demand for quantum software systems will be increased dramatically during the next years of the current decade. Besides, they mentioned nine principles as the main principles of quantum software engineering.

Garhwal et al. \cite{garhwal2019quantum} studied various high-level quantum programming languages and identified the main features of each one. They categorized quantum programming languages into five different main classes (such as Quantum Object Oriented Programming Language, Quantum Circuit Language, etc). Next, they represented that QPL and QFC are the most popular quantum programming languages. 

Moguel et al. \cite{moguelroadmap} explained that quantum software systems in quantum software engineering need processes that require methodologies, like classical software systems. They identified that classical software engineering processes such as requirements specification, architectural design, detailed design, implementation or testing can be used to carry out activities of each process in quantum software engineering effectively. While, the methodologies for each process of quantum software engineering should be adopted based on the requirements of quantum software development. The reason behind this difference is the underlying model of computing being used in each one. In classical software systems, computation is done by a sequence of instructions manipulating the data and the final state should be the output of the program. Although in quantum computing there is not a sequence of instruction. System in quantum computing has a set of states and can be in all of them at the same time. Besides, system stops when a certain subset of system states is in the desired state.

The most related work is a recent study regarding engineering quantum software \cite{zhao:2020:quantum}. In this paper, the authors introduced a quantum software life cycle and named it as \textit{quantum software engineering}. They firstly explained quantum programming as the process of designing and building an executable quantum computer program.  
Quantum Software Requirements Analysis, Quantum Software Design, Quantum Software Implementation, Quantum Software Testing, and Quantum Software Maintenance have been reported as the main stages of engineering quantum software systems. It is also mentioned that quantum software testing is more difficult than classical counterparts that are resulted from the structure of quantum computing programs and also quantum computers.
Seven different types of faults (Incorrect quantum initial values, Incorrect operations and transformations,  Incorrect composition of operations using iteration, Incorrect composition of operations using recursion, Incorrect composition of operations using mirroring, Incorrect classical input parameters, and  Incorrect deallocation of qubits) are introduced as faults related to quantum software engineering to achieve a deep understanding about the behavior of faults in quantum computer programs. To 
detect
% come up with \Foutse{you mean to detect?} 
introduced faults, an assertion has been explained for each identified fault category.

Another related work to our study is the article discussing open-source software in quantum computing carried out by Fingerhuth et al. \cite{fingerhuth:2018:openSource-quatum}. In that article, a wide range of open-source software systems focusing on quantum computing are studied and the authors introduced four paradigms 
% (Discrete variable gate-model quantum computing, Continuous variable gate-model quantum computing, Adiabatic quantum computation and Quantum simulators)
to develop quantum computing software projects, in order to ease the understanding of quantum computing systems for computer scientists and software engineers. 
The first paradigm is the discrete variable gate-model quantum computing paradigm in which bits and logical transformations are replaced by qubits and a finite set of unitary gates respectively. 
Continuous variable gate-model quantum computing is another paradigm where the qubits are replaced by qumods taking continuous values. This paradigm is mostly regarding physics aspects of quantum mechanics and particularly, quantum optics. The third introduced paradigm is Adiabatic quantum computation that uses adiabatic theorem, a phenomenon from quantum physics, to find the global optimum of a discrete optimization problem. Last but not least, they discuss quantum simulators which are application-specific quantum devices. But this paradigm is different from the simulation of quantum computations. The author of this paper just reviewed the development process of quantum computing projects from a software engineering view point. 

Shaydulin et al.~\cite{shaydulin2020making} focused on the contributors of open-source quantum computing projects hosted on GitHub. They studied data of 146 contributors and surveyed 46 of them. Based on the analysis of the collected data, they mentioned that almost all (45 out of 46) of the contributors of quantum computing software systems did not receive formal training in quantum computation. Thus, they concluded that their lack of a good understanding of quantum computing may lead to poor quality software systems. Besides, they introduced the main challenges that developers are facing in open-source quantum computing software systems.

%\Le{This subsection is pretty well written. I was just wondering if there are other related works on quantum computing? For example, [32] "Open source software in quantum computing" should be discussed, also [59,86] can also be discussed as they are mentioned in Intro.} 
%\Le{Also, does our work differ from the above works?}

\subsection{Technical Debts}\label{subsec:related-techical-debts}
Cunningham \cite{cunningham1992wycash} introduced the concept of technical debt (TD) for the first time in 1992. Since that time, a number of studies have been carried out to shed light on the different properties of TD. Ageriou et al. \cite{avgeriou2016managing} found that although TD can yield some benefit in short term, it can increase the cost of changes in a long term. Martini et al. \cite{martini2015investigating} showed that the existence of TD in software systems is inevitable. Besker et al. \cite{besker2019technical} studied the strategies to prioritize the identified TDs to be resolved and the most important factors affecting this process. Besker et al \cite{besker2018technical} carried out another study to determine the refactoring cost and negative effects of different types of TD. Besides, there have been various tools for identifying TDs. As an example, Avgeriou et al. \cite{avgeriou2020overview} conducted a comparison of 9 TD tools and highlighted the main features of each one. They showed that SonarQube can best handle multiple programming languages, which is the reason why we used this tool in our study because our subject Quantum software is written in different programming languages.

% \href{}{\texttt{}}

%\Le{The following content is move from the original Background section. The discussion on TD should belong to related work. So please check if we can use any part of the content}
Technical debts in software imply quality issues, \emph{i.e.,} code quality was ignored to achieve a faster goal instead of using a systematic approach to reach the same purpose that would take longer time~\cite{Techopedia:Technical-debt:2017}. Therefore, this may involve additional rework such as code refactoring in a later time to achieve a better quality. For this reason, tools such as 
% `Squore1'\footnote{\url{https://www.squoring.com/}}, 
\href{https://www.squoring.com/}{`Squore1'},
`SonarQube', 
% `Kiuwan2'\footnote{\url{https://www.kiuwan.com/}}, 
\href{https://www.kiuwan.com/}{`Kiuwan2'}
or 
% `Ndepend3'\footnote{\url{https://www.ndepend.com/}} 
\href{https://www.ndepend.com/}{`Ndepend3'}
have been proposed both in academia and industry to assist in identifying technical debts. Such tools are built based on a set of quality models~\cite{Samadhiya:2010} to help practitioners reduce the time it takes to synthesize and make the best decision regarding a system's future quality by providing systematic and quantifiable metrics for the industry's best practice~\cite{Reimanis:Technical-Debt:2016}. For example, SonarQube, NDepend and Squore use the SQALE model~\cite{Letouzey:SQALE:2012} that is programming language independent, whereas tools such as Kiuwan uses the concept of Checking Quality Model 
(\href{https://www.kiuwan.com/docs/display/K5/Models+and+CQM}{CQM}).

\section{Study Design} \label{sec:study-design}
%%%% ------ study design section ----

\begin{figure}[ht]
%\scalebox{0.70}{
\center
\includegraphics[width=\linewidth]{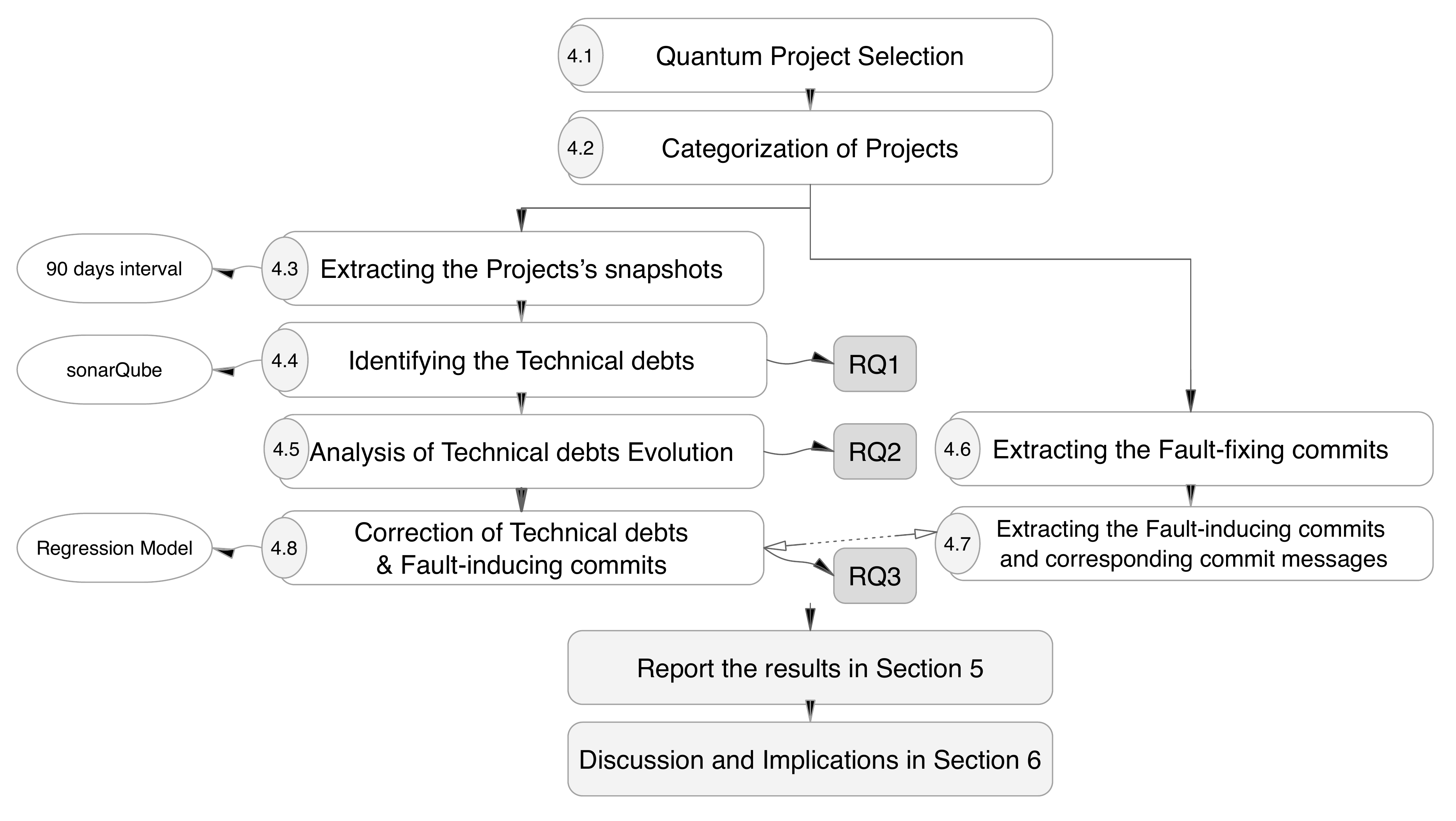}
\caption{An overview of our study methodology %\MO{Reviewer1: -Figure 1 has to be fixed since you are referring to wrong section identifiers. Section 5 reports the results while Section 4 describes the whole methodology.} \MO{ Reviewer2: In Figure 1, there are mistakes in section numbers: 3.x -> 4.x and "section 4" -> "section 5". Furthermore, MALLET is linked only to 4.10, while the library is also used in 4.9 to identify stop words.}}
%\ANLE{The image is not well rendered in my browser, please check the path
}
\label{fig:methodology}
%}
\end{figure}

%\MO{## Study Design - Looking at study design steps in Figure 1, why didn't you consider moving step 4.6 (RQ3) after step 4.8, since anyway the results from 4.8 were needed for addressing RQ3?}
This section presents the methodology we followed in this study. We used sequential mixed-methods to answer our research questions \RQone through \RQthree, from data collection, data processing, to quantitative and qualitative analyses~\cite{ivankova:2006:using}.
%collecting the required data, analysis, and applying both quantitative and qualitative data at various stages of our research process, which can help us well understand the studied problem~\cite{ivankova:2006:using}. 
%In each step, we first explored the domain to provide us with insights and then highlight the interesting cases using qualitative analysis. 
Figure \ref{fig:methodology} shows an overview of our methodology.% we followed throughout this study.

To avoid any ambiguity, we use the following terms throughout the paper to denote different kinds of defects identified by SonarQube and GitHub project's repositories.

\begin{itemize}
    \item \emph{Error}: Potential coding errors or bugs detected by SonarQube. These errors might break the code and have to be fixed as soon as possible. % that may break the code and require immediate action to be resolved.
    
%    \item \emph{SQ-issue}: Other coding problems detected by SonarQube, such as vulnerabilities and code smells.
    
    %An unexpected behavior in the software program due to the developer's coding mistake as detected by the SonarQube platform.
    
    \item \emph{Code smell}: They include poor coding decisions and any characteristics in the software source code  that indicate the possibility of a deeper problem, as reported by SonarQube \cite{brown1998antipatterns,fowler1999refactoring}.

    \item \emph{Fault}: Post-release bugs or developer/QA reported bugs. Only the GitHub issues \cite{github-issues} labeled as `bug' will be considered as faults. %\MO{check the definition.}
    
%    \item \emph{Fault-inducing}: %We used fault-inducing instead of bug-inducing to refer to the commits where the corresponding bug-fixing commits are responsible for introducing the bugs in the software system identified using developers' commit log history from GitHub projects repositories.
%    Commits where the corresponding fault-fixing commits are responsible for the introduced bugs in the software system identified using developers' commit log history from GitHub projects repositories.
    %\item \emph{fault-fixing}:  %Likewise, we used fault-fixing interchangeably with the bug-fixing to refer to the commits fixing the bugs introduced by a fault-inducing commit on GitHub repositories.
%    Commits fixing the faults introduced by a fault-inducing commit on GitHub repositories.
\end{itemize}

\subsection{Selection of Quantum Projects}

%\MO{Reviewer1: It is unclear to me the reasons why you did not use the topics assigned to projects hosted on GitHub during your project selection. Moreover, you need to justify the reasons why you looked at projects having at least 10 months of history. Why 10 months is enough for you? Probably 10 months will also include projects that are not mature enough.
%-Related to the previous point, I do not understand the reasons behind choosing projects with at least one release wince that the analysis being conducted on faults does not link faults to release. So how can you be sure that the faults being considered are post-release faults? Please clarify this aspect too.}

%\MO{Reviewer2: Step 4.1 (Project Selection)
%        - In this step, you consider three criteria for filtering repositories using GitHub API, where the third one filter non-English Readme files. How do you do that? Please include the GitHub API call in the paper, or the replication package?
%        - Moreover, you remove repositories with zero release, reasoning that such projects may not have post-release faults. But, as I understood, you never put into consideration such releases in any step. So it means, you treat all issues (including those before the first release) equally. so then what's the point of removing repositories with zero releases?}

The first step of our work is to select a list of open-source quantum computing projects. We searched quantum computing projects from GitHub, because GitHub hosts the largest collection of open-source software. Some of the famous quantum projects, such as \texttt{Qiskit} from IBM, are shared on GitHub. In the following sections, we will use this pattern \texttt{<author\_name>/<repo\_name>} to denote a GitHub project, whose URL will be \path{https://github.com/<author_name>/<repo_name>}.

Searching against the Rest API~\cite{github:api}
% \footnote{\url{https://developer.github.com/v3/}} 
provided by GitHub, we obtained a total of 1,364 repositories that: 1) contain the word `quantum' (case insensitive in either the repository name, descriptions, or project ReadME file); 2) are a mainline, not a forked repository; 3) have a ReadME or description written in English to provide us details about the project. We used the GitHub API (\texttt{:owner/:repo}) to extract the repositories descriptions and other meta-data then used to manually check if the descriptions are written in English. %\Foutse{please add details explaining how you check that projects have readme files written in English!!! A  reviewer explicitly asked about that...put the response here in the paper as well!}.  %Also, we followed the official GitHub API documentation\footnote{\url{https://docs.github.com/en/free-pro-team@latest/github/searching-for-information-on-github/searching-for-repositories}} to construct our search query.
We then use the following criteria to filter out our search results.

\begin{itemize}
	\item Following the idea of previous works~\cite{munaiah:2017:curating,Businge:2018:ICSME,Businge:2019:SANER}, we selected the repositories that have been forked more than once to reduce the chance of selecting a student's class assignment. This step removed 803 repositories and remained with 561 repositories.
	\item Our projects should contain enough history and development activities for the analysis of technical debts and faults. Thus, we only considered the repositories that have been created more than 10 months  %\Foutse{please provide a justification for this threshold of 10 months...one reviewer explicitly asked about it!  it would be nice to show the boxplot of the distribution of projects ages...may be that boxplot can help justify why 10 months makes sense!!!} 
	earlier (\emph{i.e.,} \texttt{2019-09-16})  than the date of this study. Inspired by~\cite{fingerhuth:2018:openSource-quatum}, we chose the projects that have at least 100 commits and 10 GitHub issues or pull requests. In addition, we limited our selection of projects to those with at least one release. %to further limit our studied projects to those
	These selection criteria allowed us to consider projects
	targeting the end-users because `never released projects' may not show representative data on faults experienced during the project development or by the end-users. This step further removed 417 projects and remained with 144 projects.
	\item We manually read the project descriptions and removed 13 repositories that are only related to quantum documentations or lecture notes.
We also identified and removed 9 more repositories that are not related to quantum computing but merely contain the word `quantum' in their descriptions,
such as \path{foxyproxy/firefox-extension} and \path{quantacms/quanta}. After this step, we finally obtained a list of 122 repositories.

%Using the above step, our final list of quantum computing projects repositories contains 122 repositories in total considered for the next step of analysis.
	
\end{itemize}

\subsection{Categorization of Projects} \label{subsec:category}

%\MO{Reviewer1: How are the categories being used in this study different from the ones proposed by QoSE? The only difference I see is about Quantum-Chemistry. Please rephrase explaining more the categorization process being followed.}

%\MO{Reviewer2: - Step 4.2 (Categorization of Projects)
%        - Does every project in the QoSF project list (https://qosf.org/project\_list/) included in the 118 projects you packed? If not, why?
%        - Please add the reference for "QoSF"
%        - I did not understand the definition of the "Assembly" category. What do you mean by "languages"? Does that mean the repositories in this category implement a quantum programming language?}

\nd In this step, we classified the selected quantum projects into different categories. %To construct the categories, 
We used the list of categories provided by the Quantum Open-Source Foundation 
% (QoSF)\footnote{\url{https://qosf.org/}}, 
(\href{https://qosf.org/}{\texttt{QoSF}})~\cite{QoSF:2021},
which is an initiative to promote the advancement of open-source tools for quantum computing. %During the category formation, 
To decide which category a project belongs to, we first checked whether a target project is listed by QoSF. If it is listed, we will directly use the QoSF provided category. Otherwise, two of the authors will independently read the project descriptions 
%provided in the GitHub repository 
and/or the project's official website 
%to understand what category the project should belong to.
to classify the project into one of the QoSF defined categories. The two authors will compare their results and resolve all discrepancies until reaching an agreement for all of the projects. 

During the categorization step, the authors removed four more projects either because their programming languages are not supported by SonarQube or because they are identified as experimental or toy projects, such as \path{Quantum-Game/quantum-game-2} and \path{PJavaFXpert/quantum-toy-piano-ibmq}.
%identified and removed 4 projects classified as `Fun/Game' from the list of studied repositories, because: 1) one of the projects was identified written mainly \texttt{Quantum-Game/quantum-game-2}\footnote{\url{https://github.com/Quantum-Game/quantum-game-2}} in vue programming language that our later analysis using sonarqube would not support; 2) another project \texttt{JavaFXpert/quantum-toy-piano-ibmq}\footnote{\url{https://github.com/JavaFXpert/quantum-toy-piano-ibmq}} was identified as toy-project. \mehdi{What about two other projects?}
% Our final set is created as a $10$-category list consisting of $118$ list of projects defined as follows:%\Le{We may use a table to display the categories}
The remaining 118 projects are classified into 10 categories as follows. Readers can refer to Appendix \ref{tab:example_repo} for a list of example repositories that belong to each of the categories.

%Our final list of classifying $118$ projects into $10$ categories is as below. 
%\mehdi{I am not sure that should we list a number of projects for each category as examples? We categorized and represented them as a table in Appendix}\Le{We should show the full list on GitHub or in a Google Sheet.}
\begin{itemize}
    \item \textit{Full-Stack Library or Framework:} This can be seen as all-in-one software containing all the frameworks and/or libraries required for building quantum applications. 
    
%    . An example of this category from list of our studied projects includes \texttt{ProjectQ}\footnote{\url{https://github.com/ProjectQ-Framework/ProjectQ}}, \texttt{strawberryfields}\footnote{\url{https://github.com/XanaduAI/strawberryfields}}, \texttt{Cirq}\footnote{\url{https://github.com/quantumlib/Cirq}}, \texttt{qiskit-terra}\footnote{\url{https://github.com/Qiskit/qiskit-terra}} and \texttt{pyquil}\footnote{\url{https://github.com/rigetti/pyquil}}.
    
    \item {\textit{Experimentation:}} Tools that support experimentation of quantum systems or states, such as superconducting qubit systems and parametrization of a pulse. %The list of project belonging to this category include \texttt{artiq}\footnote{\url{https://github.com/m-labs/artiq}}, \texttt{PyQLab}\footnote{\url{https://github.com/BBN-Q/PyQLab}}, \texttt{Qlab}\footnote{\url{https://github.com/BBN-Q/Qlab}}, and \texttt{pyrpl}\footnote{\url{https://github.com/lneuhaus/pyrpl}}.
    
    \item {\textit{Simulator:}} Controllable quantum systems that enable users to study the quantum systems which are hard to study on actual hardware or in laboratory.% and model by supercomputers~\cite{buluta:2009:quantum}. Example projects of this category include \texttt{qutip}\footnote{\url{https://github.com/qutip/qutip}}, \texttt{OpenFermion-Cirq}\footnote{\url{https://github.com/quantumlib/OpenFermion-Cirq}}, \texttt{QuEST}\footnote{\url{https://github.com/QuEST-Kit/QuEST}}, and \texttt{SimulaQron}\footnote{\url{https://github.com/SoftwareQuTech/SimulaQron}}.
    
    \item {\textit{Cryptography:}} This class is related to the usage of quantum mechanics' characteristics to carry out cryptography tasks. %Example projects of this category include \texttt{QRL}\footnote{\url{https://github.com/theQRL/QRL}}, \texttt{codecrypt}\footnote{\url{https://github.com/exaexa/codecrypt}}, and \texttt{supernomad/quantum}\footnote{\url{https://github.com/supernomad/quantum}}.
    %\item {\textit{Fun/Game:}} Systems of this category implements games using quantum concepts(such as photons) and quantum measurement.
    
    \item {\textit{Quantum-Algorithms:}} Systems that execute a quantum computation on a quantum model (such as a quantum circuit model). They are designed majorly to solve the classical problems in a probabilistic fashion~\cite{montanaro:2016:quantum}. %We include in our list projects such as \texttt{qiskit-aqua}\footnote{\url{https://github.com/Qiskit/qiskit-aqua}}, \texttt{OpenFermion}\footnote{\url{https://github.com/quantumlib/OpenFermion}}, \texttt{grove}\footnote{\url{https://github.com/rigetti/grove}} and \texttt{netket}\footnote{\url{https://github.com/netket/netket}}.
    
    \item {\textit{Toolkit:}} A set of libraries and tools that help interact with different components of quantum systems mainly in research settings. %Example of the projects selected in this category include  \texttt{pennylane}\footnote{\url{https://github.com/XanaduAI/pennylane}}, \texttt{quimb}\footnote{\url{https://github.com/jcmgray/quimb}}, \texttt{quantumrandom}\footnote{\url{https://github.com/lmacken/quantumrandom}} and \texttt{qtt}\footnote{\url{https://github.com/QuTech-Delft/qtt}}.
   
    \item {\textit{Quantum Annealing:}} Systems that provide meta-heuristics for finding global minimum over a very large number of possible solutions by using quantum fluctuation-based computation~\cite{finnila:1994:quantum}. 
    %\item {\textit{Web Extension:-}} This category includes the web browser extensions, mainly for Firefox Quantum. 
    \item {\textit{Quantum-Chemistry:}} Projects that focus on the use of quantum mechanics in the chemical systems experiments.
    \item {\textit{Compiler:}} When a quantum computing algorithm is implemented on actual hardware, the circuits should be compiled for the restricted topology of the particular quantum chip used for execution. This category is related to the systems translating quantum circuits to the quantum assembly format.
    \item {\textit{Assembly:}} This class is related to quantum assembly languages used for describing quantum circuits. %\Foutse{what are the projects doing concretely? are they application development frameworks?}. 
    It is used in many quantum compilation and simulation tools as the intermediate representation to describe quantum circuits \cite{cross2017open}.

\end{itemize}

\begin{figure}[ht]
\center
\includegraphics[width=\linewidth]{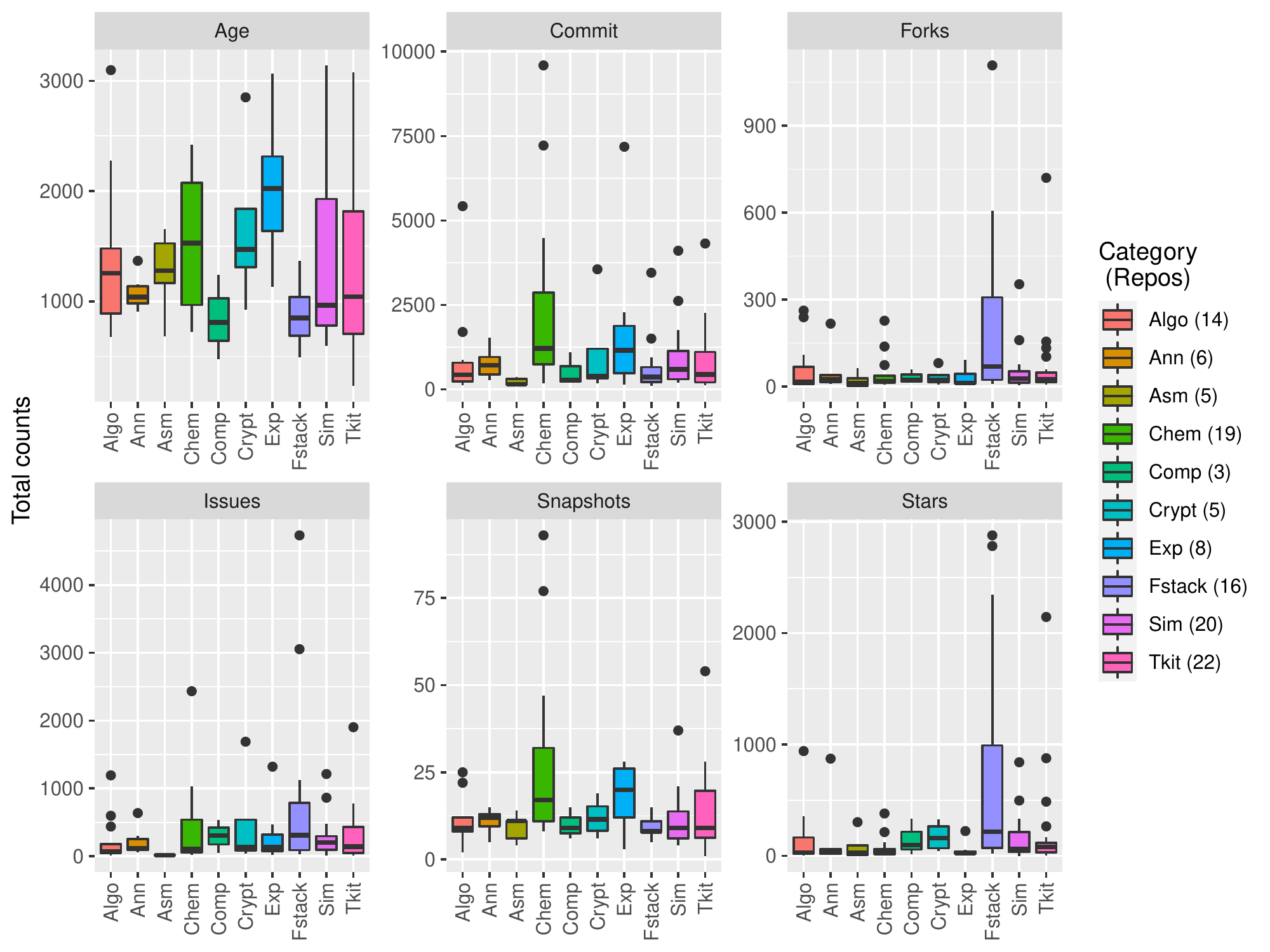}
\caption{Statistics of selected quantum projects by categories\\
(\textit{Commits}: number of commits, \textit{Age}: Time (in days) from the project creation date to its latest change, \textit{Issues}: number of issues reported on the GitHub repositories, \textit{Forks}:number of forks (popularity metric), \textit{Stars}:number of stars (popularity metric), \textit{Snaps}:number of snapshots extracted based on 90 days (extracted in Section~\ref{sec:snapshots}), \textit{Repos}: number of studied GitHub repositories in respective quantum category.)\\
   \textbf{Algo}:Algorithms, 
   \textbf{Ann}:Annealing, 
   \textbf{Asm}:Assembly,  
   \textbf{Chem}:Chemistry, 
   \textit{Comp}:Compiler, \textbf{Crypt}:Cryptography, \textbf{Exp}:Experimentation, \textbf{Fstack}:Full-stack library,
   \textbf{Sim}:Simulator,
   \textit{Tkit}:ToolKit.}
\label{fig:projects_category}
\end{figure}

The summary statistics of the selected repositories in the 10 categories are shown in Figure~\ref{fig:projects_category}.

\subsection{Extraction of Snapshots of Studied Projects}\label{sec:snapshots}

%\MO{Can you please add a little bit more information about the conjecture that there is not a huge amount of change among consecutive commits. It may be of interest to justify your conjecture while looking at a sample of your projects. -Related to the previous point, is commit-level enough considering that commits may differ based on the number of files being touched as well as the size of the changes (code churns being modified)? Please argue about this.}

%\MO{Step 4.3 (Extraction of Snapshots) - I need some clarification on Figure 2. First, how the average is calculated? Considering one category, we have several repositories. And for each repository, we have a set of numbers (which in this case are the number of commits between every X days). Is the presented "average" value, the "average of averages" or a "weighted average"? (Similar clarification is also needed for Table 1 and Table 3)}

%In this step, we extracted the studied systems' \emph{snapshots}, representing a project's history since the start of the development. 
Our study aims to analyze the change history of the selected quantum projects to investigate the maintenance effort over time. \emph{Git} allows us to take snapshots of a given project at a specific time period. We defined \emph{snapshots} as different copies of the same project. Each snapshot contains a set of changes made at that point of the project development.  We report the distribution of the commits in Figure~\ref{fig:projects_category} for the %the average total number of commits of the project %\Foutse{how do you compute this? please explain...it is average over what? number of versions? number of projects in the category? do you consider cumulative number of commits? or new commits between 2 versions? please provide details!!! I think one reviewer also asked about this!} % and thousand lines of code changes \Foutse{how do you calculate code changes lines? in the table kloc seems to be the number of lines of code not the number of code changes, please clarify this!!!} in a 
 given studied quantum category.  We use the following steps to extract the snapshots.

\begin{figure}
     \centering
     \begin{subfigure}[b]{0.80\textwidth}
         \centering
         \includegraphics[width=\textwidth]{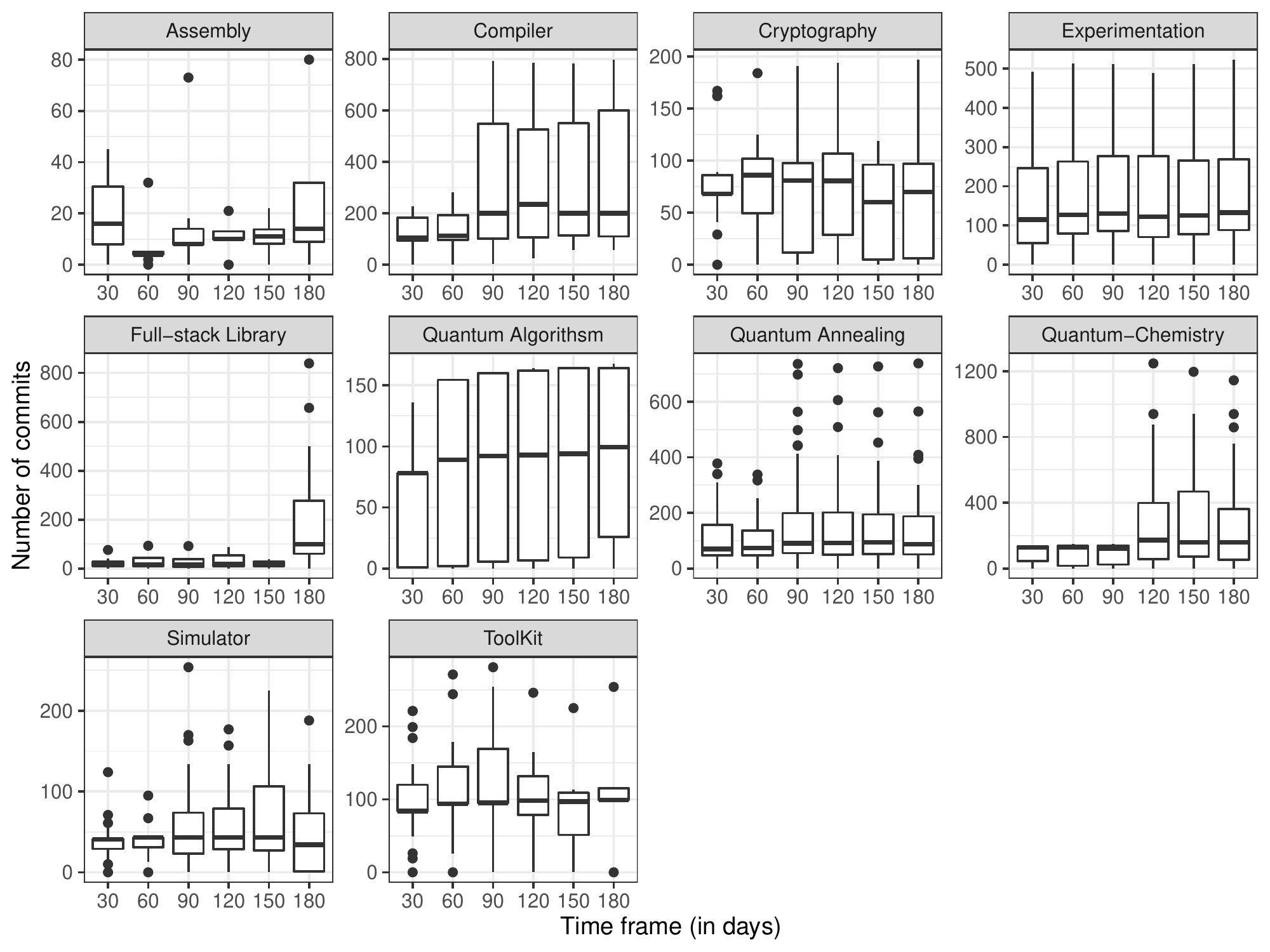}
         \caption{The distribution of commits across different time frames (in days) for the studied categories of quantum projects. The y-axis indicates the number of commits, and the x-axis indicates the time frames (in days) from 30 to 180 days with the interval of 30 days.}
         \label{fig:commits-dirstribution}
     \end{subfigure}
     \hfill
     \begin{subfigure}[b]{0.80\textwidth}
         \centering
         \includegraphics[width=\textwidth]{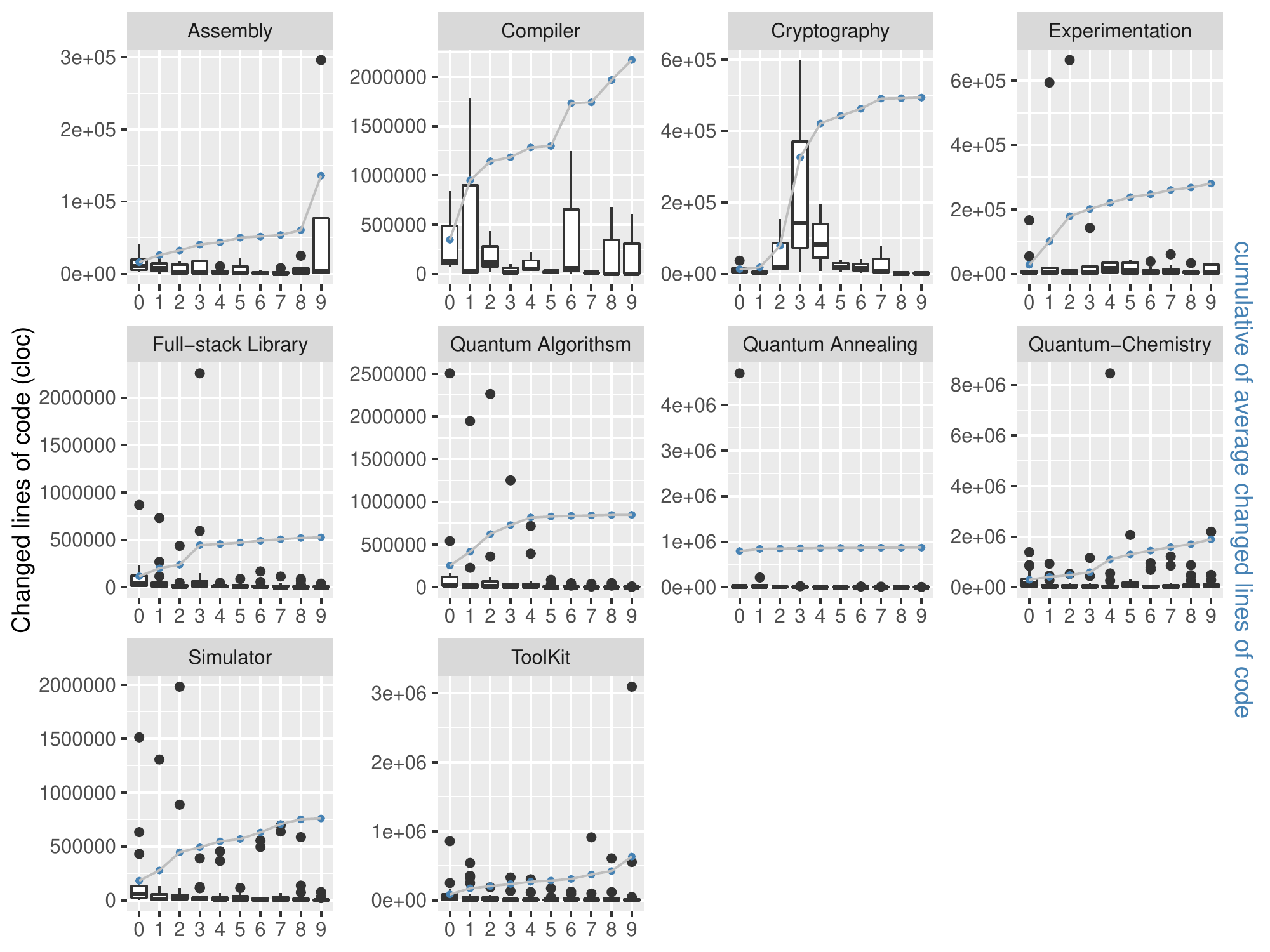}
         \caption{The distributions of lines of code for the first ten snapshot's of the studied projects, basing on the 90 days interval.}
         \label{fig:loc-dirstribution}
     \end{subfigure}
        \caption{The figures highlighting the distributions of commits and the consecutive changes. In the first part of the figures, we shows how the commits are distributions across different time frames from from 30 to 180 (in days) for the studied categories of quantum projects. The second figure indicate the changes in terms of changed lines of code (left) and the cumulative mean value of the changed lines of code (right) across the first ten snapshot's consecutive commits for 90 days interval.}
        \label{fig:commit-loc-distribution}
\end{figure}

\textbf{Choosing the Snapshot Period}. Studying snapshots from each of the commits will provide us with the most precise result, but this will also exhaust our computational resources, making it impossible to analyze all the 118 projects. Based on the idea of iterative and incremental development, we assume that developers of the studied projects did not make a large number of changes between consecutive commits. Thus, we can use a snapshot every $N$ days to represent the changes and development activities (such as fault fixes) during these days. To decide the best $N$, we investigated the distribution of commits between days interval \{30, 60, 120, 150, 180\} to be able to choose the appropriate number of days and extract project snapshots with considerable code-changes across all the studied projects. \fig\ref{fig:commits-dirstribution} shows how commits are distributed within the time frames of \{30, 60, 120, 150, 180\} days in the studied quantum categories.

For example, the box-plot corresponding to 30 indicates the number of commits every 30 days across all projects in the categories. Observing the commits distribution with the time-frames shown in the figure, we chose 90 days as the appropriate time frame, which represents the median size number of commits for most of the selected projects. This size of time frame also allows us to extract a number of snapshots from each project, which is feasible to conduct the following analyses based on our computational resources.  

\textbf{Extract Snapshots.} Using the time frame of 90 days, we identified the latest commits for every time frame (snapshots). We then used git archive (i.e., \texttt{https://github.com/<author\_name>/<repo\_name>/archive/<commit>.zip}) to create and download a zip file containing only the files under git data source from the starting of the project until the latest commit in a given snapshot.  %we executed the \texttt{git diff} command to extract the snapshots from each projects. 
Figure~\ref{fig:projects_category} shows the distribution of number of snapshots (and other metrics such as age) extracted from the projects categories. Considering that most of the projects we analyzed have at least 212 commits  (with maximum commits of 9,592) and the survival times (age) of 476 days at the point of starting this study. There is a trade-off when analyzing the number of revisions of each project %\Foutse{what do you mean by fewer to sufficient revision? please clarify!} 
of manageable size and code changes. Figure~\ref{fig:loc-dirstribution} illustrates consecutive changes %\Foutse{what is this capturing concretely? what do you mean by consecutive changes within commits? please clarify this computation and give the mathematical formula!} 
in terms of changed lines of code within the commits for the first ten snapshots based on a 90 days interval (i.e., code changes for the selected time frame). In Figure ~\ref{fig:loc-dirstribution} we also show the cumulative mean value of the changed lines of code (right) across the snapshot for the selected time frame.

%The minimum number of snapshots in our studied projects is three\Le{(project/name, which only has XX days of history)}. 

\subsection{Technical Debts in Quantum Software }

This section presents the steps carried out to identify technical debts from the snapshots extracted from the previous section. First, we give a general background of the SonarQube tool that we used to detect technical debts. Then we describe our steps for detecting the technical debts.

%the software source code's technical debts (other available tools for extracting technical debts in source-code are described in section \ref{subsec:related-techical-debts}). 

\subsubsection{SonarQube Platform} \label{subsec:background:sonar}

%\MO{Reviewer2: - Step 4.4.1 (SonarQube) - In the second paragraph, the explanations related to "tag" are very confusing. Please consider more clarification and possibly a reference to Table 6.
%        - Table 2: What is the "Impact" column?
%        - First sentence on page 10: "generated **issue**": Previously you mentioned that the terms "error" and "fault" will be used throughout the paper. So what does an "issue" here refer to?
%        - TDR: the definition provided for "devT" doesn't seem correct. I made a quick look online, and this is what I got: "devT refers to the amount of time already spent on the development of software system". Please consider revising the definition.
%        - What do you mean by "purpose-written program"?
%- Step 4.6 (Correlation of Technical Debts and Fault-Inducing)
%        - On page 11, for the first time, the notion of "code smell" and "code error" as elements of "technical debts" emerges, with no explanation. What I really miss is a subsection where all these relevant terms are explained, probably in the Background section. the distinction between code smells and error is critical as they are pivotal in the result discussion section (e.g., Section 5.1).}
SonarQube is an open-source platform used for monitoring source code quality and security. It uses static analysis to detect code smells, potential bugs (referred to as \emph{errors} in the rest of this paper), and vulnerability in software systems written with over 20 programming languages, including Python, C\#, C/C++, JavaScript, XML, and Java. It also allows creating plugins to support new programming languages. SonarQube uses a rule definition during source code analysis. An alert is raised in case a rule is broken. The properties of the created alerts include the type, severity, and effort needed to fix the alert. %Table \ref{table:sonar-severity} shows the severity rules defined by SonarQube. 

%\nd We evaluated the technical debts of quantum source code using these properties, where a type can be a code smell (maintainability) and or bug (SQ-bug) (presenting the reliability)\footnote{\url{https://docs.sonarqube.org/latest/user-guide/rules/}}.

 \begin{table*}[t]
     \caption{Severity rules define by SonarQube~\cite{sqdocumentation}.}
     \label{table:sonar-severity}
     \begin{adjustbox}{width=\columnwidth,center}
         \begin{tabular}{m{1.8cm}c c m{10cm}}\toprule
         %\Le{What does likelihood mean?}
         \textbf{Severity}  &  \textbf{Impact}& \textbf{Likelihood}&\textbf{Description} \\ \midrule
         \rowcolor{gray!10}
         BLOCKER&$\surd$ & $\surd$ & Higher impact probability to the behavior of the application, for example memory leak or unenclosed JDBC connection. The code needs to be fixed immediately.\\ %\hline
        
         %\rowcolor{gray!20}
         CRITICAL& $\surd$ &$\times$	& Either an error with a low probability to impact the behavior of the application or a security flaw such as an empty catch block or SQL injection.\\ %\hline
         \rowcolor{gray!10}
         MAJOR& $\times$ & $\surd$ & Quality issues or flaw that can highly impact the productivity of developer, for example, an uncovered piece of code, unused parameters, or duplicated blocks.\\ %\hline
        
         %\rowcolor{gray!20}
         MINOR&$\times$ &$\times$ &  Quality issues or flaw which can slightly affect the productivity of developers, for example, too long lines, or `switch' statements with fewer than 3 cases.\\ %\hline
        
         \bottomrule
     \end{tabular}
     \end{adjustbox}
 \end{table*}

%\begin{table}[t]
%    \caption{Severity rules define by SonarQube~\cite{sqdocumentation}.}
%    \label{table:sonar-severity}
%    \begin{adjustbox}{width=0.70\columnwidth,center}
%        \begin{tabular}{m{1.8cm}c m{10cm}}\toprule
%        \textbf{Severity}  &  \textbf{Impact}&\textbf{Description} \\ \midrule
%        \rowcolor{gray!10}
%        BLOCKER&$\surd$ & Higher impact probability to the behavior of the application, for example memory leak or unenclosed JDBC connection. The code needs to be fixed immediately.\\ %\hline
        
        %\rowcolor{gray!20}
%        CRITICAL& $\surd$	& Either an error with a low probability to impact the behavior of the application or a security flaw such as an empty catch block or SQL injection.\\ %\hline
%        \rowcolor{gray!10}
%        MAJOR& $\times$ & Quality issues or flaw that can highly impact the productivity of developer, for example, an uncovered piece of code, unused parameters, or duplicated blocks.\\ %\hline
        
        %\rowcolor{gray!20}
%        MINOR&$\times$ & Quality issues or flaw which can slightly affect the productivity of developers, for example, too long lines, or `switch' statements with fewer than 3 cases.\\ %\hline
        
%        \bottomrule
%    \end{tabular}
%    \end{adjustbox}
%\end{table}

SonarQube uses a risk estimation procedure to assign severity levels to the detected debt. %\Foutse{to what?}.
Table \ref{table:sonar-severity} illustrates the severity definition, extracted from the official SonarQube documentation \cite{sqdocumentation}. The columns `Impact' and `Likelihood' indicate how sonarQube assesses whether the severity of the debt is high or low. For example, evaluating if the code error can cause the application to crash or corrupt stored data (Impact) and the probability that the worst~\cite{bloch2003murphy}  %\Foutse{what is considered to be worst?}
will happen (likelihood), or assessing whether the detected code smell could lead to a error during the maintenance process. %\Foutse{what do you mean by this? what is the truth table? please clarify this sentence and provide a reference to support your statements!!!} truth table showing how SonarQube assesses the corresponding debt severity has a higher or low impact and the likelihood of the worst thing.
% \footnote{\url{https://docs.sonarqube.org/latest/user-guide}}. 
Also, sonarQube uses tags to categorize rules and debts. One or more tags are assigned to the debts inherited  %\Foutse{inherited??? it is not corresponding instead?}
from the rules that raised them. Moreover, tags denote different types of technical debts.
Also, each rule %\Foutse{each rule?}rules 
provides %\Foutse{provides?} 
an estimation function for determining the time needed to fix the %\Foutse{the corresponding error?} 
corresponding debts. %\Foutse{what is 'issue'? you said early in the paragraph that you would use the term 'error'!!! Please use a consistent terminology!}. 
The estimation %\Foutse{what is 'these' referring to? please be precise!!!!} 
usually offer either a function with a linear offset or a constant time per debt. % \Foutse{issue?}. 
For example, the rule `S3776' states that `the Cognitive Complexity should not be too high'. It generates code smells of the severity type \emph{CRITICAL} with a linear time to fix, which consists of a constant 5 minute time per issue, including one more minute for each additional complexity point over an established threshold.

Also, as part of the analysis results returned from running sonarQube, sonarQube %\Foutse{what is 'it' referring to here? SonarQube? please avoid fuzzy/imprecise terms!} 
calculates the effort required (in terms of time) %\Foutse{i.e., time required for?}) 
for fixing technical debts (TD) in a target system. Further, SonarQube %\Foutse{what is 'it'?} 
computes the technical debts ratio as $ TDR \, =\frac{Td}{devT}$; where $devT$ is the time estimated to develop the system, where a single line of code (LOC) is estimated to take 30 minutes.
%the software in which a single line of code (LOC) is estimated to \mehdi{require} 30 minutes of the development time. 
Then the $TDR$ is classified on a scaling of $A$ as the best ($TDR< 5\%$) to $E$ as the worst ($TDR \geq 50\%$). This detailed information, therefore, gives a high-level view of the system's internal quality.

The characteristics of technical debts important in our study are: \begin{inparaitem}
\item[1)] The types of debts: code smells (maintainability domain) and coding error (reliability domain). 
\item[2)] The severity of technical debts, introduced in Table~\ref{table:sonar-severity}.
\item[3)] The effort required to fix the technical debts mentioned above. \end{inparaitem} Providing these breakdowns can assist practitioners in prioritizing the allocation of resources to address critical debts.

We chose SonarQube in this study because: 1) It is broadly used for technical debts detection by over $200$ thousand users~\cite{sonarqube:official}
% \footnote{\url{https://www.sonarqube.org/}} 
and academic research setting~\cite{tan:evolution,marcilio:2019:static,saarimaki:2019:diffuseness,digkas:2017:evolution,digkas:2018:developers}; 2) The tool is based on a SQUARE quality model~\cite{Letouzey:SQALE:2012,letouzey:2010:sqale}, which is academically evaluated and published~\cite{dale:2014:impacts,letouzey:2012:managing}; 3) Our subject projects are written in different programming languages, which are supported by SonarQube.

\subsubsection{Detection and Analysis of Technical Debts}
\begin{figure}[ht]
\center
\includegraphics[width=\linewidth]{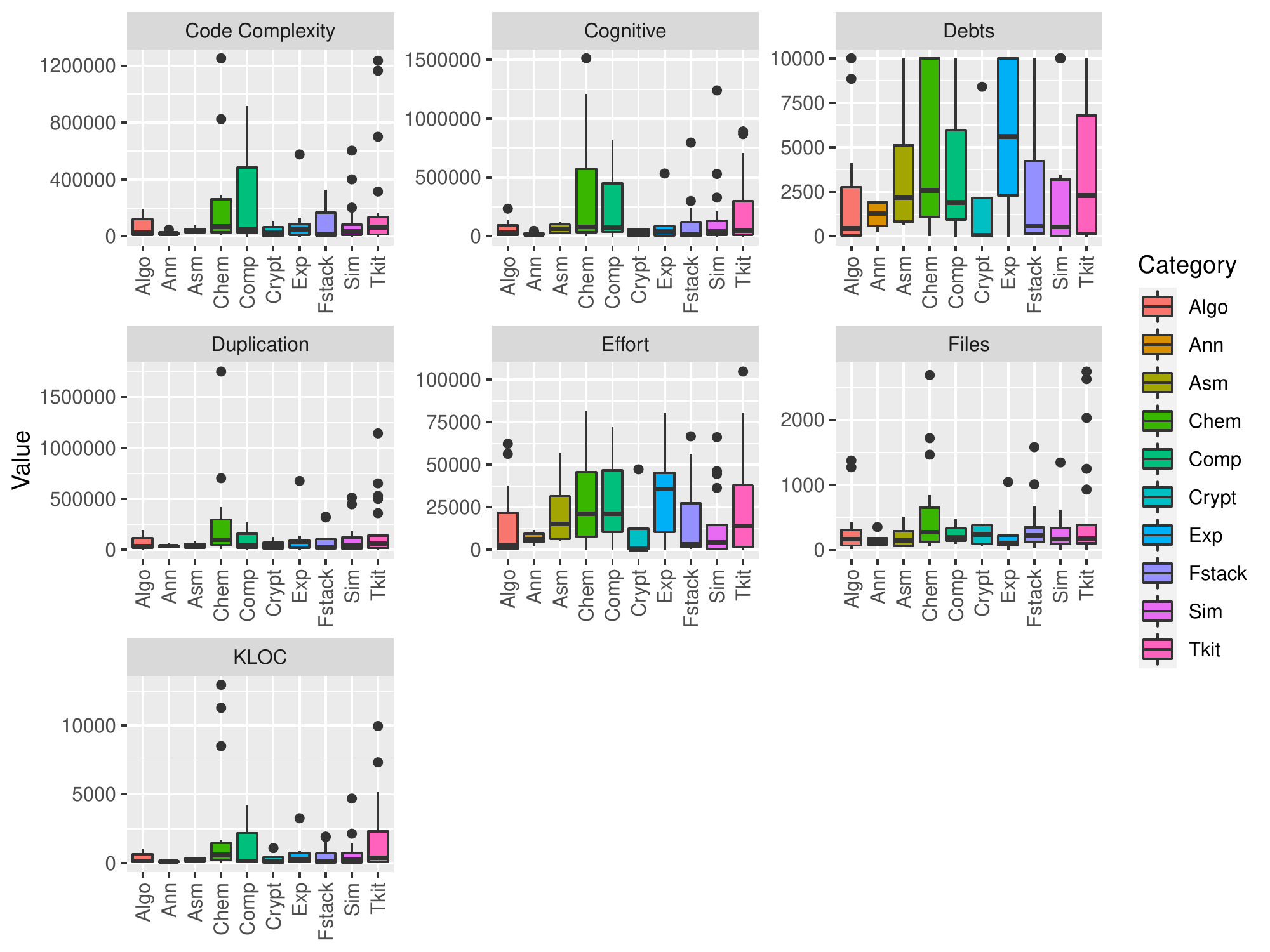}
\caption{Summary of the analysis results from SonarQube (
    \textit{Debts}: number of technical debts detected, 
    \textit{Effort}: estimated time (in minutes) to fix all reported technical debts, \textit{Complexity}: reported source-code cognitive complexity, 
    \textit{KLOC}: number of thousand lines of code  for the analyzed files, 
    \textit{Files}: number of unique files with at least one technical debt
    )\\
 \textbf{Algo}:Algorithms, 
   \textbf{Ann}:Annealing, 
   \textbf{Asm}:Assembly,  
   \textbf{Chem}:Chemistry, 
   \textit{Comp}:Compiler, \textbf{Crypt}:Cryptography, \textbf{Exp}:Experimentation, \textbf{Fstack}:Full-stack library,
   \textbf{Sim}:Simulator,
   \textbf{Tkit}:ToolKit.}
\label{fig:sonar-metrics}
\end{figure}

%To detect technical debts, we used SonarQube version 8.5. 
%We select this tool 
We ran SonarQube on every project's snapshots extracted in Section \ref{sec:snapshots}. The tool is configured on a local computer and used its web interface to monitor the results of the analysis. For data extraction and further analysis of the SonarQube results, we used a purpose-written program written with Python and Java (i.e., code written specifically to achieve the analysis goals guided by our RQ), making use of the SonarQube API and spreadsheet software as the primary storage. Figure \ref{fig:sonar-metrics} highlights the general summary of the metrics obtained after running SonarQube analysis.

To answer our \textbf{RQ1}, we investigated technical debts for every snapshot of the target project. %\Foutse{please clarify the following sentence...the correlations are between what and what? it is unclear...also please break down that sentence...it is too long with multiple 'and'!}
We used the Spearman test~\cite{zar2005:spearman} with a high correlation value ($>90$) between thousands of lines of code (KLOC) which was used as the proxy for application size and the total classes and functions for all $118$ target projects. In Section \ref{sec:composition}, we will discuss the results of this analysis. % based on the type of technical debts, severity, and the rules returned by SonarQube.

\subsection{Evolution of Technical Debts}

Previous studies~\cite{martini2015investigating, Arthur:Discovering:2017} have shown that refactoring efforts and architectural changes are some of the factors that influence technical debt in software projects compared to source code file size. Also, constraints such as time and budget have been described as some of the root causes of technical debt~\cite{Lenarduzzi:2019:Empirical-TechnicalDebt}. This step examines how technical debts (in terms of errors and code smells) evolve across the snapshots of every target project to help us understand how debt is introduced in quantum software source code. Specifically, we computed the technical debt ratio ($TDR$, described in Section \ref{subsec:background:sonar}) for the debts detected in each snapshot of the target projects. We also examined the addition of new codes (in KLOC) and how the codebase grows to identify the point in time when debts are introduced. Finally, we verify how the source code file size impacts the technical debts using a Spearman rank correlation. We present the detailed results of this analysis in Section \ref{sec:result-evolution} to answer our \textbf{RQ2}. %\MO{Reviewer1: -For addressing RQ2 you counted TDs over snapshots however it is unclear to me how the evolution has been evaluated. Specifically, have you looked at how many TDs have been removed and how many of them are instead new ones compared to the previous snapshot? If you only look at the absolute number you need to clarify this and explain the why. Moreover, you report relying on git diff but what does this mean? Have you run SonarQube only on changed files between consecutive snapshots? Why did you not use git checkout instead?}

\subsection{Identification of Fault-Fixing Commits}
%\MO{Reviewer1: -It is unclear the procedure used for identifying fault-fixing commits among the ones that are not traced in the issue tracker.}

%\MO{Reviewer2: - Step 4.7 (Identification of Fault-Fixing Commits) - Please provide details (e.g., keywords) to make the paper self-contained. - What is the final number of commits obtained? Including such commits as part of the replication package could be useful.}

One of the goals of this study is to examine activities that may introduce faults during quantum software development. 
To achieve this goal, we analyzed the fault-fixing and fault-introducing commits and their correlation with the overall technical debts discussed in the previous section. %Thus, we need a way to identify the bug-introducing and bug-fixes changes during the development of the target quantum software. 
We defined a \emph{fault-fix commit} as code changes to fix faults and a \emph{fault-inducing commit} as the code-changes that induced faults~\cite{Wen:2019:Bug-Fixing}. %The rest of this section describes the steps to identify these two kinds of commits.

%\nd To identify the bug-fixes we use two approaches: i) using the keywords (case insensitive)~\cite{Mockus:20000,Zhong:2015,Kim:2007:Cached-History} such as ``bug'' or ``fixes'' and ii) using the identifier/references to the bug reports such as \#131~\cite{Hipikat:2003,Fischer:2003,sliwerski:2005:changes-induce-fixes,Zhong:2015,Kim:2007:Cached-History}. The two approaches consider both the bugs reported in the bug report or bug-fixes that do not directly associate with the bug reports. On the other hand, a developer might fix bug in the source code not captured in bug tracking systems also known as `on-demand'~\cite{Zhong:2015}.

To identify the fault-fixing commits, we combine %\Foutse{combine?}
two approaches: 1) using a list of keywords such as ``bug'' or ``fixes'' as shown in Listing~\ref{list:bug-fixes}.  Our selection of keywords is based on %suggested in 
previous studies~\cite{Mockus:20000,Zhong:2015,Kim:2007:Cached-History,abidi2020multi} and ii) using the identifier/references to the bug reports within the commit message (i.e., references to the issue labeled as `bug') such as \#131~\cite{Hipikat:2003,Fischer:2003,sliwerski:2005:changes-induce-fixes,Zhong:2015,Kim:2007:Cached-History}. The two approaches above consider the faults that are reported in the bug tracking system (for our case GitHub issues tracker) and those that are not reported in the system, because developers might fix a fault in the source code which is not captured by such systems (also known as `on-demand')~\cite{Zhong:2015}. 

Using the two approaches described above we obtained a total of 33,917 unique fault-fixing commits that modified 49,593 unique files across all the studied projects. Next, we removed all the commits detected as fault-inducing and fault-fixing that touched only the files related to readMe/ documentation, git, or test cases. To this end, we checked if a given keyword (case insensitive) is within the file path or file name. In Listing~\ref{list:removed-bugs} we show the list of keywords used to match if the modified files by the fault-fixing commits are related to documentation, licence, test case, or git-related files. After cleaning the commits related to the documentation or test cases, the final list of the fault-fixing commits is 31,114 unique fault-fixing commits with 39,285 unique files modified.  %We eliminated the candidates with types in their commit messages and obtained a list of 47,009 fault-inducing commits.

%\Le{I have quite some worries here}

%The list of keywords similar to study~\cite{abidi2020multi} used for our first approach is shown in List \ref{list:bug-fixes}.

%%% code listing:
 \begin{figure}[ht]
 \center

 \lstset{backgroundcolor = \color{gray!4},language=Java,basicstyle=\small\ttfamily, showspaces=false, showstringspaces=false,breaklines=true}
 \begin{lstlisting} [language=Java, caption={The keywords used to extract fault-fix commits}, label={list:bug-fixes}]
 "fixed ", "fixes ", " fixed", "crash", " resolves",  "fall back",  "coverity", "reproducible", "stack-wanted", "failur", "fail", "npe ", "except", "broken", " bug", "error", "addresssanitizer", "hang ", "permaorange", "random orange", "intermittent", "steps to reproduce", "crash", "assertion",  "failure", "leak", "stack trace", "heap overflow", "freez",  "fix ",  " problem", " overflow",  " issue",  "workaround ", "break ",  "stop"
 \end{lstlisting}
 \vspace{-0.45cm}
 \end{figure}

 \begin{figure}[ht]
 \center

 \lstset{backgroundcolor = \color{gray!4},language=Java,basicstyle=\small\ttfamily, showspaces=false, showstringspaces=false,breaklines=true}
 \begin{lstlisting} [language=Java, caption={The keywords used to removed the fault-fixing and fault-inducing commits related to documentation or test case}, label={list:removed-bugs}]
 `readme', `doc', `test', `.png', `.pdf',  `.jpeg', `.jpg', `.gif', `licence',
 `.txt', `.md', `.git', `changelog', `.rst', `.bib', `.json', `tox.'
 \end{lstlisting}
 \vspace{-0.45cm}
 \end{figure}

\subsection{Extracting the Fault-Inducing Commits} \label{sub:bug-inducing}

This step aims to identify the fault-inducing commits and use the information to check for the correlation with the debts in the quantum software systems. We may not directly know how faults are introduced, but we can extract characteristics of the fault-inducing commits. We used the SZZ algorithm~\cite{kim:2006:automatic-bug-introducing} to identify fault-inducing commits using the fault-fixing commits.

Given the fault-fixing commit, SZZ will track the fault-inducing predecessor lines to lines modified in the fixing commit within a software repository. As a step in SZZ, for every fault-fixing commit, SZZ will identify all previous commits that changed the same lines of code using \texttt{git blame} command, resulting in a set of a fault-inducing commits that might have introduced the fault. Next, the identified commit time is compared to the time when the corresponding bug reports were submitted to determine if it should be ruled out as fault-introducing or not. The commit created later than the submission time is inducing commit if it is either a partial fix-  %\Foutse{please consider revising the following sentence..do you mean that it didn't fully resolved the bug?} 
didn't fully resolve the fault as evident in the later fault-fixing commit for the same issue or is responsible for another fault. Hence, separate fault-fixing commits may originate from similar fault-inducing commits that have modified related files. Many previous studies such as~\cite{Bernardi:2012:Bug-Mozilla,Asaduzzaman:2012:Bug-Android,Bavota:2012:Bug-Refactoring,Canfora:2011:Bug-Survive,Ell:2013:Bug-networks,Eyolfson:2011:Bugginess,Kamei:2013:Bug-Quality-Assurance,Kim:2006:Bug-Fix,Kim:2008:Buggy,Rahman:2011:Bugs-Ownership, undefinedliwerski:2005:Risk,tufano:2017:empirical,Wen:2016:Locating-Bugs,Rongxin:2018:ChangeLog,yin:2011:fixes} have also leveraged the SZZ algorithm to detect fault-inducing commits from a fault-fixing commit. Also, many tools have been proposed implementing the SZZ algorithm such as Pydriller~\cite{Spadini:PyDriller:2018}, Commit Guru~\cite{Rosen:Commit-Guru:2015} or SZZ Unleashed~\cite{Borg:SZZ-Unleashed:2019}. In this study, we used the Pydriller framework to detect the fault-inducing commits  %due to its easy to adapt and customized in python programming language.
because it is very convenient for mining software repositories and provides a set of APIs to extract important historical information regarding commits \cite{Spadini:PyDriller:2018}. %\Foutse{one reviewer asked how ambiguous cases were handled...we didnt respond to that here! did you assessed the accuracy of this heuristics? you could check a sample and report the accuracy obtained on that sample!}}
 
To verify the accuracy of the extracted fault-inducing commits, we performed a manual analysis on the sampled fault-inducing commits. We randomly selected 382 fault-inducing commits based on 95\% confidence interval and manually checked if the changes in the sampled fault-inducing commits are indeed related to the modifications performed in the corresponding fault-fixing commits.  Two authors discussed and assigned each commits with tags as either ``True'', ``False'', or ``Unclear''. In this case, the tag ``True'' was assigned for a situation where the authors were convinced that the change performed in fault-fixing was indeed related to the changes applied in fault-inducing. While ``False'' was used for the case where the changes are not related. Finally, a tag ``Unclear'' was attached in situations where the authors could not completely relate to a tag ``True'' or ``False''. The two authors discussed any conflicts until reaching an agreement. We calculated the precision considering only the True and False tags for data emerging from our manual validation and found a precision scores of 84.8\%.
 
Next, we used the GitHub API (\texttt{:owner/:repo/commits/:id}) to extract the commit messages and other related information such as commit author from the candidate fault-inducing commits. We eliminated the fault-inducing commits where the modified files are not the actual source code but are related to documentation, licence, test case, or git-related files shown in Listing~\ref{list:removed-bugs} and obtained a list of 51,269 unique fault-inducing commits that modified a total of 56,003 unique files across all the studied projects.

\subsection{Correlation of Technical Debts and Fault-Inducing}\label{subsec:predict-step}

%\MO{Reviewer1: -Section 4.6 has to be improved a lot. The metrics reported in Section 4.6 differ from the one used while reporting your results where you show types of TDs as reported by SonarQube. Moreover, you never report about which of the independent variables have been removed due to collinearity. -It is unclear which data points you used for building regression models. Is a data point corresponding to a snapshot? Please clarify.}

\nd This step aims to investigate how technical debts and fault-inducing changes are correlated. Examining the relation between them can show us the potential impact of technical debts. %gives us a clear information about the main factors leading to bugs in quantum software systems. 
This information can also provide explanatory answers regarding the potential relationship between maintenance and reliability activities.
% more clear understanding of what factors lead to bugs in the quantum software system that may be controlled and gives explanatory answers to the relation between the maintenance and reliability activities. 
To predict the occurrence of faults due to technical debts, we built multiple regression models on the set of technical debt metrics derived from the results of \textbf{RQ1}, which will be described in Section \ref{sec:composition}. We also used the number of fault-inducing commits (described in Section \ref{sub:bug-inducing}) as the dependent variable. Each data point is derived at the files level of a given project's snapshot. The metrics used in the regression model include:

\begin{itemize}

    \item \emph{File\_ID:} A unique identifier is sequentially assigned to every new file added to the target project since the beginning of the development. We used this identifier to map the file history activities, such as a renamed file or deleted files. To assign the identifier, we first identified all code changes at the beginning of the project development and sequentially assigned a number (from 1) to all unique added files within the commit. We then checked for every new commit added to the project repository, and used the \texttt{git diff} command to detect any code modifications. For a renamed file, we mapped all file paths with the original file path and assigned the previous identifier to the file. For an added file, we gave a new sequential number. 
    The \emph{File\_ID} was used to map the technical debts of the same files that may have been renamed. 
    
    \item \emph{File\_size:} number of lines of code in a file as computed by SonarQube.  This metric was used as the control variable when examining the relationship between %\Foutse{ratio?} 
    technical debts and fault inducing occurrence. %\Foutse{isnt it instead being used as control variable when examining the relationship between technical debt and fault occurrence?}
     %\Foutse{Moses is that what you meant?}%Sum of the added and removed lines of code to a given file in the target project. This metric was used as the control variable. \MO{We actually used KLOC, thousands of LOC returned by sonarqube as mentioned in the upper-scope.}
    
    %\item \emph{Total\_debts:} number of code smells and errors.%/bugs \Foutse{you meant errors right? since you said 'fault' would be used for post release bugs!} reported in a file of the target project by SonarQube.
    
    %\item \emph{Smells:} Total number of code smells reported in a file of the target project by SonarQube.
    
    %\item \emph{Errors:} Total number of potential errors reported in a file of the target project by SonarQube.
        
    \item \emph{Is\_smelly:} A dummy value $1$ if at least one smell was reported in the file, otherwise $0$.
    
    \item \emph{Is\_erroneous} A dummy value $1$ if at least one error was reported by SonarQube in the given file, otherwise $0$.
    
    %\item \emph{Severity:} Total number of technical debts classified as critical, major, minor and blocker reported in a given file by SonarQube. 
    
   \item \emph{Tag:} Number of different types of technical debts in a specified file. The full list of types is presented in Table \ref{table:minimun-tags} such as `accessibility', `brain-overload', `clumsy', `redundancy'. 
\end{itemize}
%\Le{I changed the name of the metrics. We need to change the metric names in the result section accordingly}

\paragraph{\textbf{Approach:}} We built Multiple Linear Regression (MLR) models for individual categories and a model for all combined quantum projects using the independent variables discussed above and the total number of fault-inducing commits (identified in the following Section \ref{sub:bug-inducing}) as the dependent variable. We used a separate model to examine how the technical debt metrics impact each category of the target projects. The combined category model was used to investigate 
the effect of each metric on the faults for all the target quantum projects.
% each metric's impact on bugs for all the target quantum projects.

\paragraph{\textbf{Model Building and Evaluation:}} We follow two steps suggested in previous studies~\cite{shihab:2010:understanding,taba:2013:predicting} to build our MLR models: removing the independent variables and analysis of multicollinearity.

\nd \textbf{The independent variable removal} step eliminates the independent variables that are not statistically significant based on the dependent variable. For this case, we built a multiple regression model using the aforementioned metrics as independent variables and %whether a commit is fault-inducing\MO
the number of fault-inducing commits as the dependent variable. We repeated this process and removed independent variables that have $p$-value $\geq$ 0.1, representing statistically non-significant variables.

\nd \textbf{Multicollinearity} is a phenomenon that happens when two or more independent variables used in a regression model are highly correlated. This is an issue because it increases the variance of the regression coefficients, making them unstable. To check for multicollinearity in the models, we used the variance inflation factor (VIF). The VIF score of an independent variable represents how well the variable is explained by other independent variables. Therefore, the higher the value of VIF for an independent variable, the higher is the multicollinearity with that particular variable. 

%The VIF score of each variable indicates the extent to which the o what extend the collinearity explains its variance with other variables. 
In this study, we follow previous works~\cite{taba:2013:predicting} and filter out %considered that a pair of 
independent variables for which %the are not correlated if their 
VIF $>$ 2.5. 
%We set a VIF threshold similar to the previous study!\cite{taba:2013:predicting}. 
We used the \texttt{vif} function of the \texttt{car} package in R~\cite{fox:2011:multivariate} to calculate VIF scores. 

Next, we built the final model on the final set of independent variables. %For model evaluation, 
One of the parameters we used to assess MLR model's performance is the adjusted $R^{2}$ value instead of $R^{2}$. The adjusted $R^{2}$ is recommended to be more accurate for model assessment~\cite{hastie:2009:elements} because it reflects the complexity of models. Besides, additional independent variables with lower explanatory power can directly reduce the value of $R^{2}$, which can be mitigated by using the adjusted $R^{2}$.
We used the $F$-test \cite{wiki:ftest} to measure whether any of the independent variables are significant in the model. 
Another metric we used is Akaike’s Information Criteria (AIC), which aims to add a penalty to the low importance variables of the model. Likewise, for the model selection, we considered the models with lower AIC values. 
We will discuss the results of this analysis in detail in Section \ref{sec:result-correlation} to answer \textbf{RQ3}.
% In section \ref{sec:result-correlation} we discuss into details the results of this analysis to answer our research question \textbf{RQ3}.

\section{Study Results}\label{sec:results}

This section presents the results of the analysis described in Section \ref{sec:study-design} answering our proposed questions \textbf{RQ1} through \textbf{RQ3}. %\MO{Reviewer2:  - Though a bit too low-level, I really enjoyed the level of detail provided in the discussion. Do you have any explanation on what are the factors that make projects in one category less issue-prone than another? In the "Compiler" category example, for instance, you bring up the prevalent issues related to that category (error handling, graph generation). Can we say something similar about other categories?}

\subsection{\textbf{RQ1: What Are the Characteristics of Technical Debts in Quantum Software?}}\label{sec:composition}
%\MO{Reviewer1: -While reporting RQ1 results in Table 4 you focus on effort to fix TDs what about the total number of TDs? -For RQ1, have you looked at the last snapshot of each project in your dataset? Please explain.}

\begin{figure}[ht]
\center
\includegraphics[width=\linewidth]{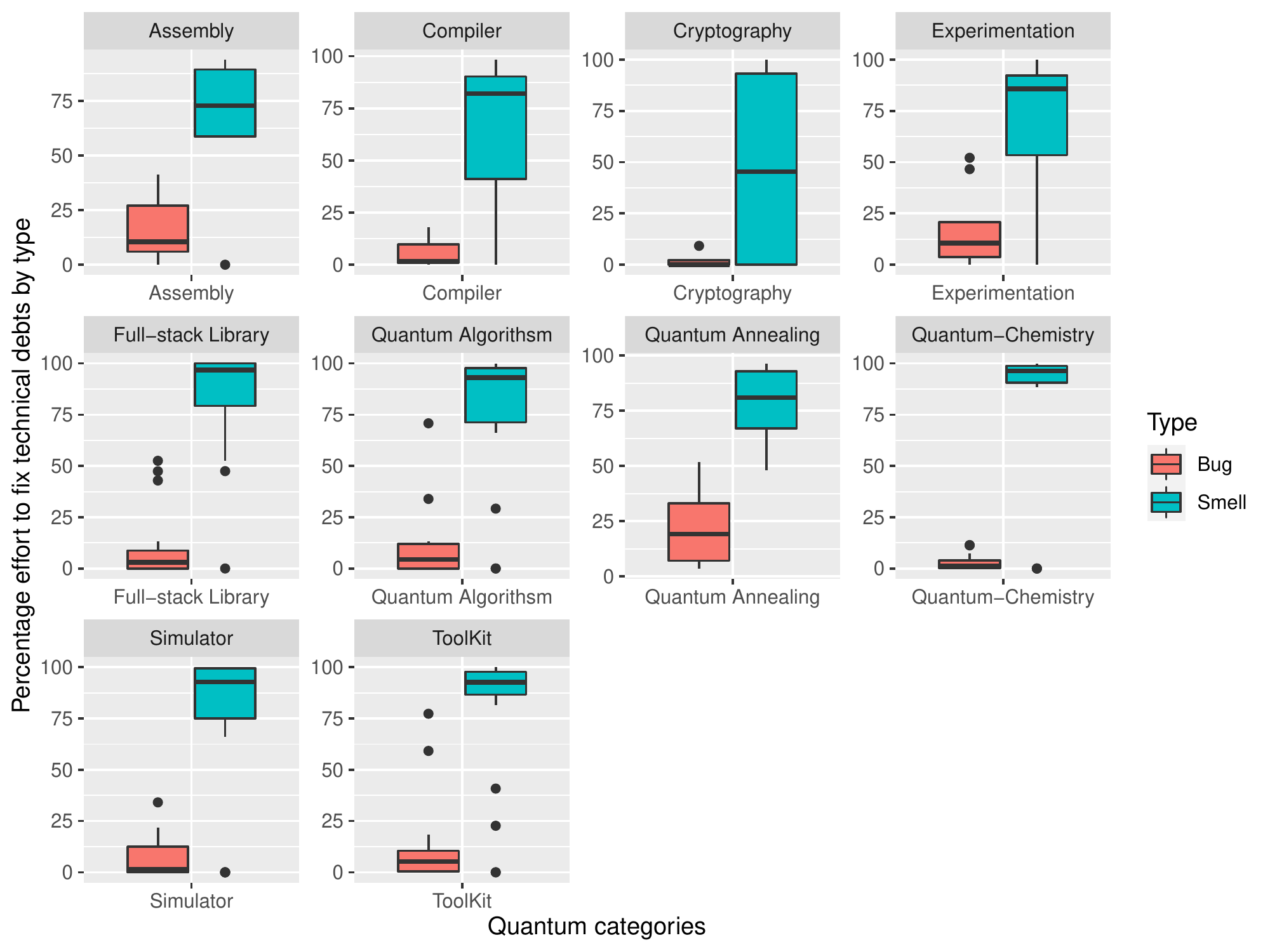}
\caption{Effort required to fix technical debts in terms of the debt types (coding errors and code smells)}
\label{fig:debt-type}
\end{figure}

\begin{figure}[ht]
\center
\includegraphics[width=\linewidth]{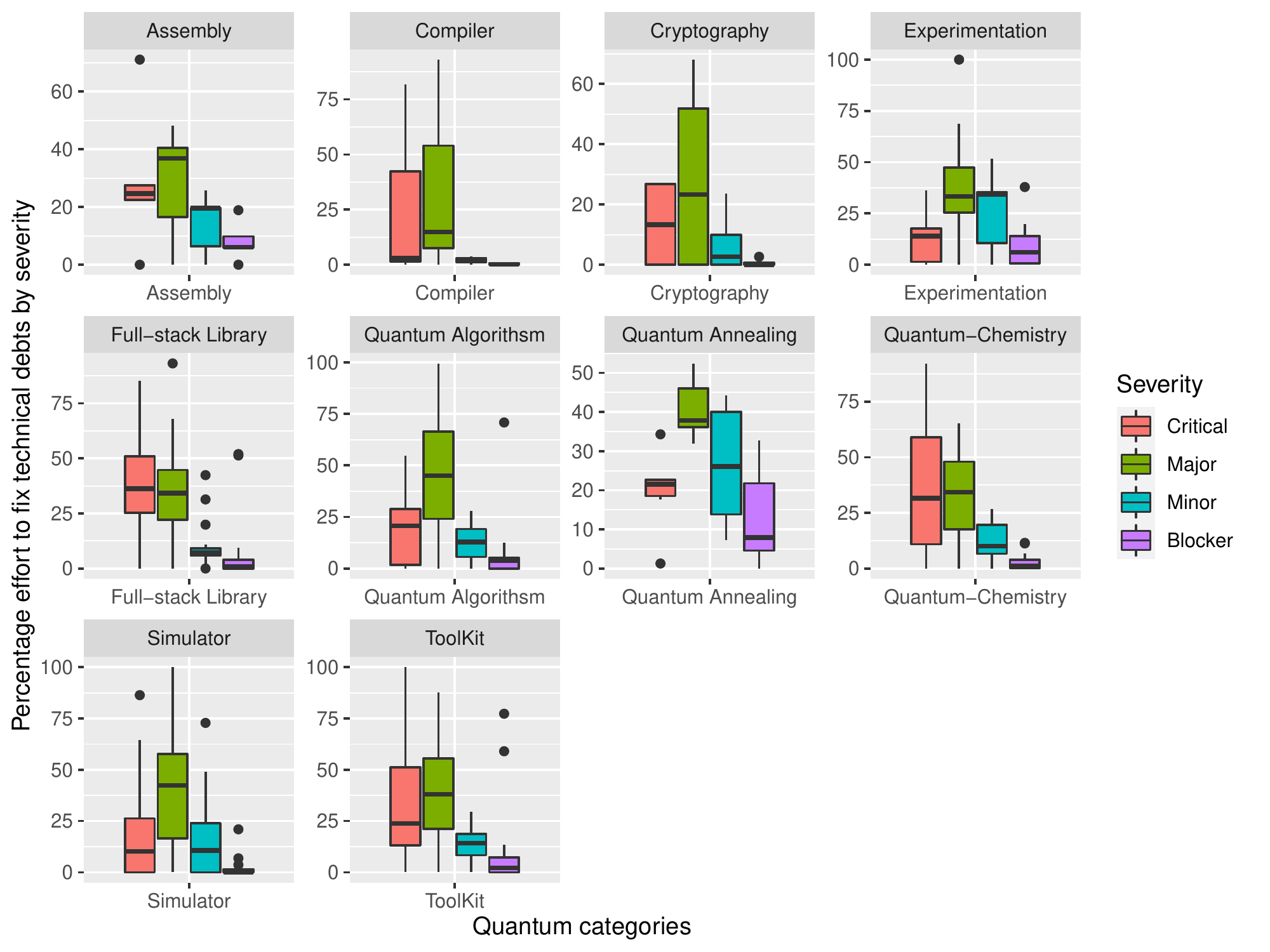}
\caption{Effort required to fix the technical debts in terms of types}
\label{fig:debt-severity}
\end{figure}

We computed the average time to fix each debt type and separately debt severity across the snapshots of a given project. Figure~\ref{fig:debt-type} and \ref{fig:debt-severity} present the composition of technical debts by quantum software category. We show the percentage of the estimated time required to fix the technical debts (errors and code smells) and the severity assessed based on the total fixing effort in each of the project's snapshots in a quantum category. 

As shown in Figure \ref{fig:debt-type}, technical debts reported as code smells would require more fixing effort than errors (potential bugs). It implies that developers need more effort to fix maintenance-related issues (\emph{e.g.,} code smells) than reliability-related issues (\emph{e.g.,} coding errors). 
We observed a low value of the standard deviation ($\sigma$) across the target systems (from 2\% to 25\%), which indicates the consistency of required fixing effort in different quantum projects from different categories.
%Our results also show that coding errors in different categories do not have the same amount of effort to fix.
Our results also show that coding errors in different categories do not require the same fixing effort.
For example, errors in the categories of `Quantum-Annealing', `Experimentation', and `Toolkit' require the most fixing effort while errors in the categories of `Cryptography', `Compiler', and `Quantum-Chemistry' require the least fixing effort.

As indicated by~\cite{tan:evolution,molnar:2020:long,saika:2016:developers}, technical debts can increase development overhead. Some of our findings can concretely explain this. For example, in 
\href{https://github.com/BBN-Q/QGL}{\texttt{BBN-Q/QGL}}
% \texttt{BBN-Q/QGL}\footnote{\url{https://github.com/BBN-Q/QGL}} 
(an experimental quantum application for domain-specific language embedded to specify pulse sequences), SonarQube detected about 2,000 coding errors and 20,000 code smells, which respectively estimated to require about 42 days and 186 days to fix. 
Another example is that, in certain programming languages (such as Python), function names should follow a lowercase convention. Generally, the naming convention is considered necessary in a project with a shared team for effective collaboration. Renaming each of the functions in \texttt{BBN-Q/QGL} will take an average of two minutes.

%Therefore, going contrary to a function named pattern raises a bad convention issue by SonarQube and is estimated to have two minutes to rename each of these functions in \texttt{BBN-Q/QGL}. 

%deemed refactored with each estimated to cost about two minutes to rename these functions found in the project \texttt{BBN-Q/QGL}. 
%Among the issues detected is the usage of more than the authorized function naming convention, where function naming like `isPhyiscal', `isLogical' and variable `channelShift' are all detected as naming convention issues, and the expected naming should follow a lowercase convention and each are estimated to take 2 minutes to fix. The naming convention is generally essential in a project with a shared team for effective collaboration, and therefore, SonarQube expects the standard language naming convention. 
We also observed multiple coding errors with the critical severity in the project 
% \texttt{Quantomatic/pyzx}\footnote{\url{https://github.com/Quantomatic/pyzx}} 
\href{https://github.com/Quantomatic/pyzx}{\texttt{Quantomatic/pyzx}}
(a compiler project for rewriting quantum circuit and optimization).  
%critical SQ-bugs were reported by SonarQube, such as 
For example, a single file 
% \texttt{`pyzx/circuit.py'}\footnote{\url{https://raw.githubusercontent.com/Quantomatic/pyzx/23bf9018e9e2a6a99d1bf8c980d4d440c9f3ef56/pyzx/circuit.py}} 
\href{https://raw.githubusercontent.com/Quantomatic/pyzx/23bf9018e9e2a6a99d1bf8c980d4d440c9f3ef56/pyzx/circuit.py}{\texttt{pyzx/circuit.py}}
for representing a quantum circuit has up to $78$ cognitive complexity. 
Cognitive complexity is a measure of how hard to understand the control flow in a function, and the high Cognitive Complexity will be challenging to maintain~\cite{campbell:2018:cognitive}.
The maximum authorized cognitive complexity of the file is 15, indicating that the file needs about one hour to get fixed.
%according to SonarQube, and a single file is estimated to cost about 1hour8minutes\mehdi{?!?!, can we used 68 Min instead} refactoring effort. 
In another error with the critical severity, the constructor method uses 12 parameters. A long parameter list indicates that the function or method is doing too many things, or that a new structure should be created to wrap the numerous parameters \cite{sonarsourcerules}
% \footnote{\url{https://rules.sonarsource.com/python/RSPEC-107}}
% \Le{This should go to references}. \Le{Who, ref?} 
suggested that functions or methods should use less than 7 parameters to make the code maintainable. Refactoring each of these functions or methods requires about 20 minutes.

In Figure \ref{fig:debt-severity}, we further observed that the technical debts with the major and critical severity, require the most fixing effort, while debts with the blocker severity require the least fixing effort. Compared to other severity types, the fixing effort of critical debts has a higher standard deviation, which indicates that errors with this level of severity do not require a similar amount of time to get fixed. In addition, we observed that debts in the categories of `Compiler' and `Full-stack Libraries' require the most fixing effort.

Overall, our findings are consistent with previous studies~\cite{molnar:2020:long,digkas:2018:developers,marcilio:2019:static} where people studied technical debts in traditional software. 
For instance, Digkas et al.~\cite{digkas:2018:developers} studied technical debts in 57 software projects developed by the Apache ecosystem. They presented that more than 68\% of identified technical debts were related to the major category. 
Marcilio et al.~\cite{marcilio:2019:static} mined 426 different projects and reported that 68\% of all types of technical debts belong to the major class. 
Molnar et al.~\cite{molnar:2020:long} carried out research on technical debts in three Java-based open-source projects. They showed that the technical debts with major severity account for 57.5\% in FreeMind, 74.5\% in jEdit, and 61.8\% in TextGuitar. 
In other words, their result shows that a high percentage of technical debts have a major severity.

\begin{table}[]
    \centering
    \caption{Rules generating about 80\% or more of technical debts, indicating the mean percentage of occurrence in overall application snapshots, severity, type of technical debts, and the descriptions of the rule (as indicated by SonarQube). The highlighted rows indicate the technical debts with the CRITICAL severity. %\MO{Reviewer2: - There is no definition for "tags" in Table 6. The reference for tags is #32 which refers to the homepage of SonarQube and it is not useful.}
    }
     \label{tab:tags-rules}
     \begin{adjustbox}{scale=0.70}
            %\scalebox{0.58}{
% \begin{adjustbox}{width=\columnwidth,center}
% \begin{sidewaystable}{width=7in,center}
% \begin{adjustbox}{center}
% \begin{tabular}{p{2.0cm}|  p{0.5cm} p{0.7cm} p{0.4cm} p{0.6cm} p{0.5cm} p{0.8cm} p{0.5cm} p{0.5cm} p{0.5cm} p{0.6cm} |p{2.0cm} p{5.0cm}}\toprule
\small
\begin{tabular}{m{2.0cm}|  r r r r r r r r r r |p{2.0cm} p{5.0cm}}\toprule

\textbf{rule(severity)}&\textbf{Cryp}&\textbf{Exp}&\textbf{Ann}&\textbf{Fstack}&\textbf{Tkit}&\textbf{Chem}&\textbf{Comp}&\textbf{Ass}&\textbf{Alg}&\textbf{Sim}&\textbf{tag~\cite{sonarsourcerules}}&\textbf{Example Descriptions}\\ \midrule

S117(minor)&17.5&51.73&29.85&13.01&18.42&11.55&2.07&26.64&25.91&36.7&convention&parameter naming convention\\ %hline

S1542(major)&32.98&6.09&6.59&6.32&6&8.15&4.17&10.54&4.35&14.01&convention&function naming convention\\ %hline
\rowcolor{gray!10}
S3776(critical)&8.92&1.98&5.13&6.43&4.48&8.64&21.88&8.64&8.36&1.96&brain-overload&Cognitive Complexity of function/ method shouldn't be too long\\ %hline
\rowcolor{gray!10}
S1192(critical)&7.85&1.99&2.79&8.61&14.12&13.69&2.46&2.88&10.56&7.19&design&define a constant instead of duplicating this literal \\ %hline

S1481(minor)&1.77&1.99&12.65&5.89&7.98&5.56&6.46&7.59&6.72&8.25&unused&Remove this unused declaration should be removed\\ %hline

S125(major)&0.61&4.33&4.69&1.85&5.15&10.67&19.12&6.65&5.38&4.88&unused&Remove this commented out code\\ %hline
PrintStatement Usage(major)&0.54&1.34&4.5&6.04&3.9&18.55&0&0.67&11.75&1.33&python3,obsolete&Replace print statement by built-in function.\\ %hline

S3827(blocker)&0.16&3.95&6.57&1.12&1.47&1.42&0.94&6.49&8.71&0.81&undefined usage. \\ %hline

S1117(major)&0.04&0&0&8.26&1.05&0.08&13.95&0.56&0.06&1.58&confusing,pitfall, suspicious&Two or more field should not be declared with same name\\ %hline
\rowcolor{gray!10}
S5754(critical)&2.01&0.9&0.45&4.43&7.12&1.59&3.64&1.88&0.58&1.94&confusing,unpredictable, error-handling&Specify an exception class to catch or reraise the exception\\ %hline

S100(minor)&2.46&3.8&3.27&4.32&1.66&1.85&0.44&1.54&0.31&0.97&convention&method naming convention.\\ %hline

S112(major)&4.19&0.61&0.77&1.46&3.32&1.52&5.61&1.6&0.64&0.3&error-handling,cwe&replace this generic exception class with a more specific one.\\ %hline

S1827(major)&0&0.88&0.96&0.21&3.09&0.74&0&1.55&0.57&6.55&user-experience,html5, obsolete&Remove this deprecated attribute\\ %hline

S107(major)&0.32&1.04&2&1.48&2.77&1.37&1.15&0.76&2.4&0.82&brain-overload&parameter list of a function should not be greater than the 7 authorized.\\ %hline

S101(minor)&0.23&0.65&1.94&2.13&1.06&1.01&1.57&1.28&1.2&0.22&convention&class naming convention\\ %hline
S1066(major)&1.04&0.38&0.41&2.28&1.36&1.37&1.66&0.68&1.82&0.25&clumsy&Merge this if statement with the enclosing one.\\ %hline

S116(minor)&2&2.93&0.19&1.0&2.75&0.84&0.27&0.01&0.21&0.44&convention&field naming convention\\ %hline
BoldItalicTags Check(minor)&0.02&0.47&0.91&2.64&0.84&0.2&0.34&3.27&0.99&0.81&accessibility&Replace this <i> tag by <em>.\\ %hline
\rowcolor{gray!10}
S1186(critical)&1.38&0.39&0.18&2.32&1.48&0.33&0.38&0.57&0.34&2.86&suspicious&Add a nested comment explaining why this method is empty, or complete the implementation.\\ %hline

\bottomrule
\rowcolor{gray!15}
&\textbf{84.02}&\textbf{85.45}&\textbf{83.85}&\textbf{80.0}&\textbf{88.02}&\textbf{89.13}&\textbf{86.11}&\textbf{83.8}&\textbf{90.86}&\textbf{91.87}&&\\ %hline

\bottomrule
\end{tabular}
% \end{adjustbox}
% \end{sidewaystable}
%}
\vspace{10pt}
%\end{table}
    \end{adjustbox}
\end{table}

\begin{table}[t]
%\Huge
\caption{Types of technical debts that contribute to greater than 80\% of the overall technical debts across different categories of the quantum projects (\emph{s} for code smells and \emph{e} for coding errors).}
\label{table:minimun-tags}
%\scalebox{0.75}{
    \begin{adjustbox}{width=0.85\columnwidth,center}
    \begin{tabular}{l l | r r r r r r r r r r}\toprule
    \rowcolor{gray!15}
    \textbf{Tag~\cite{sonarsourcerules}}&\textbf{Type}&\textbf{Cryp}&\textbf{Exp}&\textbf{Ann}&\textbf{Fstack}&\textbf{Tkit}&\textbf{Chem}&\textbf{Comp}&\textbf{Ass}&\textbf{Alg}&\textbf{Sim}\\  \midrule
    convention&s, e&48.31&56.9&36.92&18.49&21.66&17.65&6.14&37.99&26.44&40.71\\ %hline
    \rowcolor{gray!15}
    unused&s, e&2.62&6.74&15.62&8.03&10.63&13.52&18.87&14.61&12.45&11.56\\ %hline
    
    brain-overload&s&7.62&2.64&6.25&5.45&5.16&7.48&15.67&8.14&8.96&2.16\\ %hline
    \rowcolor{gray!15}
    design&s&6.44&1.73&2.48&6.13&10.05&10.28&1.96&2.63&8.93&5.8\\ %hline
    
    obsolete&s, e&0.45&4.12&6.38&4.83&6.82&14.75&0&3.09&10.9&6.19\\ %hline
    \rowcolor{gray!15}
    
    cwe&s, e&3.77&2.61&1.33&3.45&3.95&3.04&5.5&2.71&3.08&0.54\\ %hline
    
    confusing&s, e&3.1&1.16&0.94&11.28&6.38&1.63&13.74&2.87&1.13&3.53\\ %hline
    \rowcolor{gray!15}
    error-handling&s, e&5.08&1.42&1.09&4.48&7.46&2.77&6.33&2.95&1.08&2.43\\ %hline
    
    accessibility&s, e&6.93&0.74&1.02&3.83&2.33&3.17&0.47&4.31&1.05&3.18\\ %hline
    \rowcolor{gray!15}
    suspicious&s, e&2.65&1.07&1.09&9.2&2.33&1.19&11.03&1.92&1.24&3.78\\ %hline
    
    python3&s, e&0.45&1.18&3.96&4.67&2.78&14.15&0&0.57&9.91&1.04\\ %hline
    \rowcolor{gray!15}
    user-experience&s, e&0&3.03&2.44&1.49&4.05&0.69&0&2.78&1.21&5.43\\ %hline
    
    html5&s, e&0&2.95&2.44&0.16&4.01&0.66&0&2.52&1.2&5.15\\ %hline
    \rowcolor{gray!15}
    unpredictable&s&1.64&0.85&0.4&3.3&5.04&1.35&2.51&1.59&0.48&2.1\\ %hline
    
    wcag2-a&s, e&6.92&0.33&0.21&2.02&1.69&3.02&0.24&1.53&0.23&2.52\\ %hline
    \rowcolor{gray!15}
    clumsy&s&1.59&0.43&0.6&1.96&1.65&1.51&2.45&1.15&1.61&0.34\\ %hline
    \bottomrule
    % \rowcolor{gray!20}
    \textbf{Total}&&\textbf{97.57}&\textbf{87.9}&\textbf{83.17}&\textbf{88.77}&\textbf{95.99}&\textbf{96.86}&\textbf{84.91}&\textbf{91.36}&\textbf{89.9}&\textbf{96.46}\\ %hline
    \bottomrule
    
    \end{tabular}
    \end{adjustbox}

\end{table}

% \begin{table}[]
% \begin{minipage}{1.0\linewidth}
% 	\caption{Rules generating about 80\% or more of technical debt, indicating the mean \%ge of occurrence in overall application snapshots, severity, tag name,  and the Descriptions of the rule (as indicated by sonarQube). The  highlighted rows indicates the technical debt of severity type `\textbf{CRITICAL}'}
% 	\label{table:student}
% 	\centering
% 	\input{tables/subTable/tbl_rules_v2}
% \end{minipage}    
% \end{table}

% \Le{Mehdi, help me put all the footnotes to references in this paragraph}
In Table \ref{tab:tags-rules} and Table \ref{table:minimun-tags}, we examined the types of technical debts in the target projects based on the set of rules and types defined by SonarQube~\cite{sonarsourcerules}.
% \footnote{\url{https://docs.sonarqube.org/latest/user-guide/built-in-rule-tags/}}. 
Table \ref{tab:tags-rules} shows the set of rules contributing to 80\% and more of the overall technical debts in the target quantum categories. The highlighted rows indicate the debts classified as CRITICAL. From Table \ref{table:minimun-tags}, we found the most prevalent types of debts:
%basing on the defined types of technical debts and the corresponding rule details. 
%As the initial observation, 
%We found that about 80\% of the detected debts can be classified into the two main groups, \emph{i.e.,} maintainability and reliability shown in Table \ref{table:minimun-tags}. 
%fewer tags related to maintainability and reliability, shown in Table \ref{table:minimun-tags}. 
`convention' (problem with coding convention such as formatting, naming, white-space),
% \footnote{\url{https://rules.sonarsource.com/java/tag/convention}}
`unused' (unused code, \emph{e.g.,} commented out code or unused private variable),
% \footnote{\url{https://rules.sonarsource.com/java/tag/unused}},  
%\Mona{Please check the code complexity part as requested by reviewer 1}
`brain-overload' (related to cognitive complexity, there is too much code to keep in the head at once),
% \footnote{\url{https://rules.sonarsource.com/java/tag/brain-overload}}, 
`design' (the design of the code is questionable, \emph{e.g.,} duplicate string literals),`cwe' (relates to a rule in the Common Weakness Enumeration~\cite{cwe:mitre}),
% \footnote{\url{http://cwe.mitre.org/}}, 
and clumsy (unnecessary steps was used on something that should be straightforward)~\cite{sonarsourcerules}.
% \footnote{\url{https://rules.sonarsource.com/java/tag/clumsy}}. 
According to our findings in Table \ref{table:minimun-tags}, the type `code convention' dominates in at least 50\% of the studied categories, followed by the high rate of `unused code', `design smells' and `brain-overload'. We also observe that `Quantum Chemistry' shows a higher percentage of code deprecation and design issues. In addition, a high percentage of code in the `Compiler' and `Full-stack Library' categories are characterized as confusing and suspicious.  
%We also observed a high deviation between the highly dominating tag and least dominating in most selected quantum categories compared to the non-quantum type. 
%This is an indication that quantum software is impacted more by the minimum set of issues.  %\Le{Any suggestions to developers based on the observation?} 

%\nd We manually examine the source files of some projects with higher rate of issues.  For example, Quantum game project \texttt{HuangJunye/QPong}\footnote{\url{https://github.com/HuangJunye/QPong}} where sonar detected many suspicious codes in the source-code files (consisting of both python and HTML scripts), that handle input parameters from the button/ keyboard and Toffoli gates in which multiple functions was associated with the similar block of code similarity containing multiply nested if statement parameters. Consequently, some files were categorized as critical due to their high cognitive complexity. 
%\Le{What do we want to show in this paragraph? and how developers can learn or benefit from our discussion?}
In our manual investigation of source codes of two projects from `Compiler' category (\path{Quantomatic/pyzx} and \path{QE-Lab/OpenQL}) 
% in the `Compiler' category, 
we found that most of the technical debts were related to handling and generating graphs, benchmarks, unreachable code (due to jump statements), and unexpected expressions. 

\nd Also, we manually investigating the source code for the project \path{artiste-qb-net/qubiter}  of Assembly category. We observed most of the code smells and coding errors detected (such as suspicious code, design, unused, brain-overload) are related to the source code for implementing in-memory storage of the circuits. In \path{artiste-qb-net/qubiter}, circuits are stored entirely in memory, essentially as Python lists of gates. Lists of this type can be sliced, combined, etc. When using such lists, it is easy to randomly access any gate of the circuit at will, for instance, when doing circuit optimizations (e.g., replacing the circuit with an equivalent shorter one). \path{artiste-qb-net/qubiter} uses a Python list of the lines stored as strings of the circuit's English file. For example, we observed multiple empty blocks of codes detected as `suspicious' in the function for generating the Python list of gates of Hermitian conjugate. These features are particular in quantum software, while in traditional software, the most frequent technical debts are related to different groups. Marcilio et al.~\cite{marcilio:2019:static} explained that the problems regarding packages and exception handling contain the biggest proportion of technical debts in their studied projects. Digkas et al.~\cite{digkas:2018:developers} also reported that the most frequent technical debts are related to resource management, null pointer, and exception handling problems.

%and I/O associated with unreachable code due to jump statements (e.g., break, raise) and unexpected expression instead of assignment. 
We also observed that the files containing debts have high cognitive complexity due to a large number of code lines, loops, and if-else statements. For example, a function used for computing the echelon form of a matrix has the cognitive complexity of 155, which exceeds the maximum allowed cognitive complexity of the function (\emph{i.e.,} 140). Moreover, we found 86.6\% of the code duplication on 70,000 lines in \texttt{Quantomatic/pyzx} and 88.9\% code duplication on 1.3M lines in \texttt{QE-Lab/OpenQL}.

%similarly 86.6\% of the codes duplication on 70,000 lines for \texttt{Quantomatic/pyzx}, and 88.9\% duplicate code on 1.3M lines for \texttt{QE-Lab/OpenQL} as reported by SonarQube analysis. 
% \texttt{Quantomatic/pyzx} 
%\href{https://github.com/Quantomatic/pyzx}{\texttt{Quantomatic/pyzx}}
%is a python library for optimisation and rewriting quantum circuit using ZX-calculus
% \footnote{\url{https://github.com/Quantomatic/pyzx}} 
%while
% \texttt{QE-Lab/OpenQL} 
%\href{https://github.com/QE-Lab/OpenQL}{\texttt{QE-Lab/OpenQL}}
%provides a compiler for compiling and optimizing quantum code written in C++/Python.
% \footnote{\url{https://github.com/QE-Lab/OpenQL}}.

We will further discuss the implication of our findings in Section \ref{sec:discussion}.

\begin{tcolorbox}
On the one hand, more than half of technical debts in all kinds of software systems (quantum and non-quantum) are classified as major from the severity point of view. 
On the other hand, the most frequent types of technical debts are different in various software types. With respect to this fact, we found that
the most frequent technical debts in quantum computing software systems are related to the `code convention', `design issue', `brain-overload', and `error-handling' problems.
\end{tcolorbox}
% \Le{Mehdi, your comparison with traditional software is pretty good, could you put your discussion again in a summary box here?}

%\begin{tcolorbox}
%\textit{\textbf{Summary of findings (4)}}:- At least 80\% of the  SQ-bugs are categorised within a minimum set of tags relating to both maintenance and reliability SQ-bugs in the order of dominance `code convention', `unused',  `brain-overload', `design issues', `obsolete' `cwe', `confusing' and `error-handling'.
%\end{tcolorbox}

%\begin{tcolorbox}
%\textit{\textbf{Summary of findings (5)}}:- In particular,  ``Compilers'' categories are characterized by source-code that i) `looks suspicious and needs to be examined' (suspicious), and ii) `hard to understand' (confusing) and ii) `working source-codes that is likely to fail in future' (pitfall).
%\end{tcolorbox}

%\nd We manually examine the source code of some projects,  for example, Quantum game project \texttt{HuangJunye/QPong} where sonar detected many suspicious codes in the source-code files (consisting of both python and HTML scripts), that handle input parameters from the button and keyboard click in which multiple functions was associated with the similar block of code similarity containing multiply nested if statement parameters. Consequently, the files most issues in the files were categorized as critical due to their high cognitive complexity.

\subsection{\textbf{RQ2: How Do Technical Debts Evolve Over Time?}} \label{sec:result-evolution}
%\MO{-I understand that computing RQ2 on the whole set of 118 projects requires time and effort but you need to explain the reasons why you did not investigate evolution on the whole set of projects and on the whole categories. Moreover, you need to detail how did you select the 2 projects for each category.}

%\MO{Reviewer1: -In reporting the main finding for RQ2 you mention LOC and code complexity but you never look at complexity. From where the statement comes from?}

%\nd This step examines how the technical debt evolves with time across the snapshots of every target project. In particular, 
%\Le{Note: this is moved from Section 3}
%In this research question, we investigate how code smells and SQ-bugs evolve over time. We also examine the addition of new codes and how the code base grows to identify the point in time when a debt is introduced or fixed. We present the detailed results of this analysis in section \ref{sec:result-evolution} to answer our \textbf{RQ2}.

This section discusses how technical debts evolve in the target quantum projects. From the 10 identified quantum software categories, we selected the 8 most representative categories, \emph{i.e.,} `Assembly', `Quantum-Annealing', `Experimental', `Full-stack Library', `Quantum-chemistry', `Toolkit', `Simulator', and `Quantum-Algorithms'. Figure~\ref{fig:evolution-general} (\ref{fig:sub1-assembly1} to \ref{fig:sub1-Algorithsm}) provides the general summary of how the technical debts (in terms of code errors on the left and code smells on the right) evolve in eight of the studied categories. Each line represents the evolution of technical debt ratio (TDR, described in Section \ref{subsec:background:sonar}) on the y-axis and the number of days based on 90 days intervals (snapshots) on the x-axis, for a single project. In the following, we report our results from manually analyzing two projects in each of the categories and discuss the evolution of technical debts in detail.

\definecolor{light-blue}{HTML}{A1E0E6}
\definecolor{light-yellow}{HTML}{FFCD72}
\definecolor{strong-green}{HTML}{219521}
%\begin{adjustbox}{width=0.75\columnwidth,center}
\begin{figure}
%\begin{adjustbox}{width=0.75\columnwidth,center}
    \scalebox{1.0}{  
        \begin{subfigure}[b]{0.4\textwidth}
            \includegraphics[width=\linewidth]{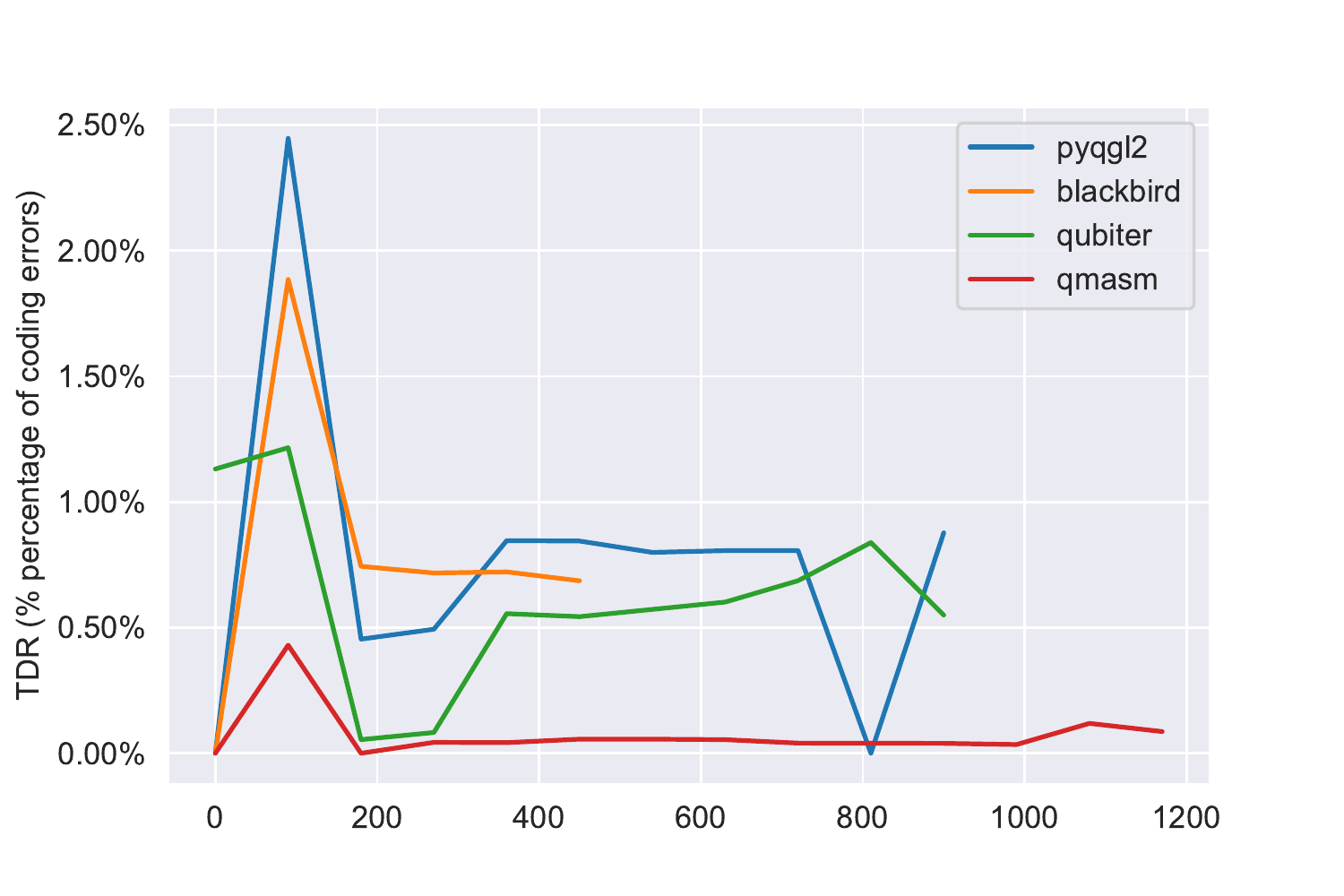}
            \caption{Assembly (code errors)}
            \label{fig:sub1-assembly1}
        \end{subfigure}%
        \begin{subfigure}[b]{0.4\textwidth}
            \includegraphics[width=\linewidth]{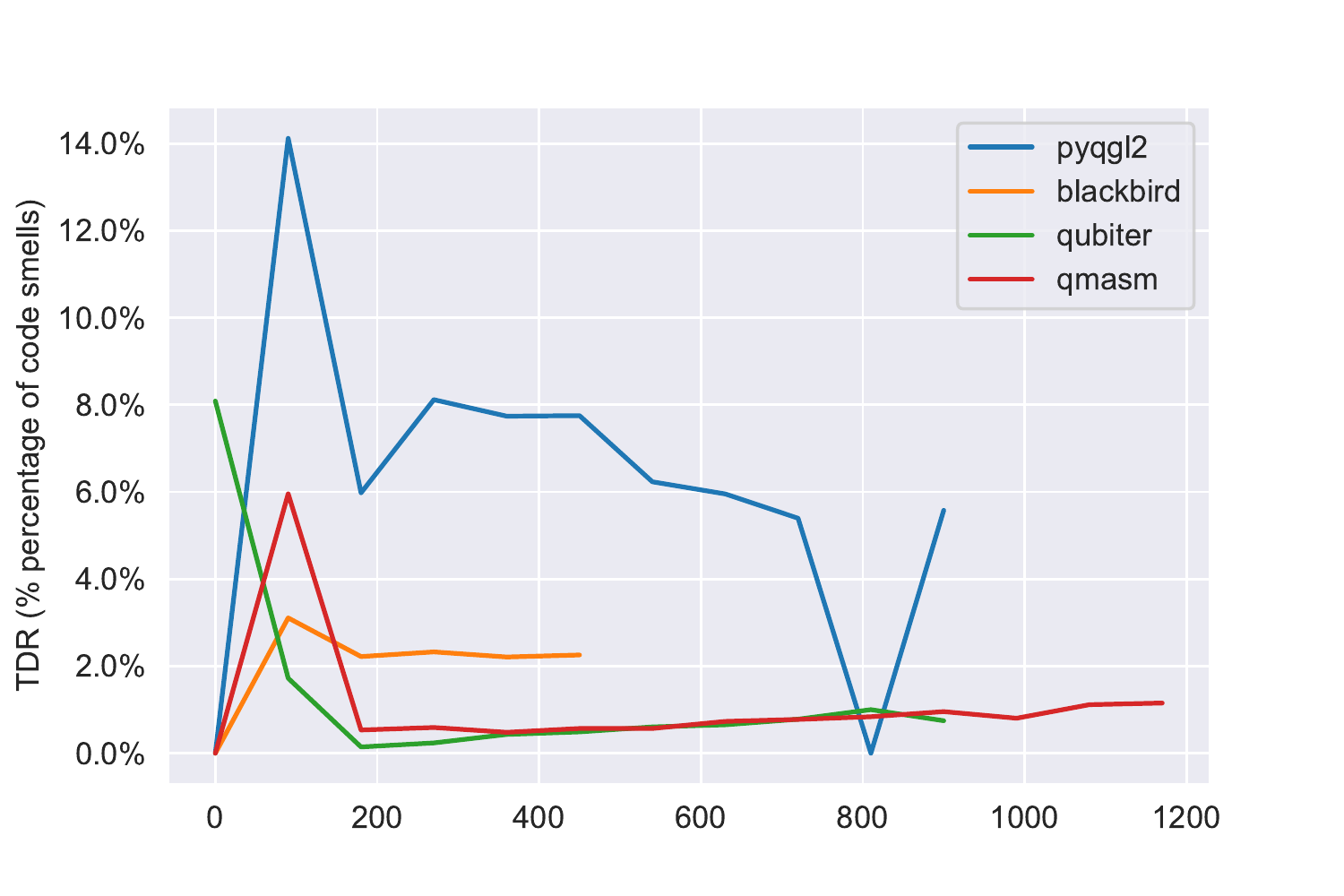}
            \caption{Assembly (code smells)}
            \label{fig:sub1-assembly2}
        \end{subfigure}%
        \hfill
    %\caption{Pictures of animals}\label{fig:animals}
    }

	\scalebox{1.0}{
    \begin{subfigure}[b]{0.4\textwidth}
        \includegraphics[width=\linewidth]{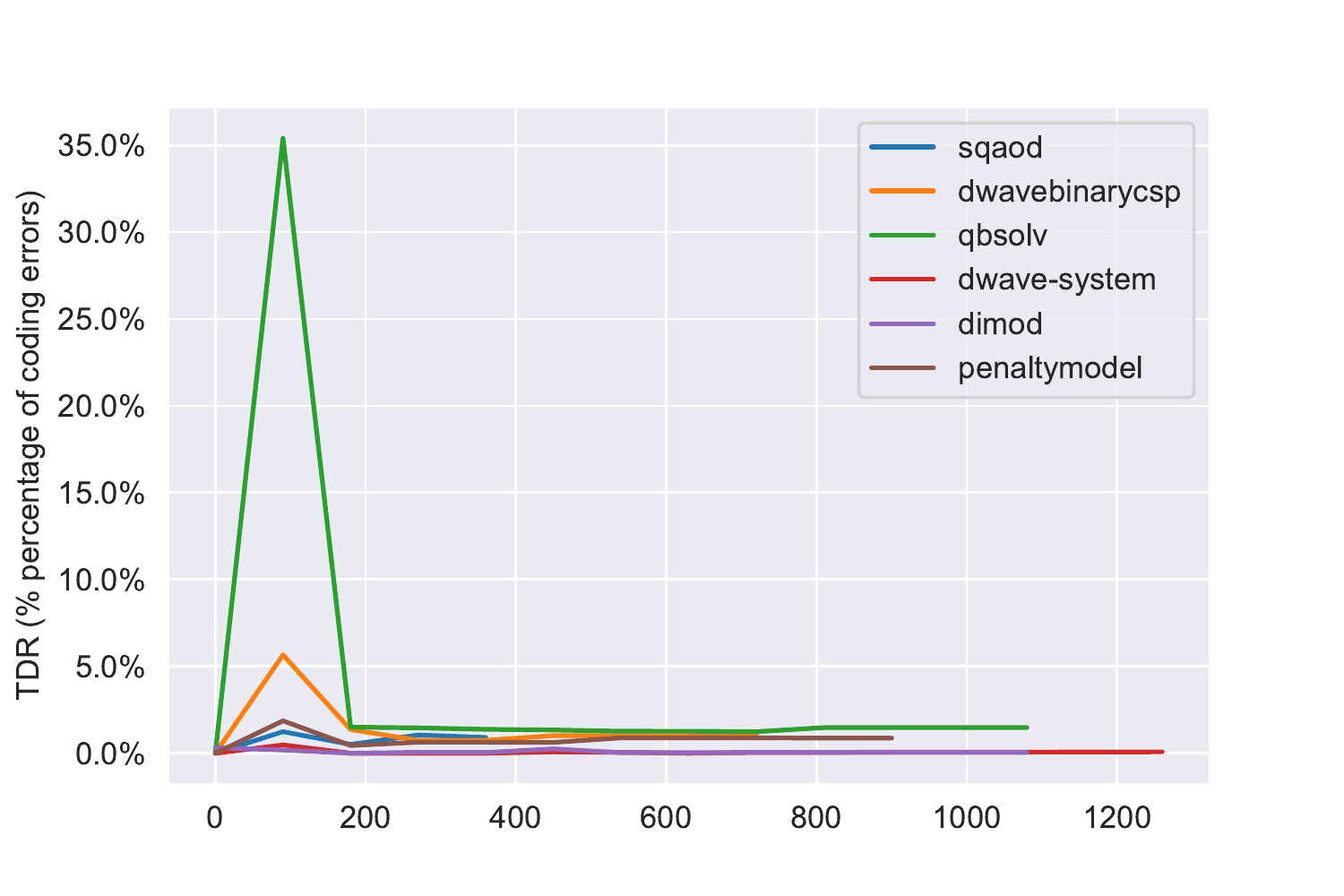}
        \caption{Quantum Annealing (code errors)}
        \label{fig:sub1-Annealing1}
    \end{subfigure}%
    \begin{subfigure}[b]{0.4\textwidth}
        \includegraphics[width=\linewidth]{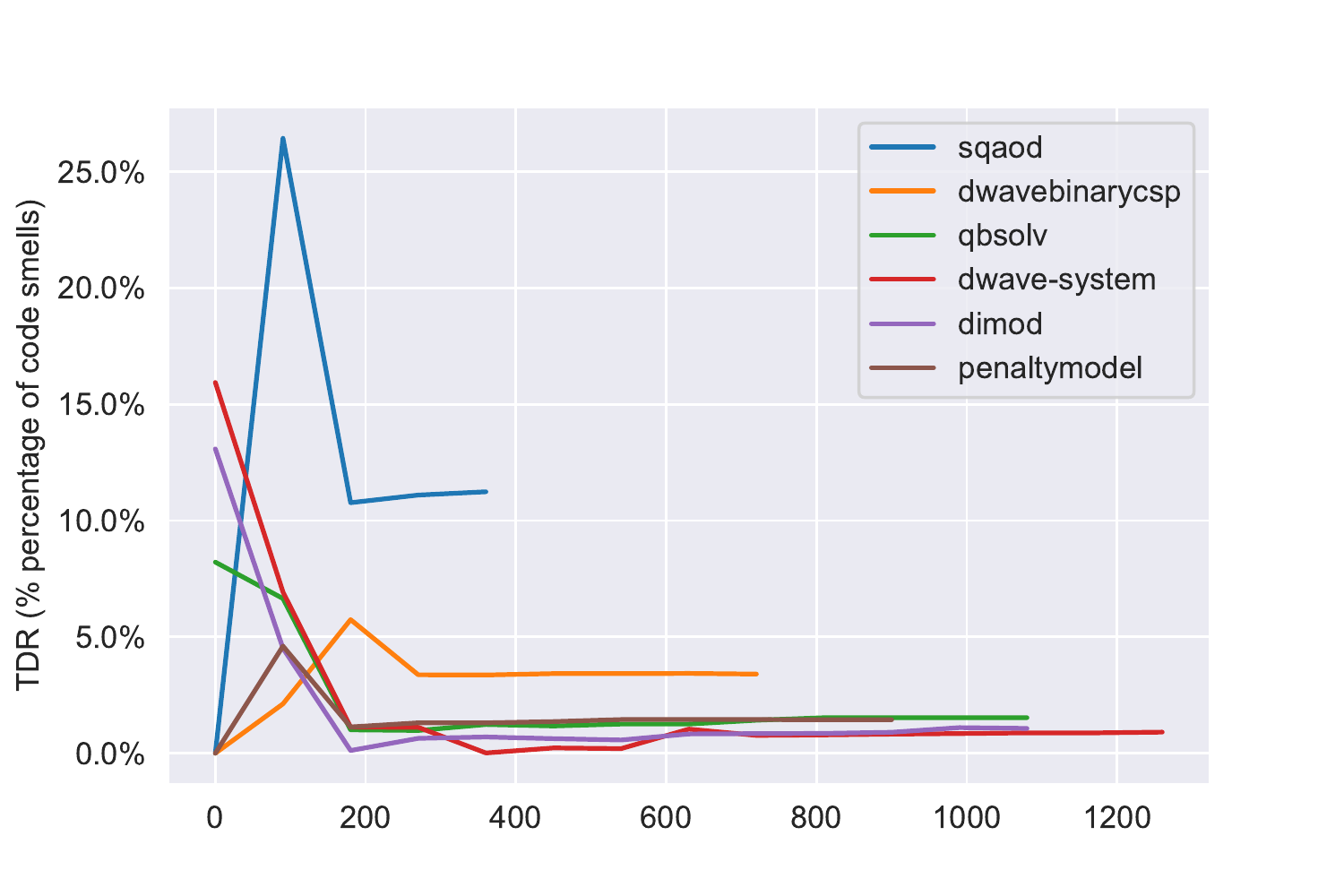}
        \caption{Quantum Annealing (code smells)}
        \label{fig:sub1-Annealing2}
    \end{subfigure}
    \hfill
    }
        %%% row2
       % 
        \scalebox{1}{ 
         \begin{subfigure}[b]{0.4\textwidth}
                \includegraphics[width=\linewidth]{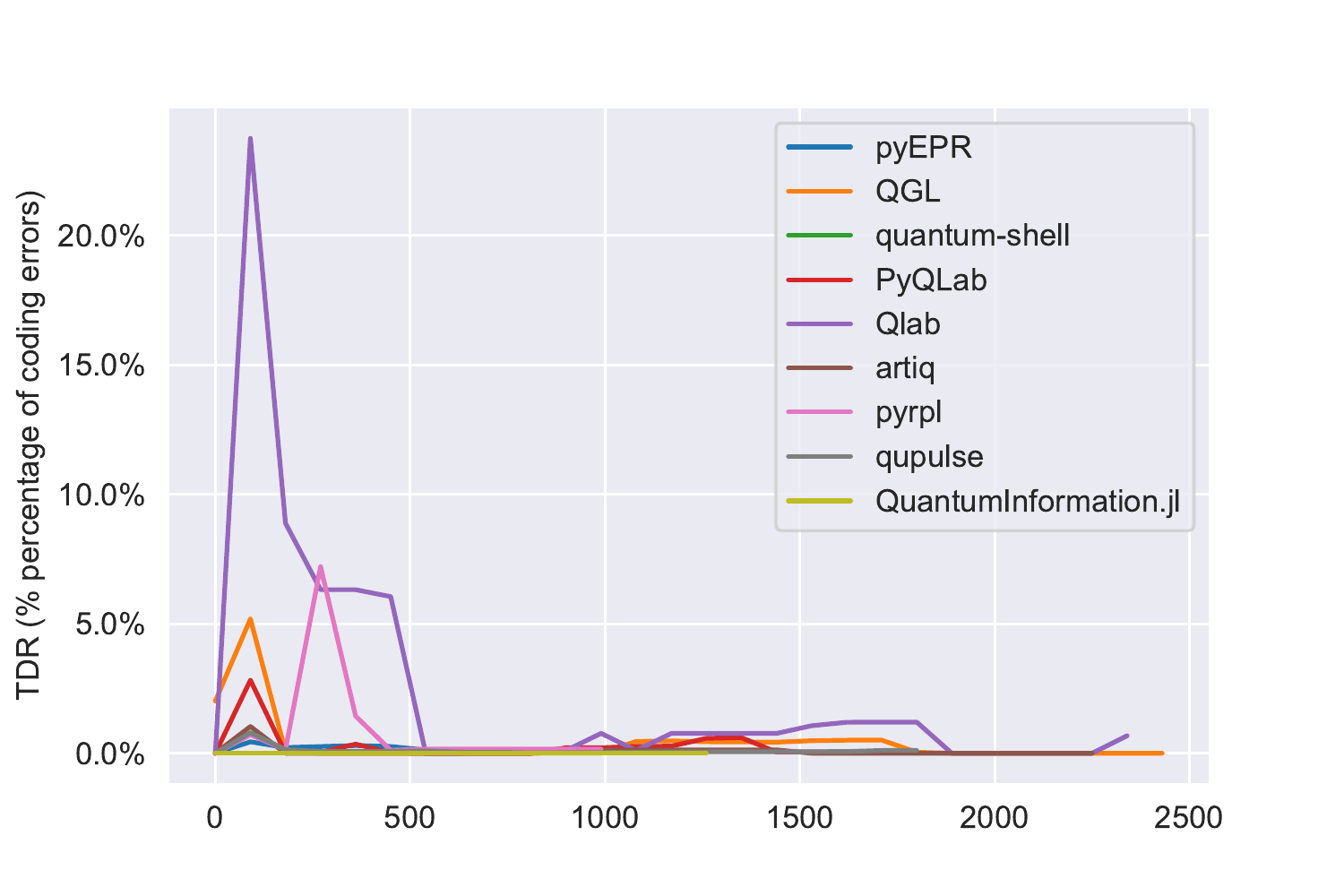}
                \caption{Experimentation (code errors)}
                \label{fig:sub1-Experimentation1}
        \end{subfigure}%
        \begin{subfigure}[b]{0.4\textwidth}
                \includegraphics[width=\linewidth]{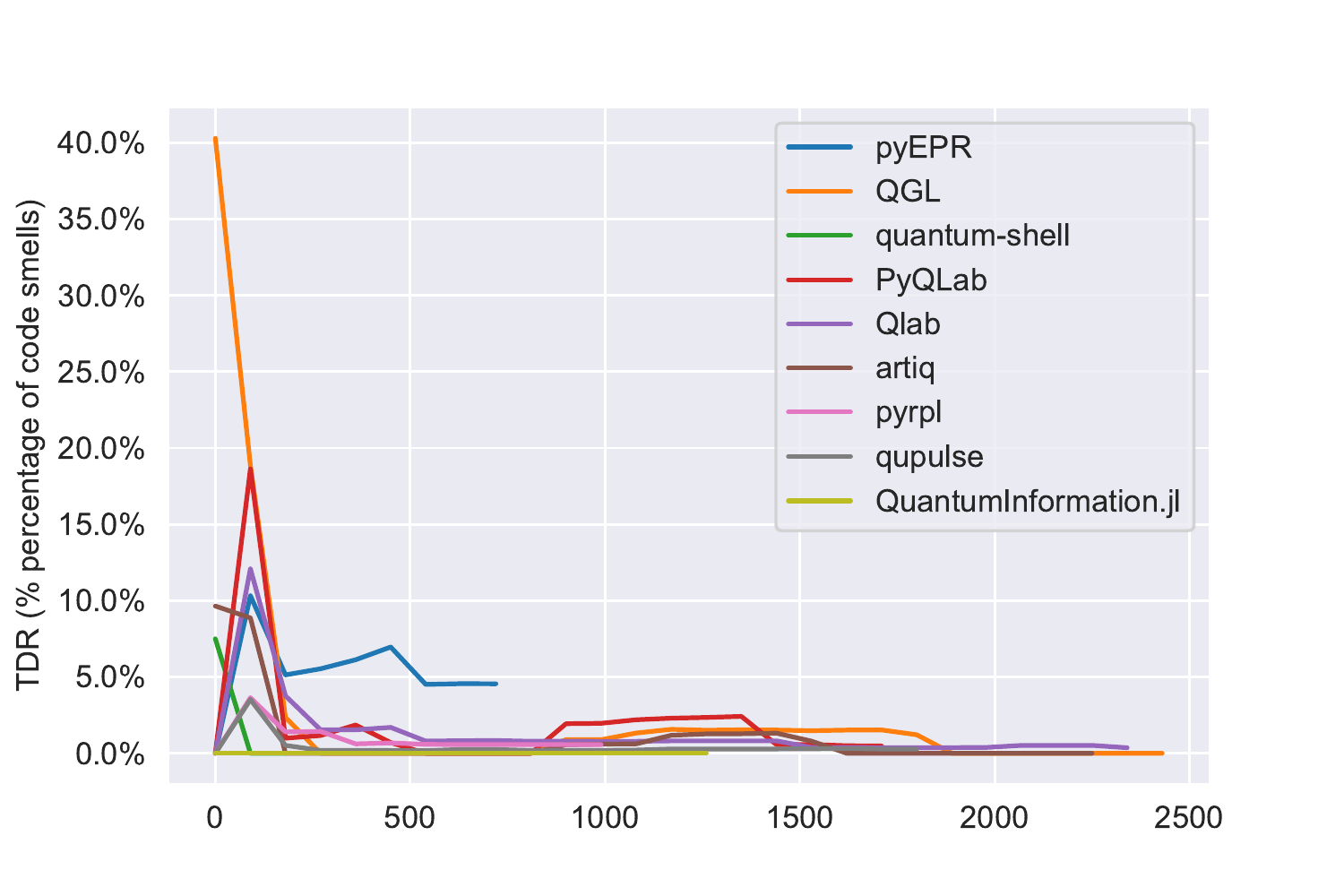}
                \caption{Experimentation (code smells)}
                \label{fig:sub1-Experimentation2}
        \end{subfigure}%
        }
        %\hfill
        \scalebox{1}{ 
        \begin{subfigure}[b]{0.4\textwidth}
                \includegraphics[width=\linewidth]{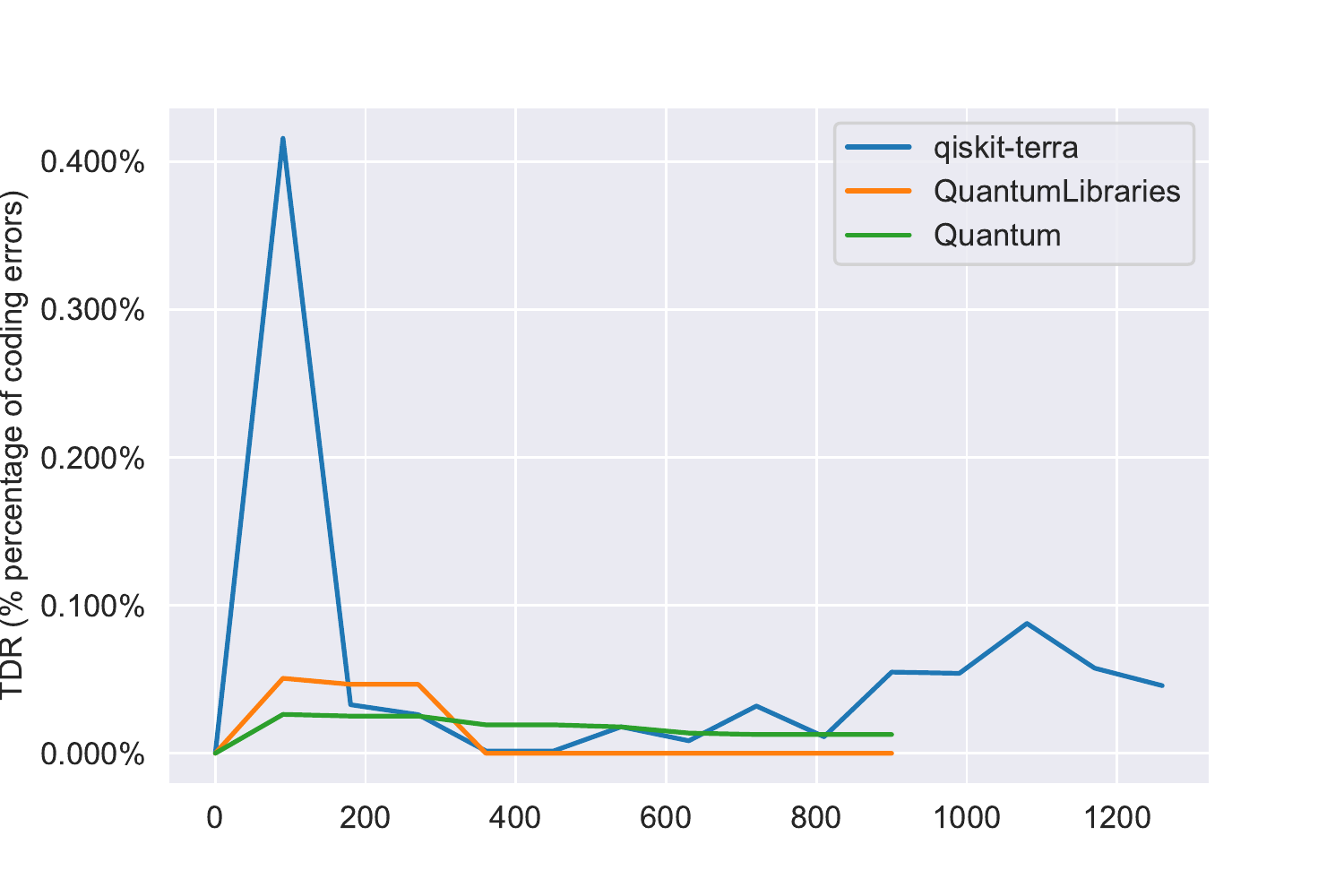}
                \caption{Full-stack Library (code errors)}
                \label{fig:sub1-Full-stack1}
        \end{subfigure}%
        \begin{subfigure}[b]{0.4\textwidth}
                \includegraphics[width=\linewidth]{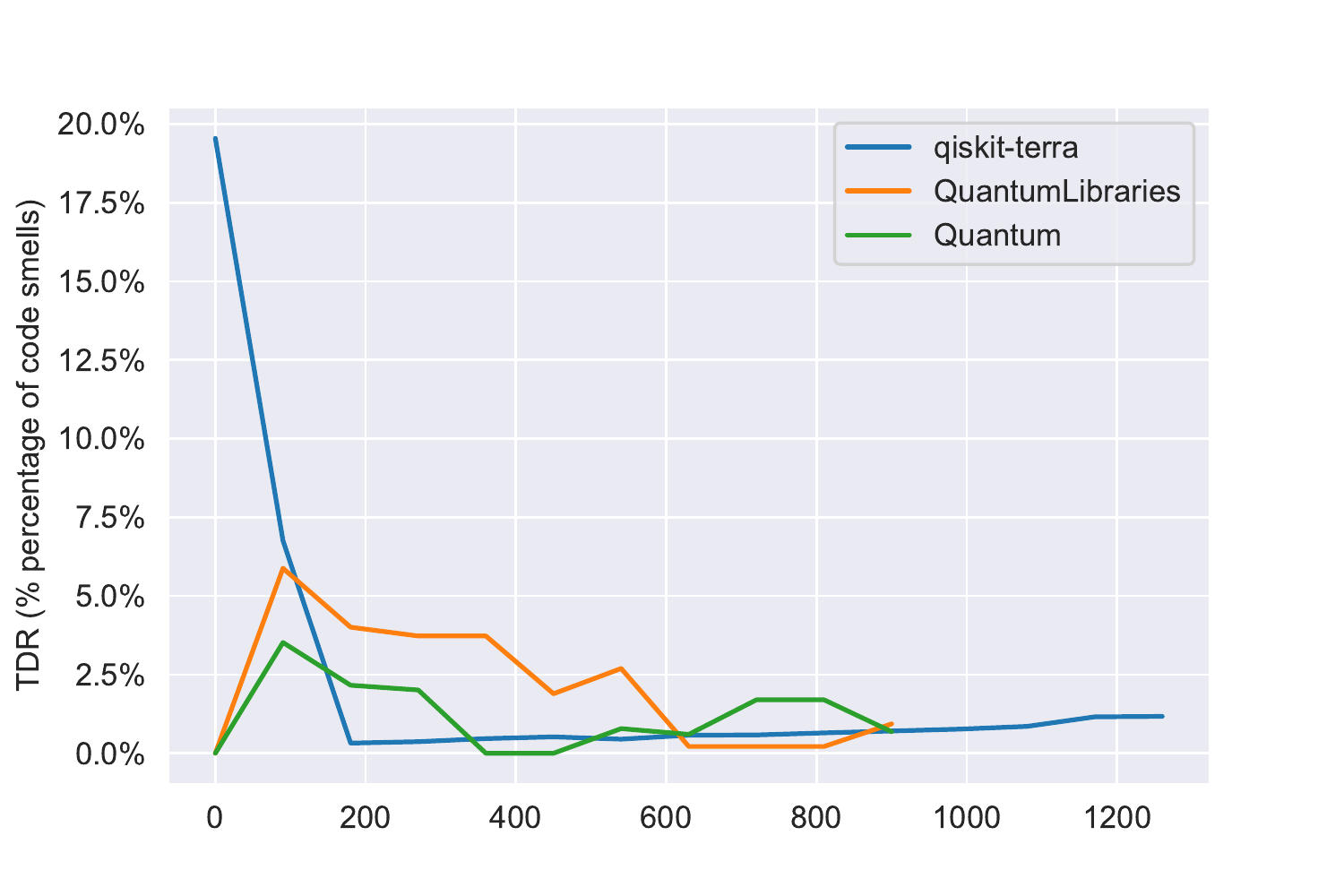}
                \caption{Full-stack Library (code smells)}
                \label{fig:sub1-Full-stack2}
        \end{subfigure}
          }
        %%%%%% row 3

       \end{figure}
    %\end{adjustbox}
    \begin{figure}\ContinuedFloat

        %\hfill
        \scalebox{1.0}{ 
         \begin{subfigure}[b]{0.4\textwidth}
                \includegraphics[width=\linewidth]{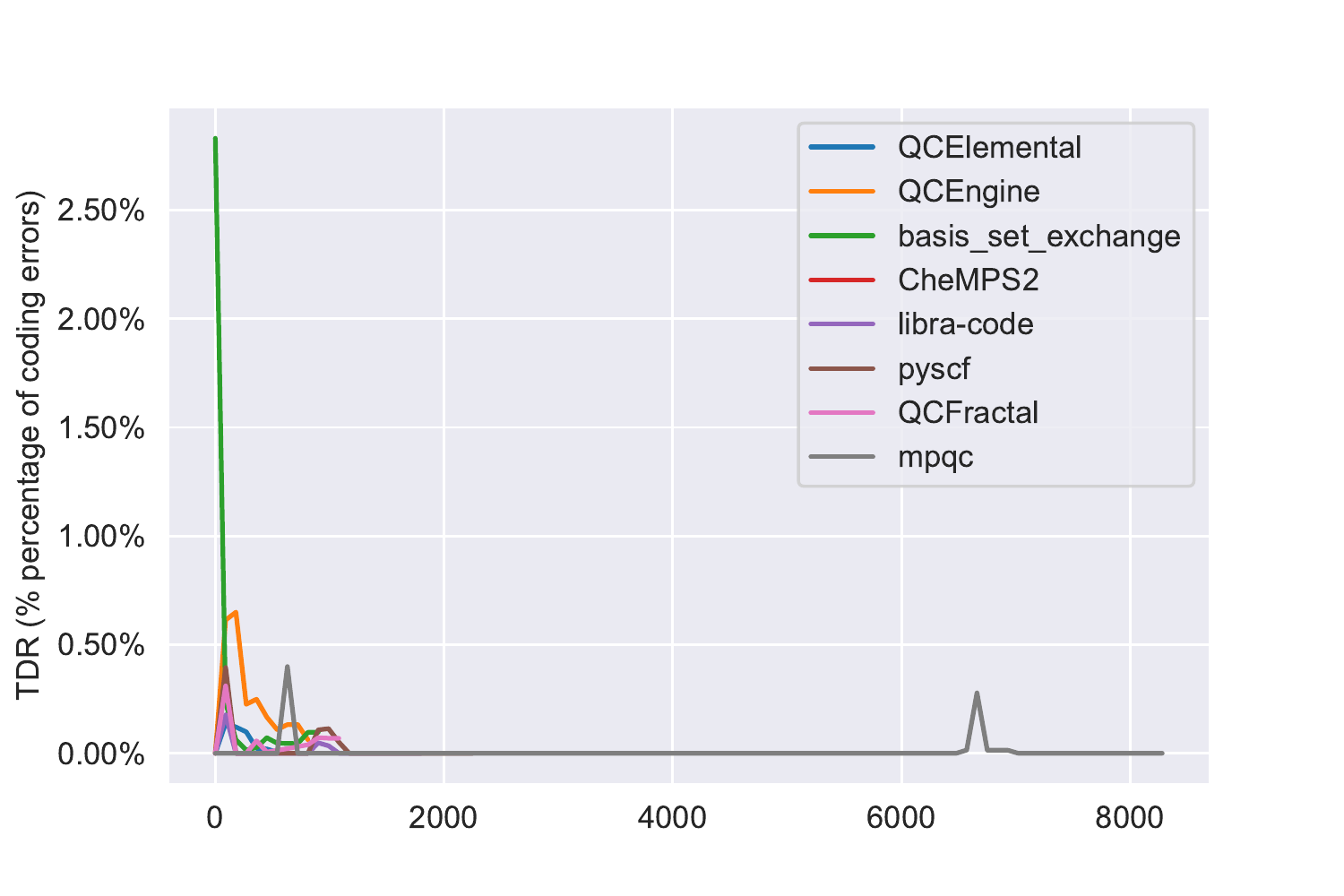}
                \caption{Quantum Chemistry (code errors)}
                \label{fig:sub1-Chemistry1}
        \end{subfigure}%
        \begin{subfigure}[b]{0.4\textwidth}
                \includegraphics[width=\linewidth]{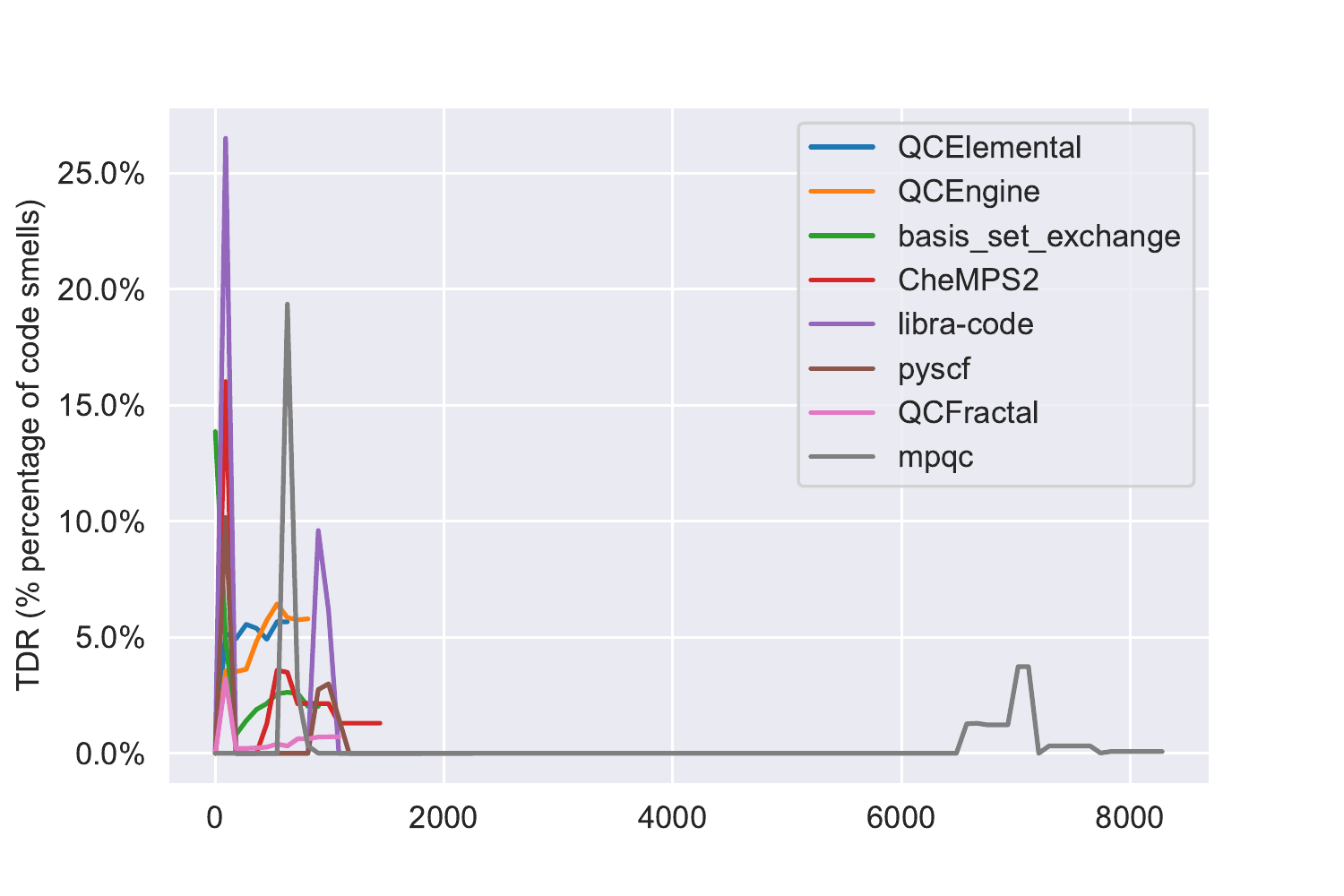}
                \caption{Quantum Chemistry (code smells)}
                \label{fig:sub1-Chemistry2}
        \end{subfigure}%
        }

        \scalebox{1}{ 
         \begin{subfigure}[b]{0.4\textwidth}
                \includegraphics[width=\linewidth]{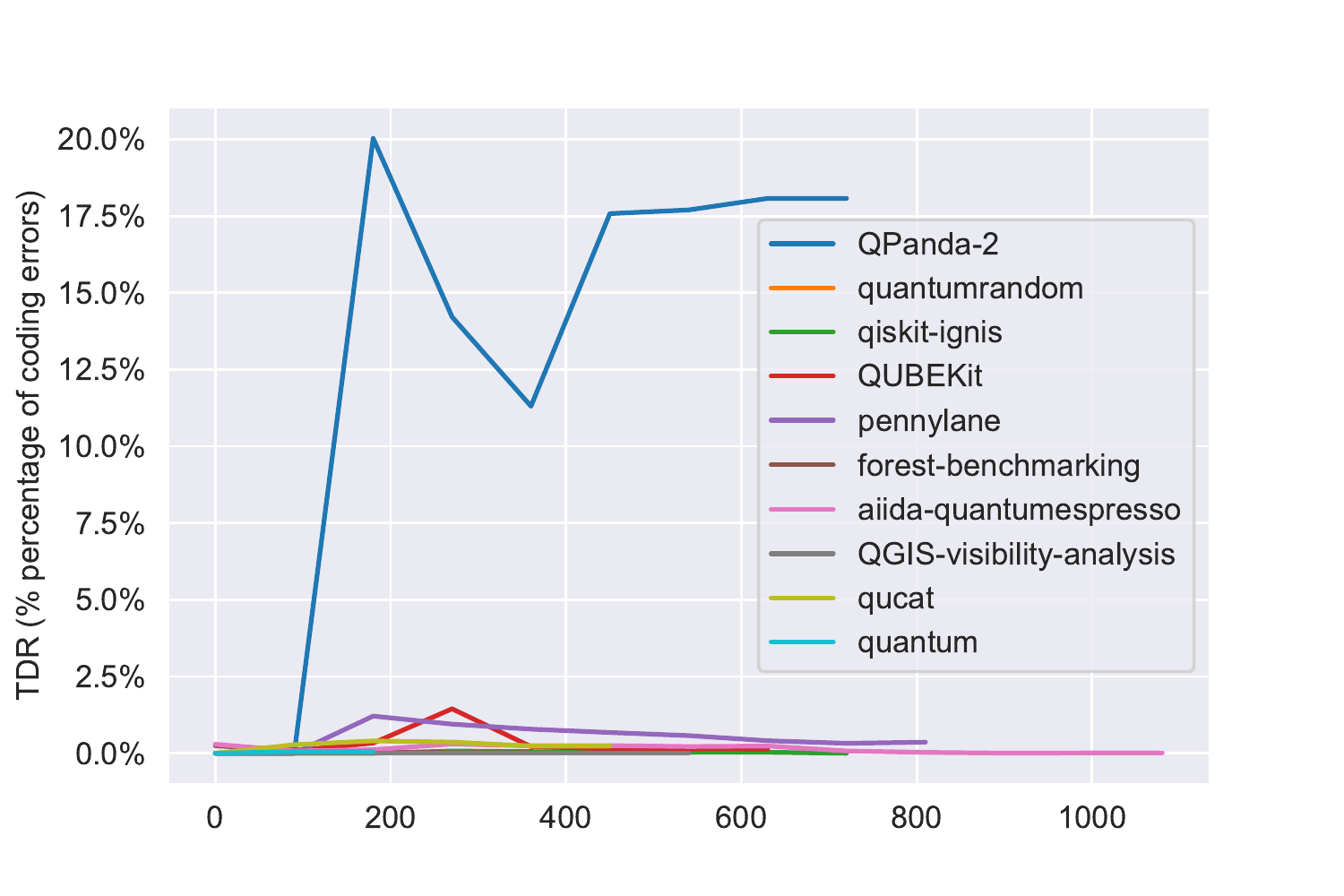}
                \caption{ToolKit (code errors)}
                \label{fig:sub1-toolkit1}
        \end{subfigure}%
        \begin{subfigure}[b]{0.4\textwidth}
                \includegraphics[width=\linewidth]{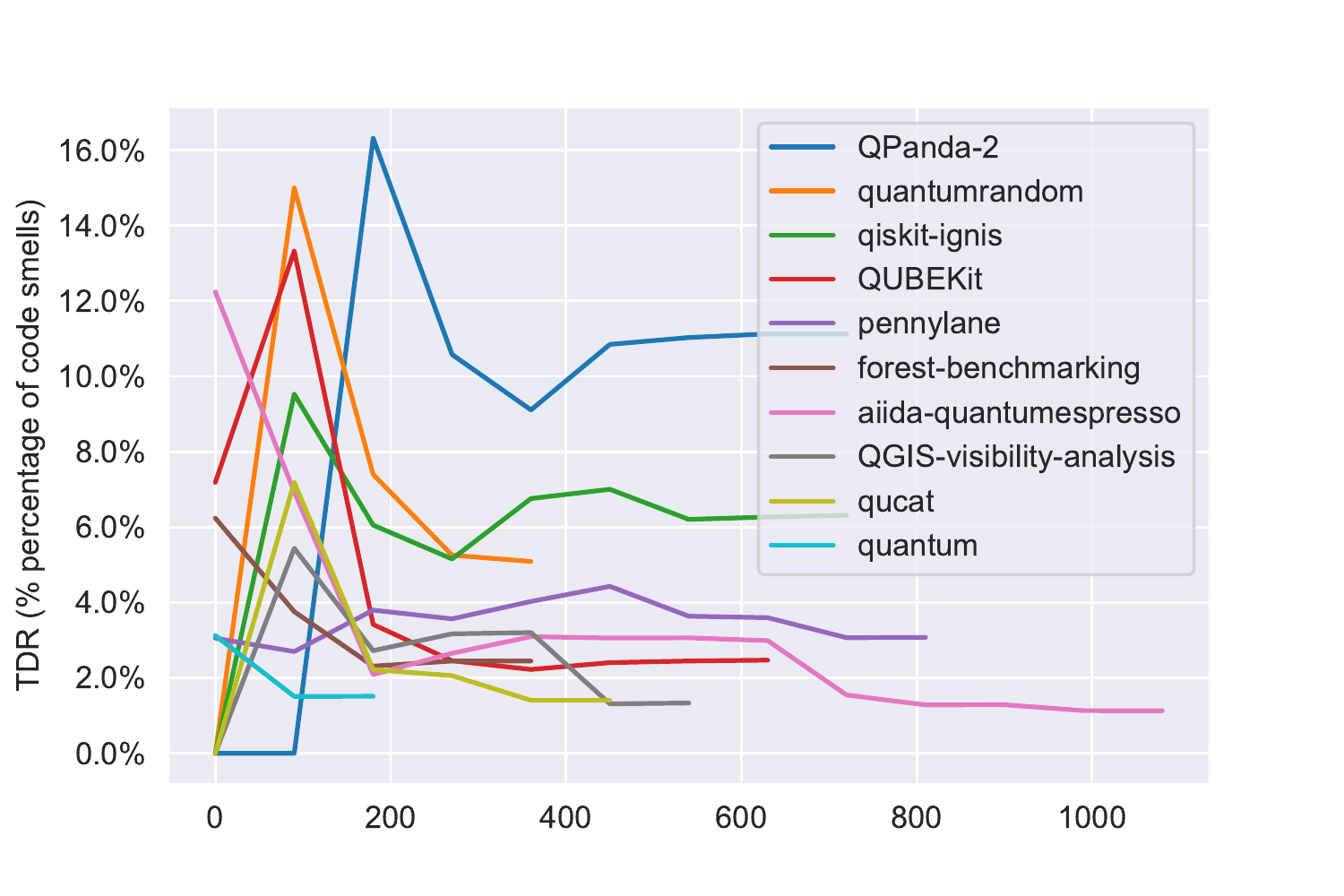}
                \caption{ToolKit (code smells)}
                \label{fig:sub1-toolkit2}
        \end{subfigure}%
        }
        %\hfill
        \scalebox{1}{ 
        \begin{subfigure}[b]{0.4\textwidth}
                \includegraphics[width=\linewidth]{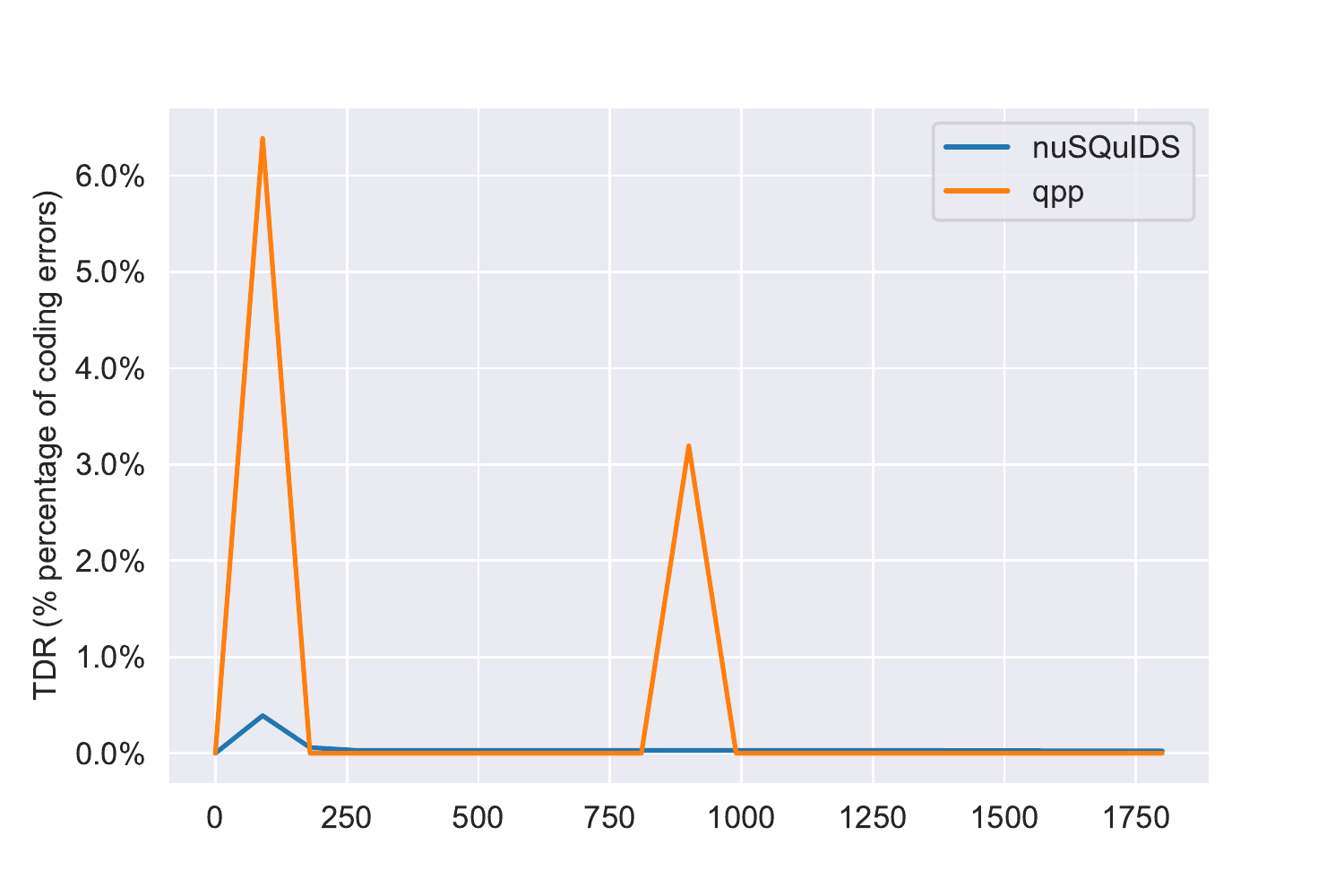}
                \caption{Simulator (code errors)}
                \label{fig:sub1-simulator1}
        \end{subfigure}%
        \begin{subfigure}[b]{0.4\textwidth}
                \includegraphics[width=\linewidth]{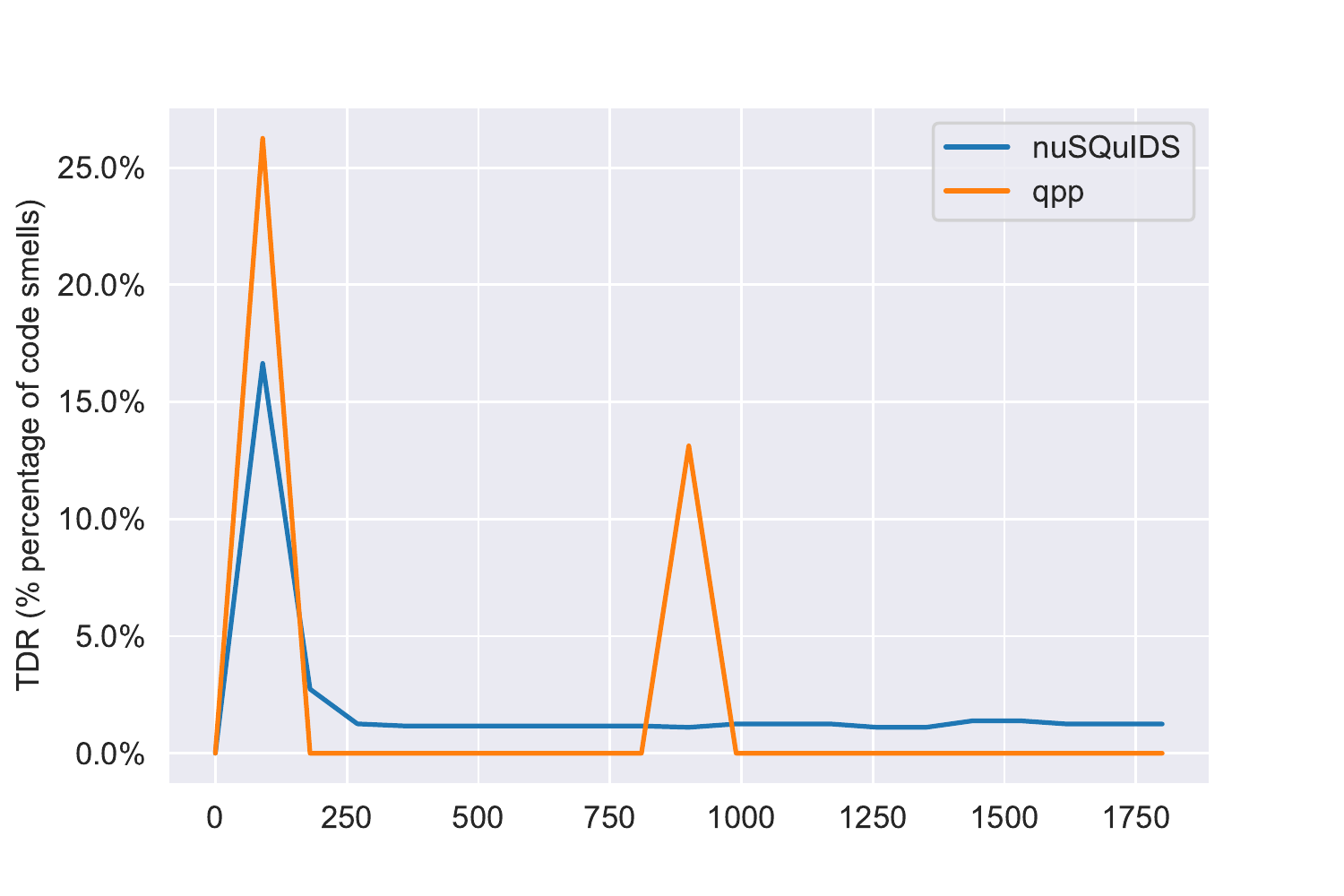}
                \caption{Simulator (code smells)}
                \label{fig:sub1-simulator2}
        \end{subfigure}
          }
        %%%%%% row 3
        
        %\hfill
        \scalebox{1.0}{ 
         \begin{subfigure}[b]{0.4\textwidth}
                \includegraphics[width=\linewidth]{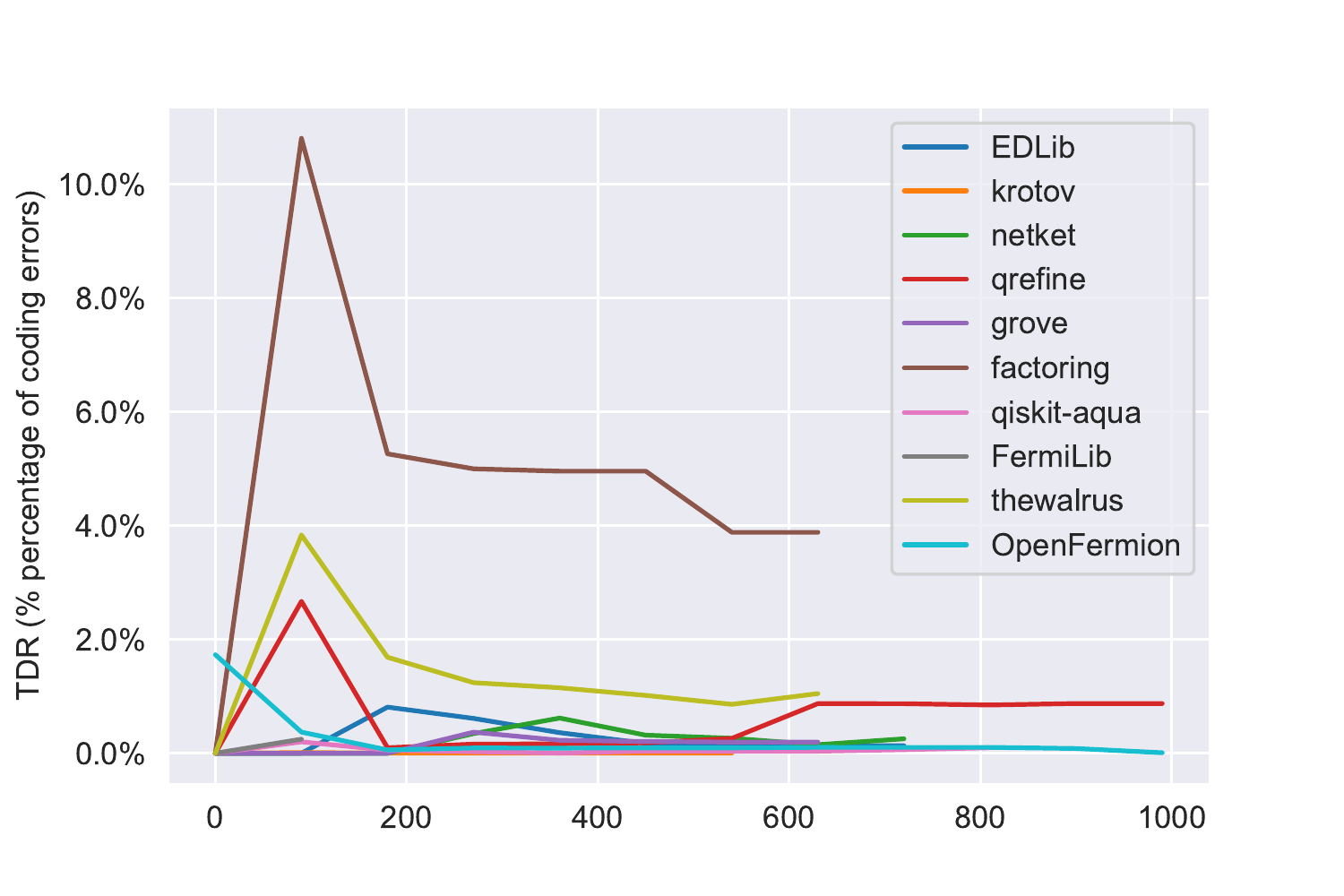}
                \caption{Quantum Algorithms (code errors)}
                \label{fig:sub1-Algorithsm}
        \end{subfigure}%
        \begin{subfigure}[b]{0.4\textwidth}
                \includegraphics[width=\linewidth]{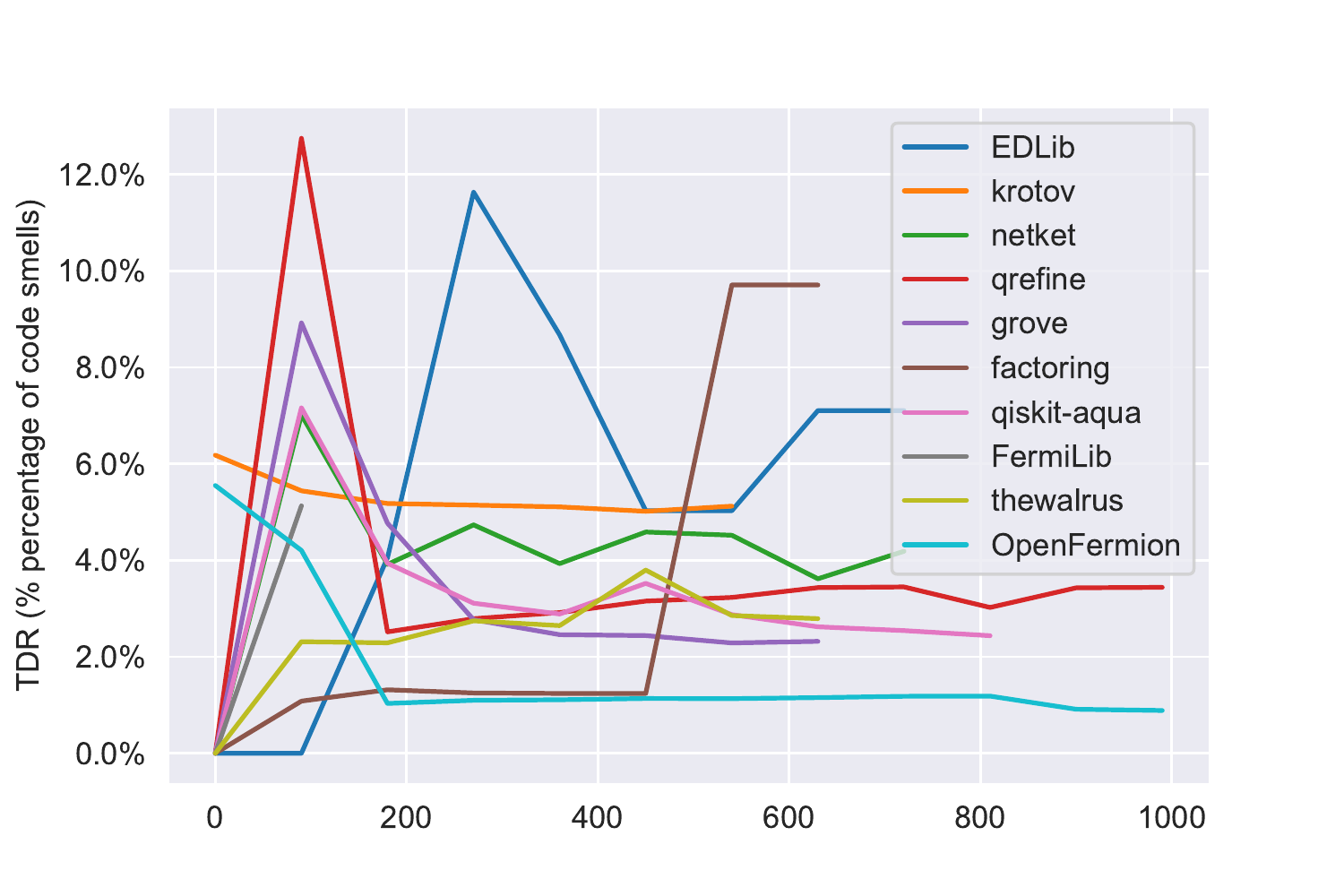}
                \caption{Quantum Algorithms (code smells)}
                \label{fig:sub1-Algorithsm}
        \end{subfigure}%
        }

        \end{figure}
         \begin{figure}\ContinuedFloat
        %%% Compiler and Cryptography
         \scalebox{1}{ 
        \begin{subfigure}[b]{0.4\textwidth}
                \includegraphics[width=\linewidth]{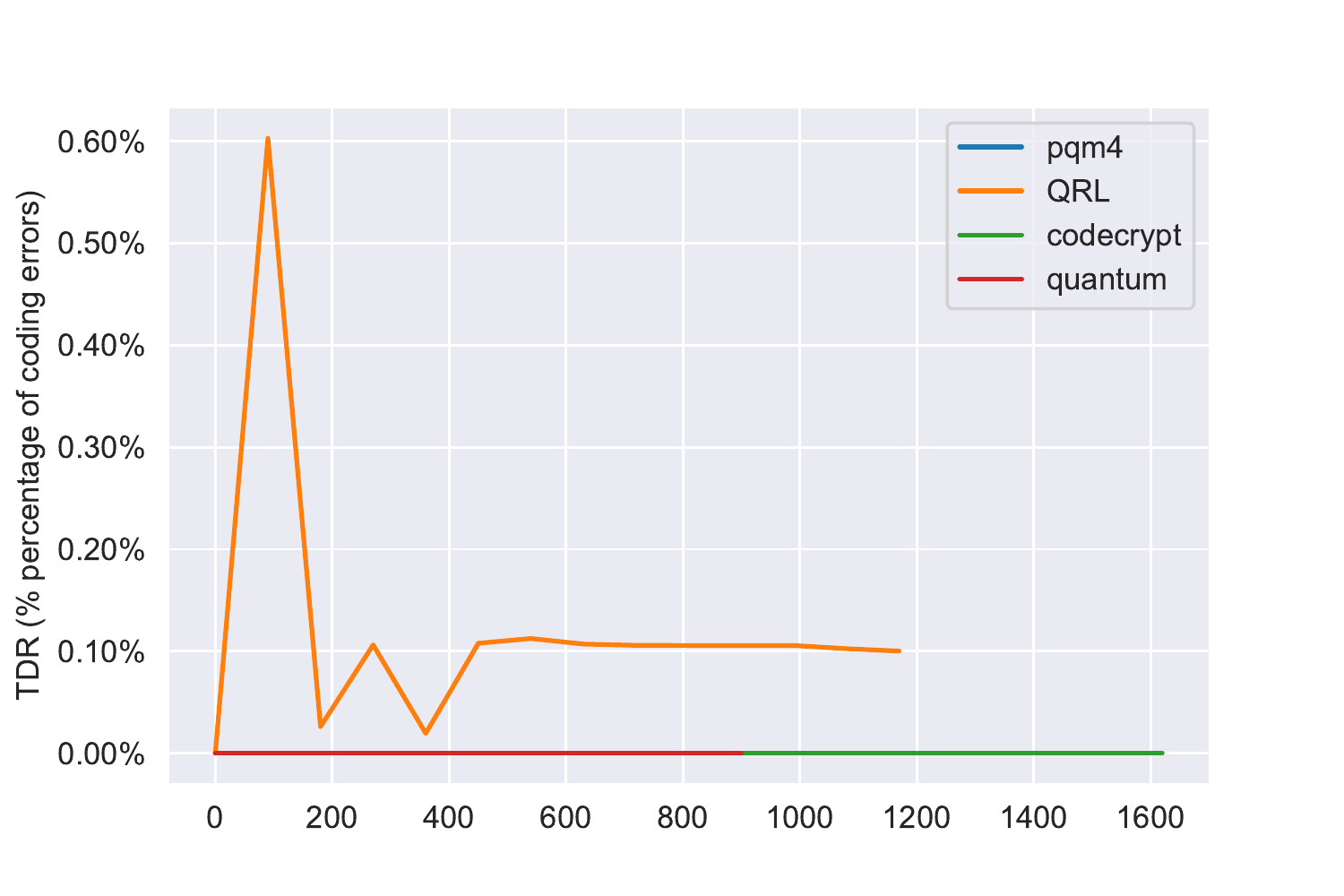}
                \caption{Cryptography (code errors)}
                \label{fig:sub1-simulator1}
        \end{subfigure}%
        \begin{subfigure}[b]{0.4\textwidth}
                \includegraphics[width=\linewidth]{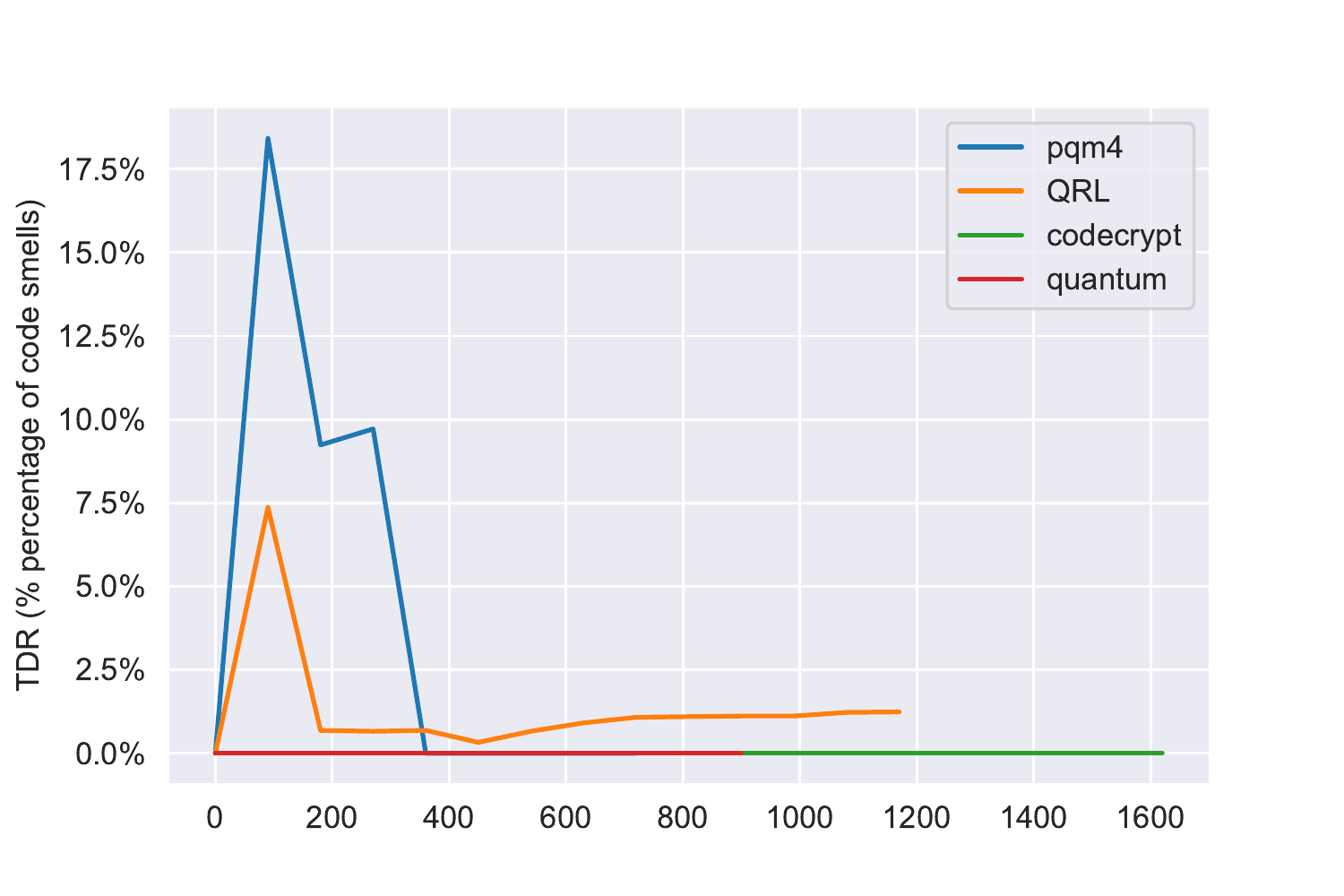}
                \caption{Cryptography (code smells)}
                \label{fig:sub1-simulator2}
        \end{subfigure}
          }
        %%%%%% row 3
        
        %\hfill
        \scalebox{1.0}{ 
         \begin{subfigure}[b]{0.4\textwidth}
                \includegraphics[width=\linewidth]{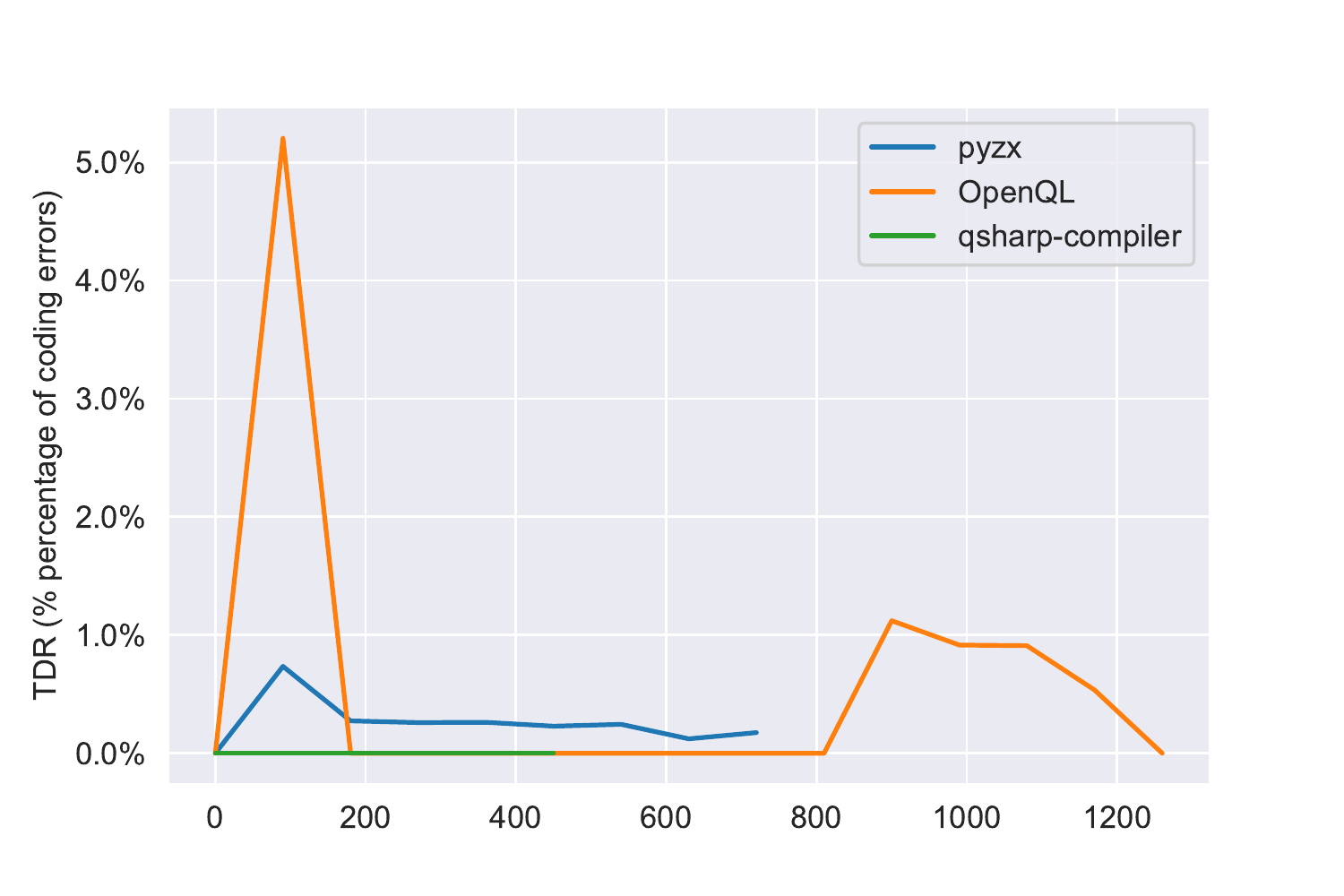}
                \caption{Compiler (code errors)}
                \label{fig:sub1-Algorithsm}
        \end{subfigure}%
        \begin{subfigure}[b]{0.4\textwidth}
                \includegraphics[width=\linewidth]{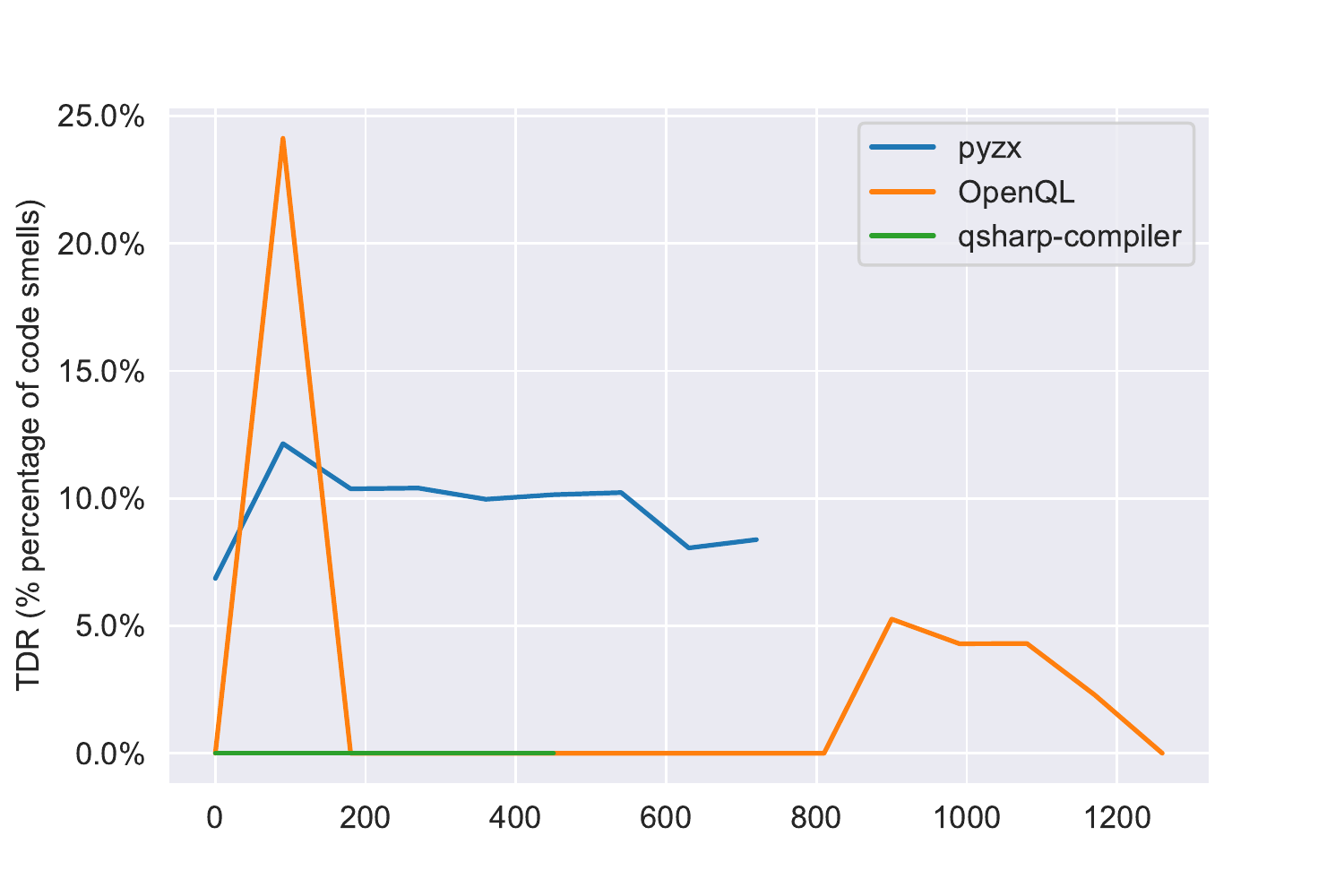}
                \caption{Compiler (code smells)}
                \label{fig:sub1-Algorithsm}
        \end{subfigure}%
        }

	\caption{
% 	\Mehdi{we need to explain the meaning of colored areas and green line} \MO{Not sure how the color keys disappeared! these were the keys;- light blue: code smells, yellow: the coding errors(bugs), green line: the KLOC;} \MO{Also the axis of the figures should be labelled, y-axis left is TDR, y-axis on the right is KLOC, x-axis are the snapshot num (from 0 to latest or we could label x-axis as 0, 90,180,270,.. based on 90days interval)}
	How the technical debts (coding errors on the left and code smells on the right) evolve for ten randomly selected projects in each of the eight studied categories with high rate of technical debts. Each line represents the evolution of technical debt ratio (TDR, described in Section \ref{subsec:background:sonar}) on the y-axis and the number of days based on 90 days interval (snapshots) on the x-axis, for a single project.}\label{fig:evolution-general}
\end{figure}

\definecolor{light-blue}{HTML}{A1E0E6}
\definecolor{light-yellow}{HTML}{FFCD72}
\definecolor{strong-green}{HTML}{219521}
%\begin{adjustbox}{width=0.75\columnwidth,center}
\begin{figure}
%\begin{adjustbox}{width=0.75\columnwidth,center}
    \scalebox{1.0}{  
        \begin{subfigure}[b]{0.4\textwidth}
            \includegraphics[width=\linewidth]{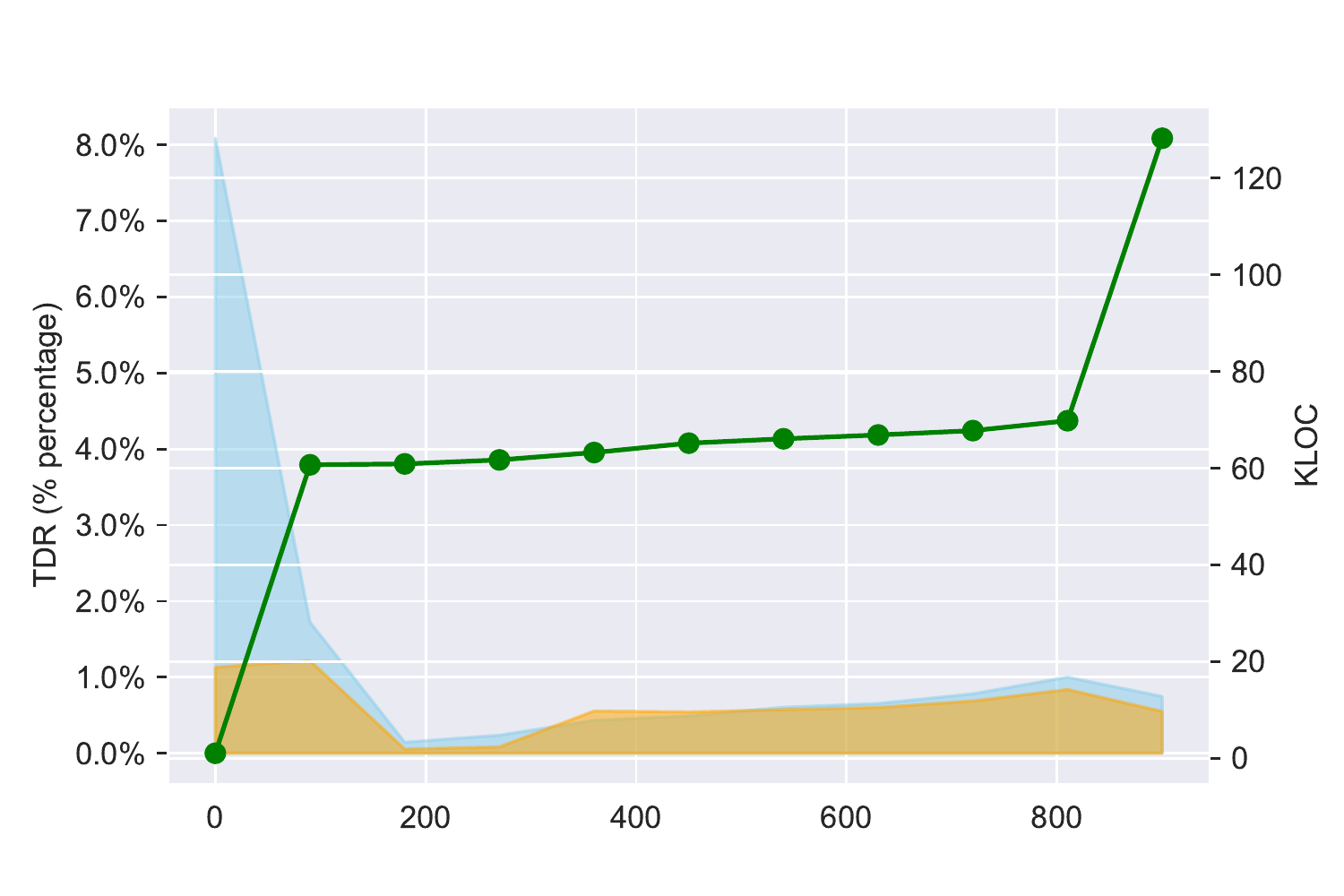}
            \caption{Assembly (qubiter)}
            \label{fig:sub-assembly1}
        \end{subfigure}%
        \begin{subfigure}[b]{0.4\textwidth}
            \includegraphics[width=\linewidth]{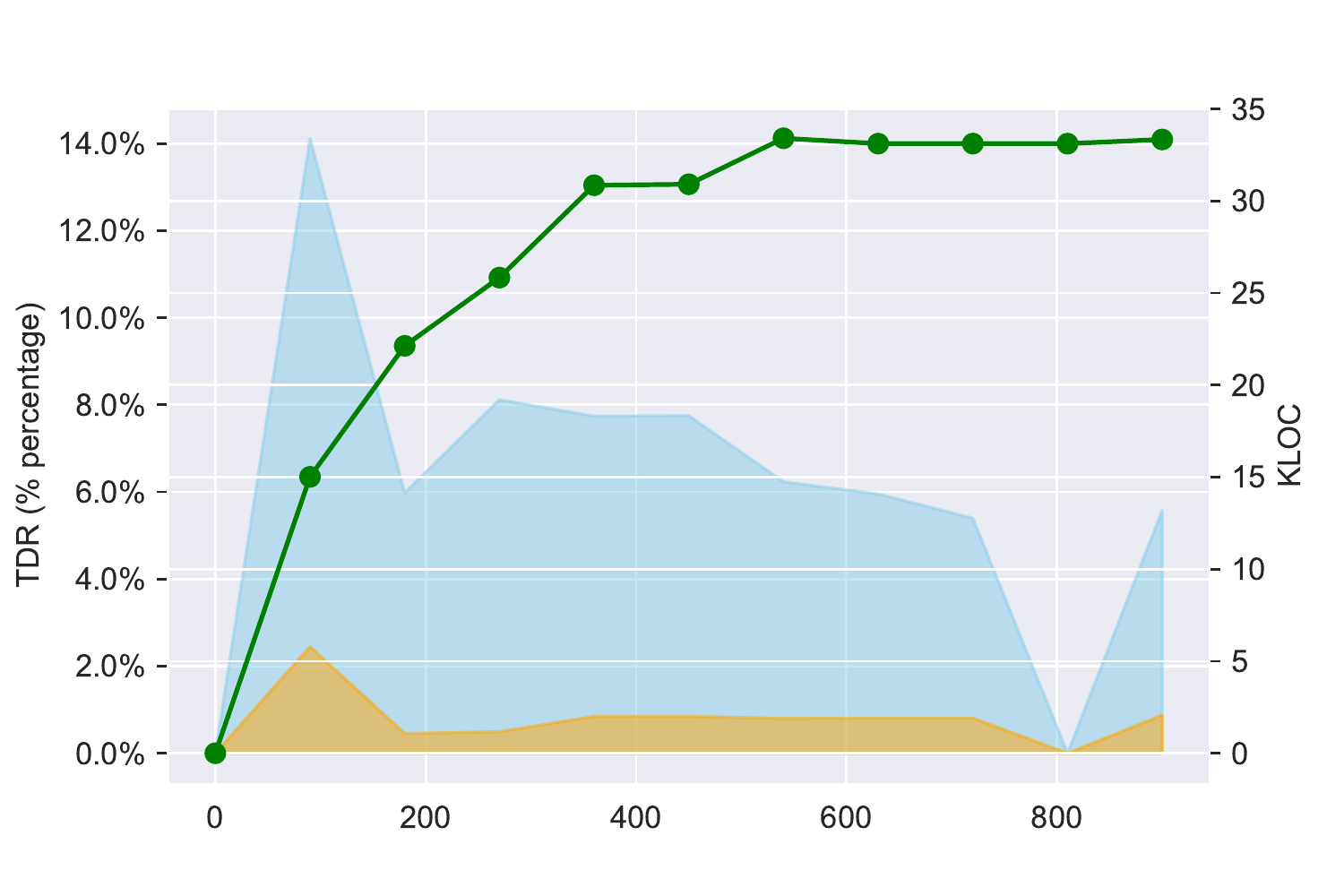}
            \caption{Assembly (pyqgl2)}
            \label{fig:sub-assembly2}
        \end{subfigure}%
        \hfill
    %\caption{Pictures of animals}\label{fig:animals}
    }

	\scalebox{1.0}{
    \begin{subfigure}[b]{0.4\textwidth}
        \includegraphics[width=\linewidth]{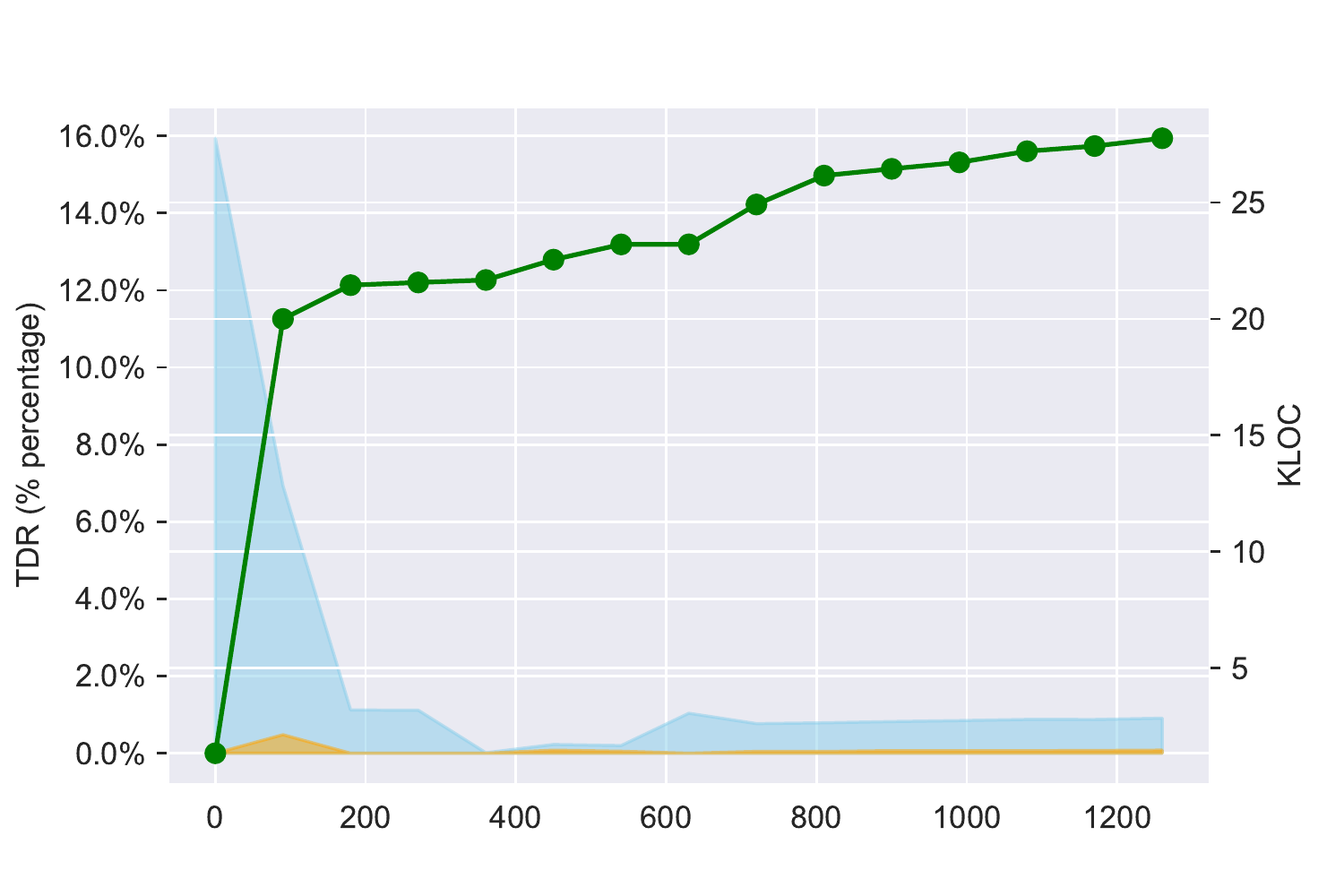}
        \caption{Quantum Annealing (dwave-system)}
        \label{fig:sub-Annealing1}
    \end{subfigure}%
    \begin{subfigure}[b]{0.4\textwidth}
        \includegraphics[width=\linewidth]{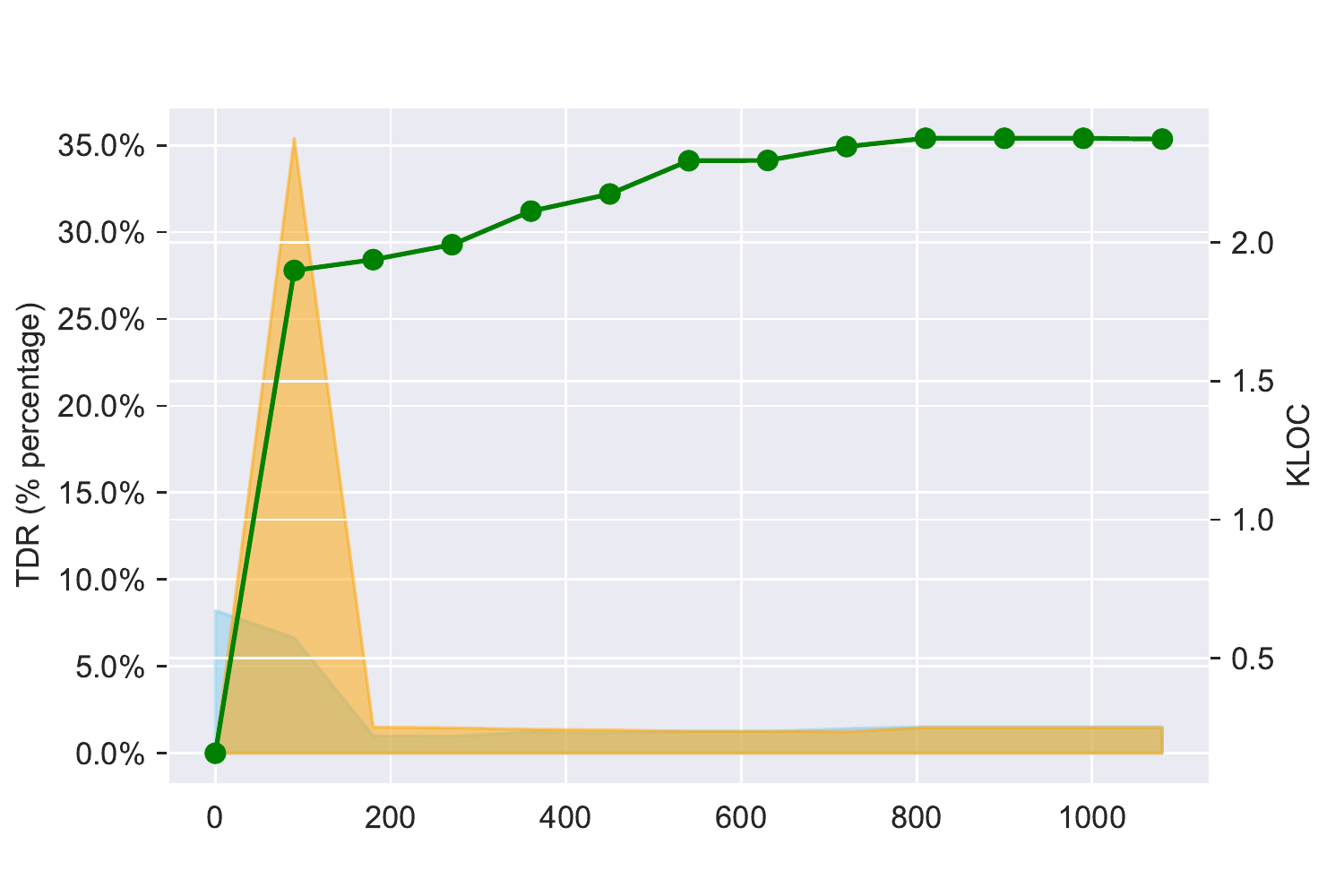}
        \caption{Quantum Annealing(qbsolv)}
        \label{fig:sub-Annealing2}
    \end{subfigure}
    \hfill
    }
        %%% row2
       % 
        \scalebox{1}{ 
         \begin{subfigure}[b]{0.4\textwidth}
                \includegraphics[width=\linewidth]{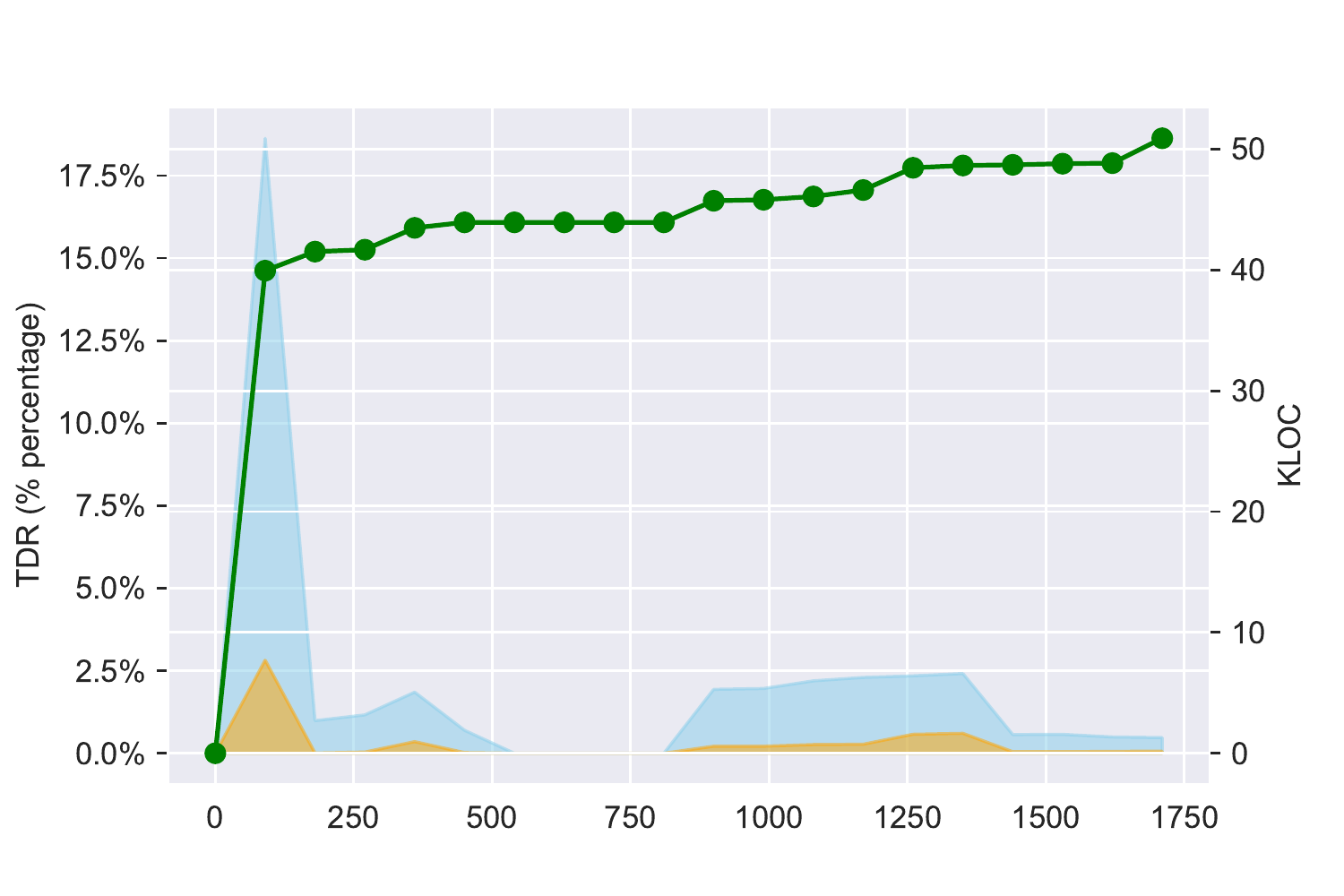}
                \caption{Experimentation (PyQLab)}
                \label{fig:sub-Experimentation1}
        \end{subfigure}%
        \begin{subfigure}[b]{0.4\textwidth}
                \includegraphics[width=\linewidth]{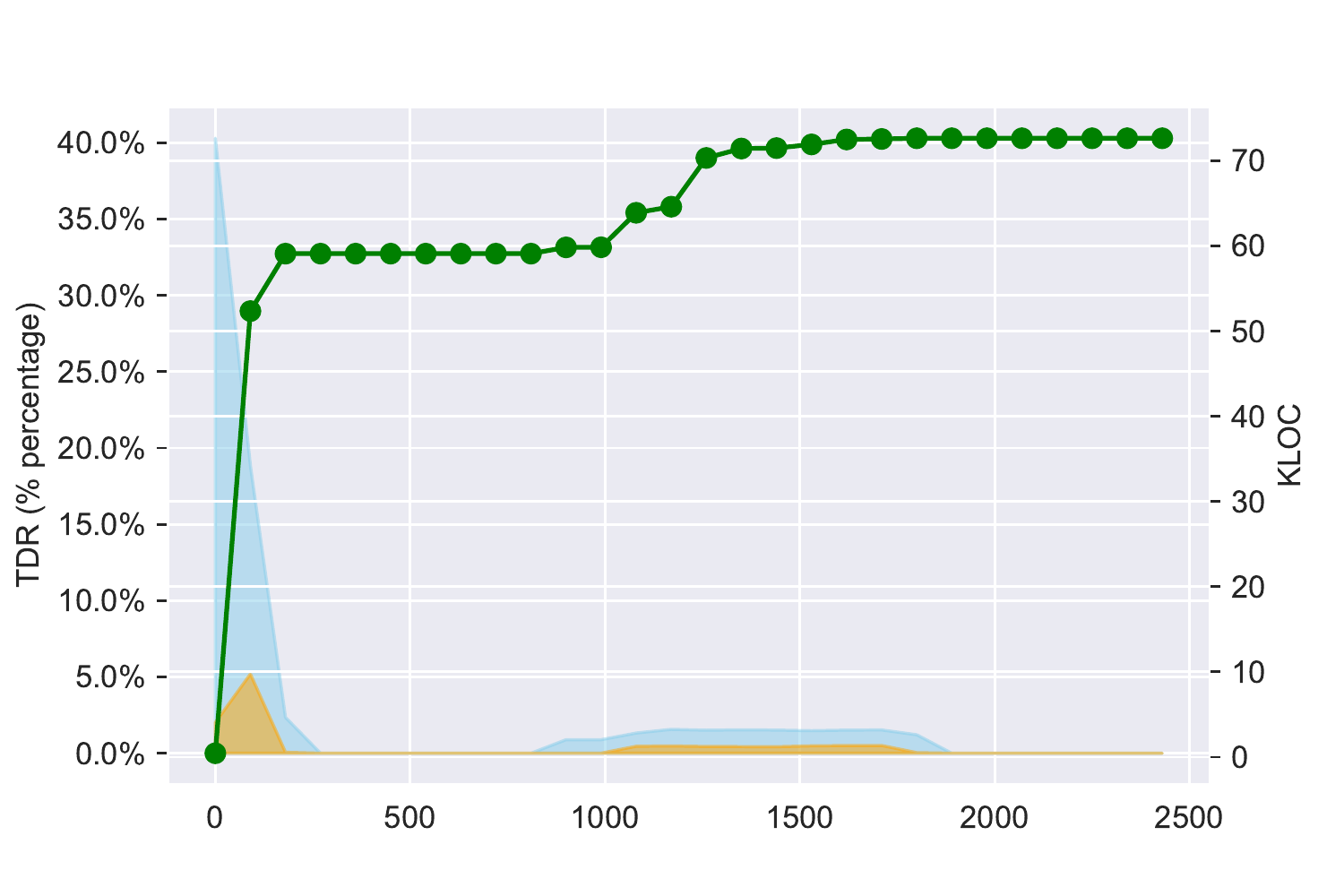}
                \caption{Experimentation (QGL)}
                \label{fig:sub-Experimentation2}
        \end{subfigure}%
        }
        %\hfill
        \scalebox{1}{ 
        \begin{subfigure}[b]{0.4\textwidth}
                \includegraphics[width=\linewidth]{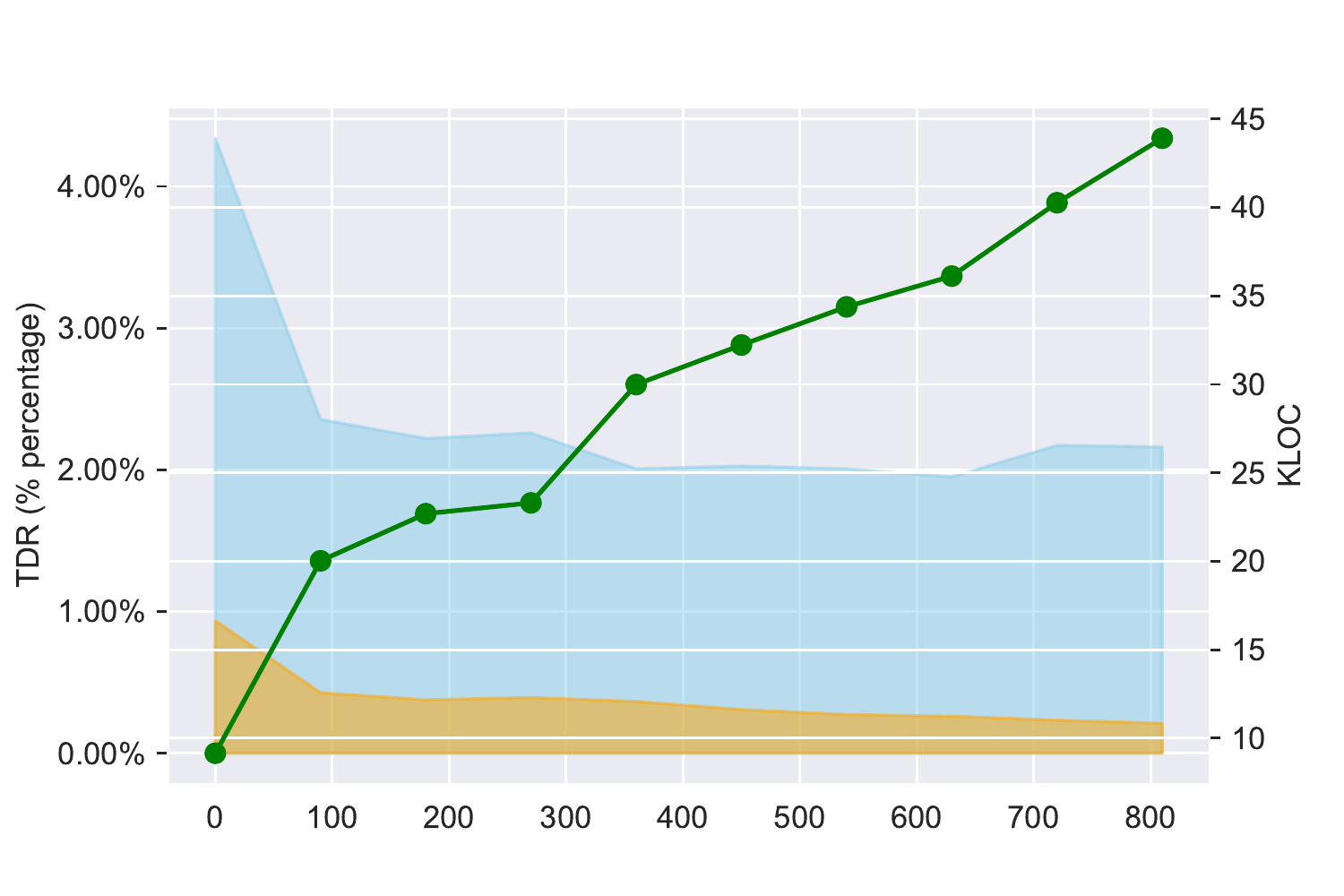}
                \caption{Full-stack library (strawberryfields)}
                \label{fig:sub-Full-stack1}
        \end{subfigure}%
        \begin{subfigure}[b]{0.4\textwidth}
                \includegraphics[width=\linewidth]{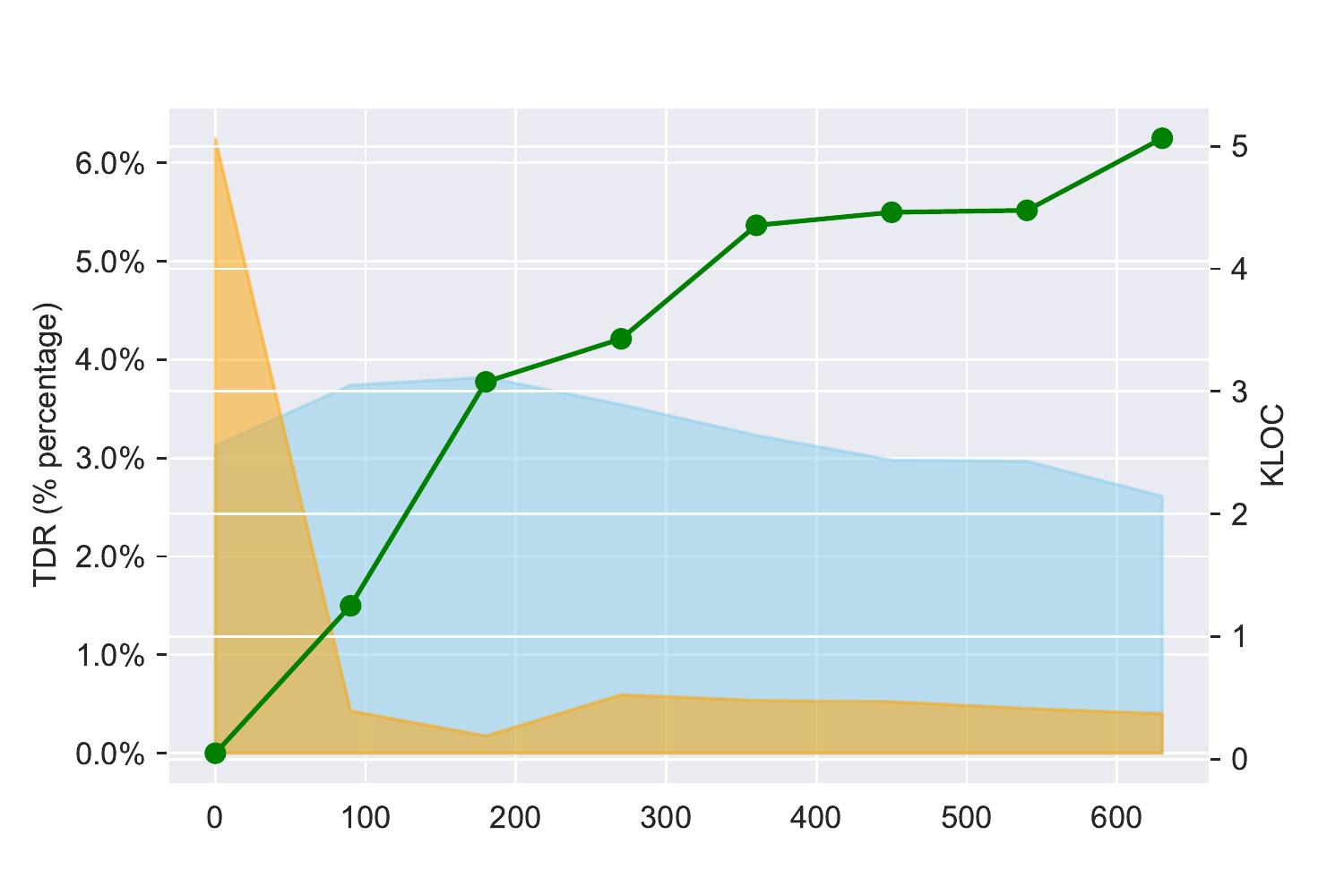}
                \caption{Full-stack library (Blueqat)}
                \label{fig:sub-Full-stack2}
        \end{subfigure}
          }
        %%%%%% row 3
        
        %\hfill
        \scalebox{1.0}{ 
         \begin{subfigure}[b]{0.4\textwidth}
                \includegraphics[width=\linewidth]{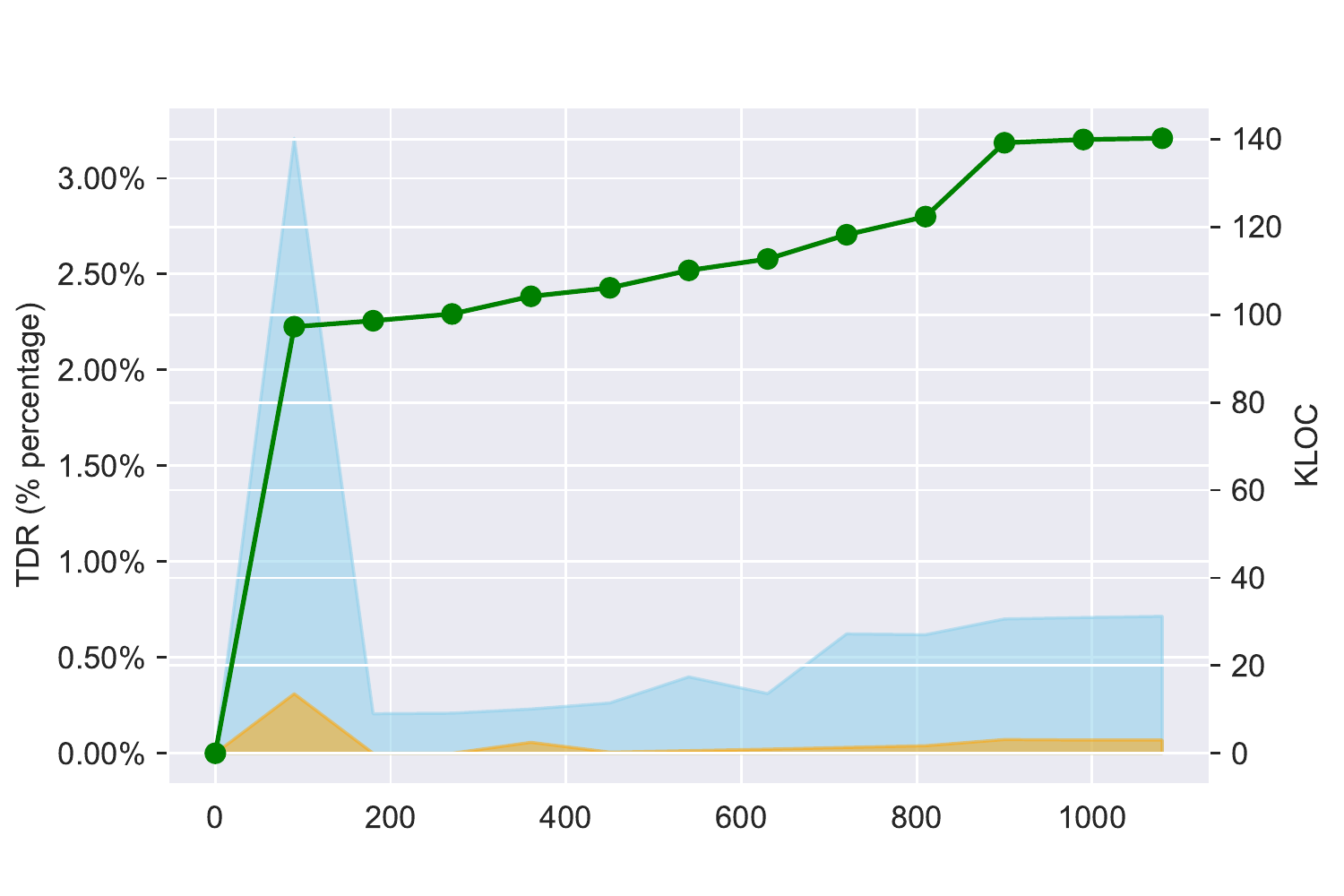}
                \caption{Quantum-chemistry (QCFractal)}
                \label{fig:sub-Chemistry1}
        \end{subfigure}%
        \begin{subfigure}[b]{0.4\textwidth}
                \includegraphics[width=\linewidth]{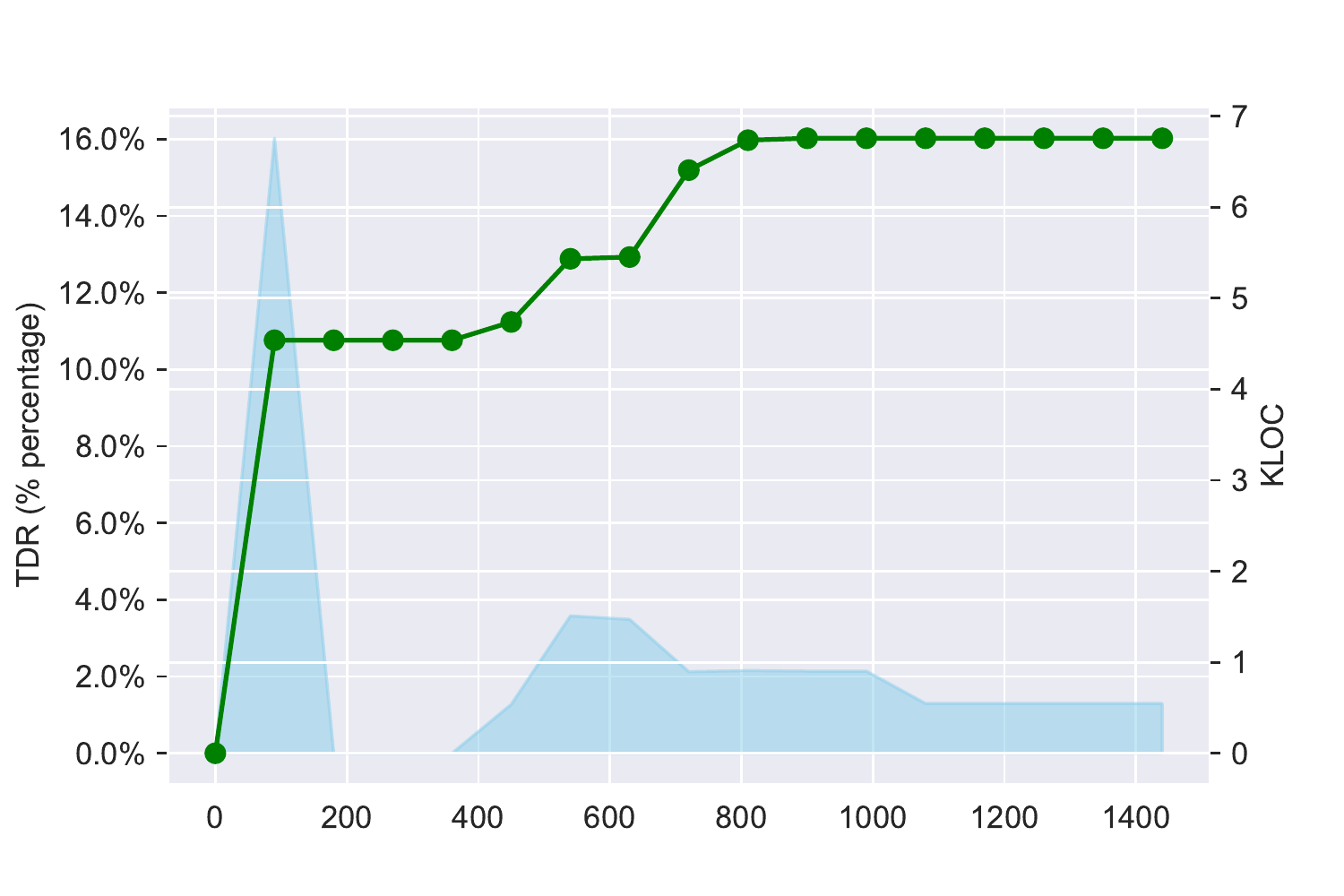}
                \caption{Quantum-chemistry (CheMPS2)}
                \label{fig:sub-Chemistry2}
        \end{subfigure}%
        }
       
    \end{figure}
    %\end{adjustbox}
    \begin{figure}\ContinuedFloat
        \scalebox{1.0}{ 
        \begin{subfigure}[b]{0.4\textwidth}
                \includegraphics[width=\linewidth]{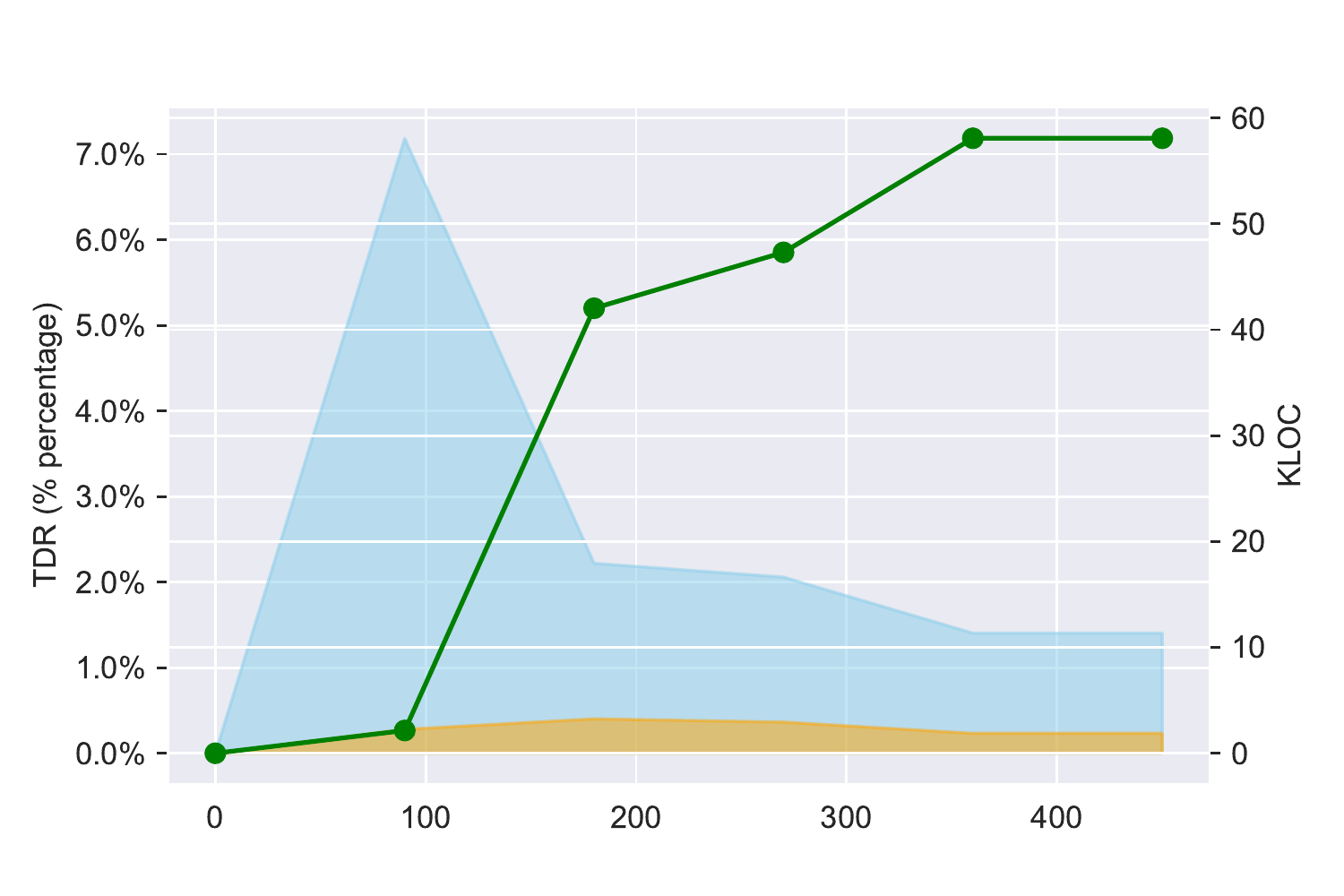}
                \caption{Toolkit (qucat)}
                \label{fig:sub-Toolkit1}
        \end{subfigure}%
        \begin{subfigure}[b]{0.4\textwidth}
                \includegraphics[width=\linewidth]{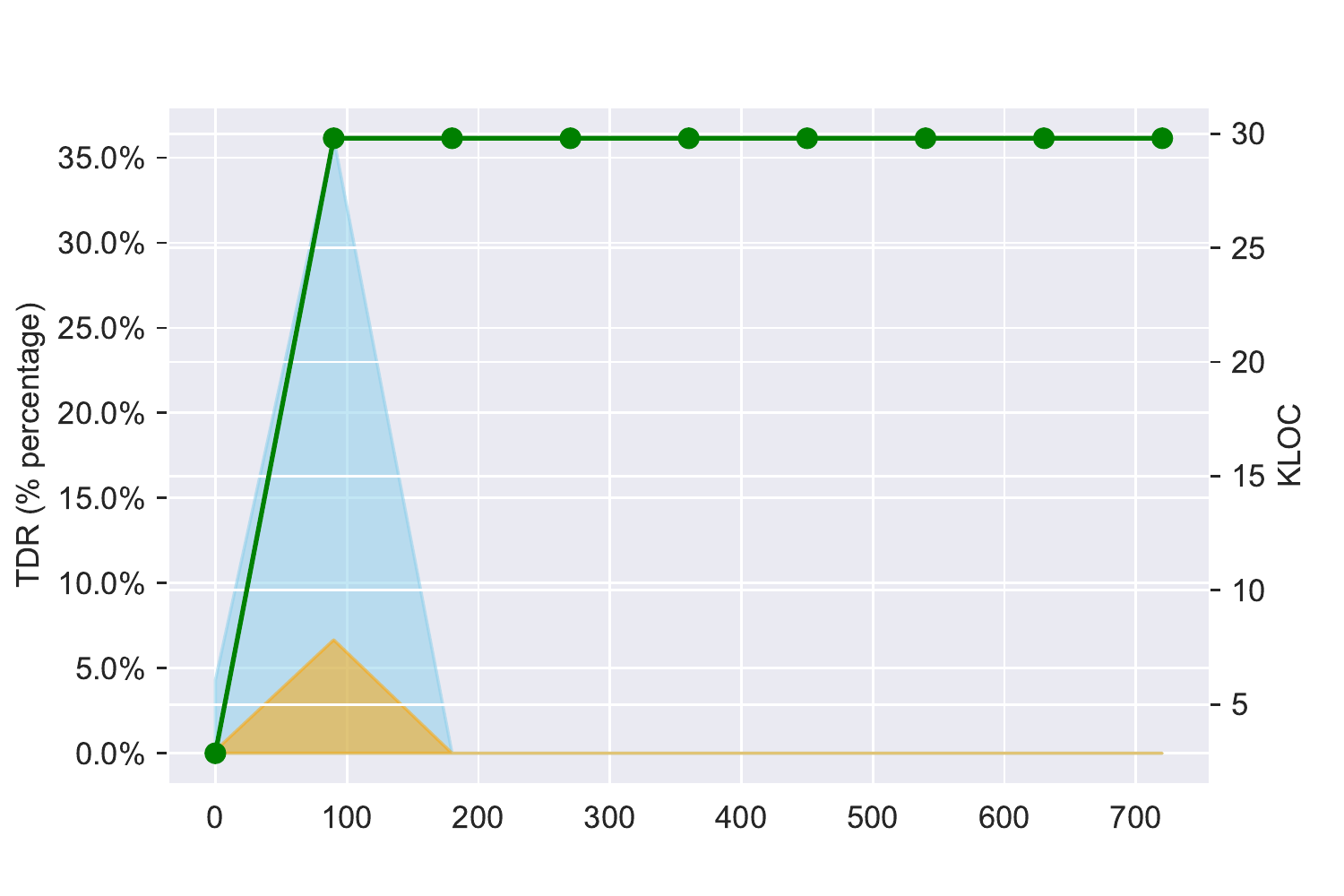}
                \caption{Toolkit (bloomberg)}
                \label{fig:sub-Toolkit2}
        \end{subfigure}
          }
        
         %%%%%% row 4
        
        %\hfill
        \scalebox{1.0}{ 
         \begin{subfigure}[b]{0.4\textwidth}
                \includegraphics[width=\linewidth]{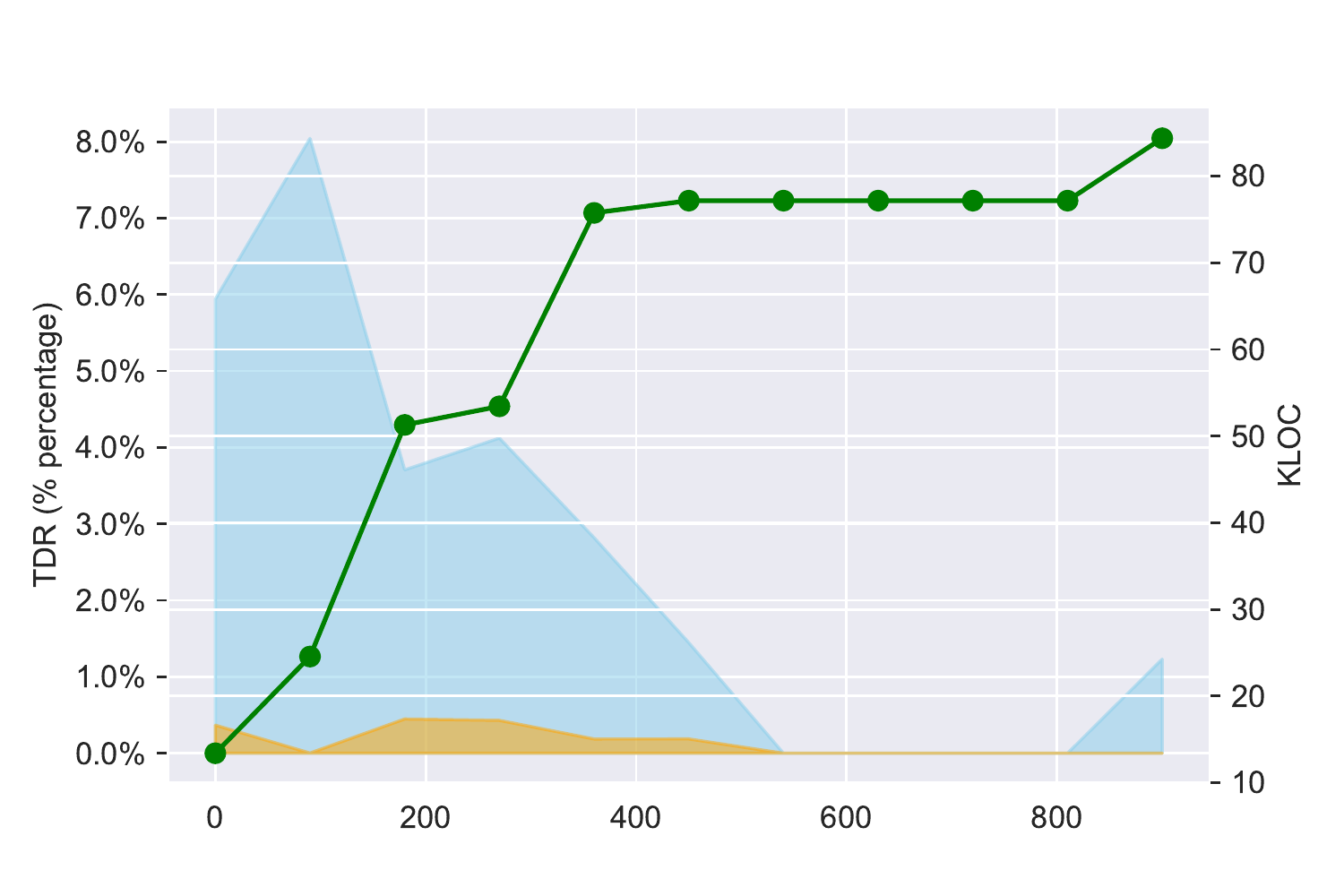}
                \caption{Simulator (SimulaQron)}
                \label{fig:sub-simulator1}
        \end{subfigure}%
        \begin{subfigure}[b]{0.4\textwidth}
                \includegraphics[width=\linewidth]{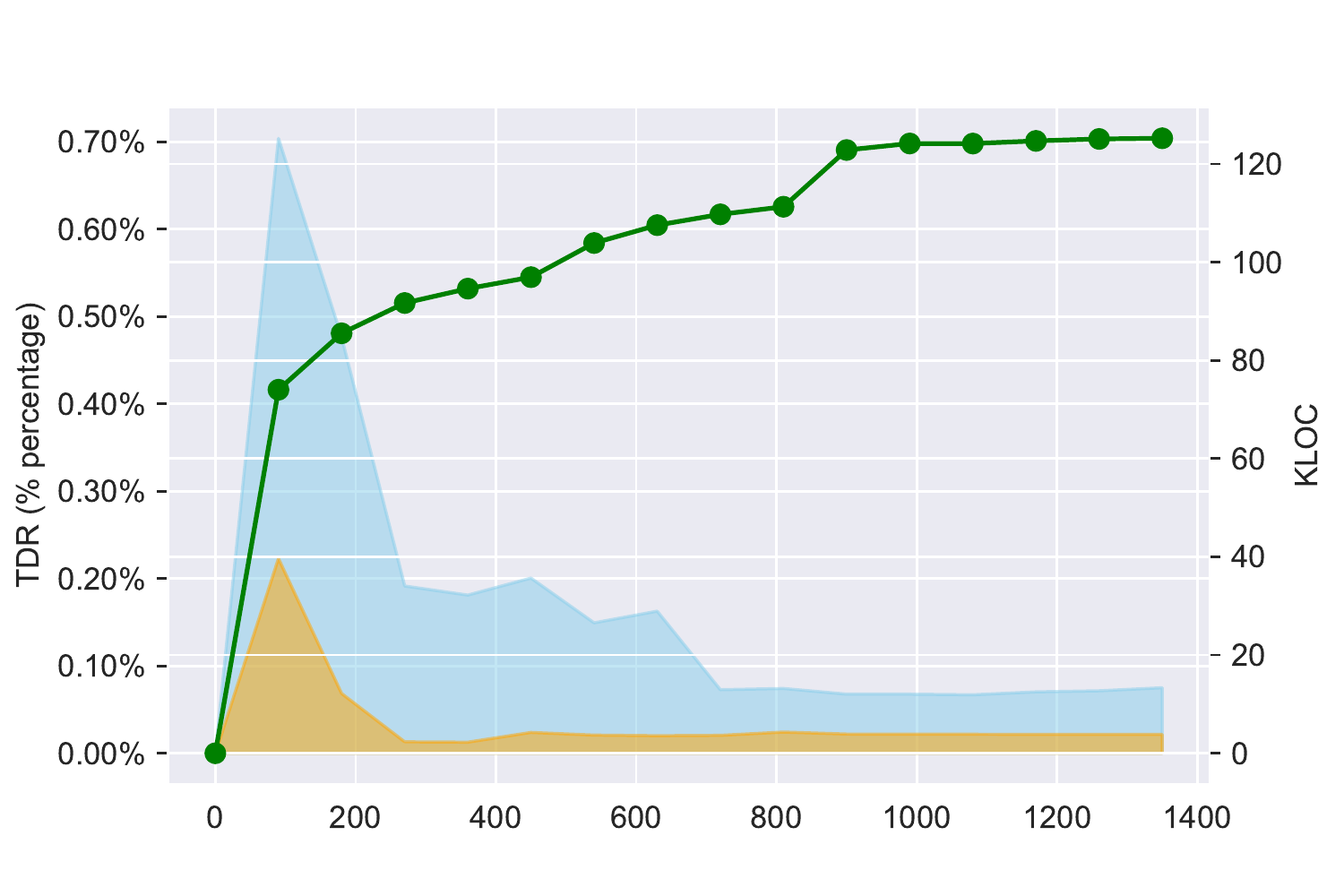}
                \caption{Simulator (Quirk)}
                \label{fig:sub-simulator2}
        \end{subfigure}%
        }
         \hfill
        \scalebox{1.0}{
        \begin{subfigure}[b]{0.4\textwidth}
                \includegraphics[width=\linewidth]{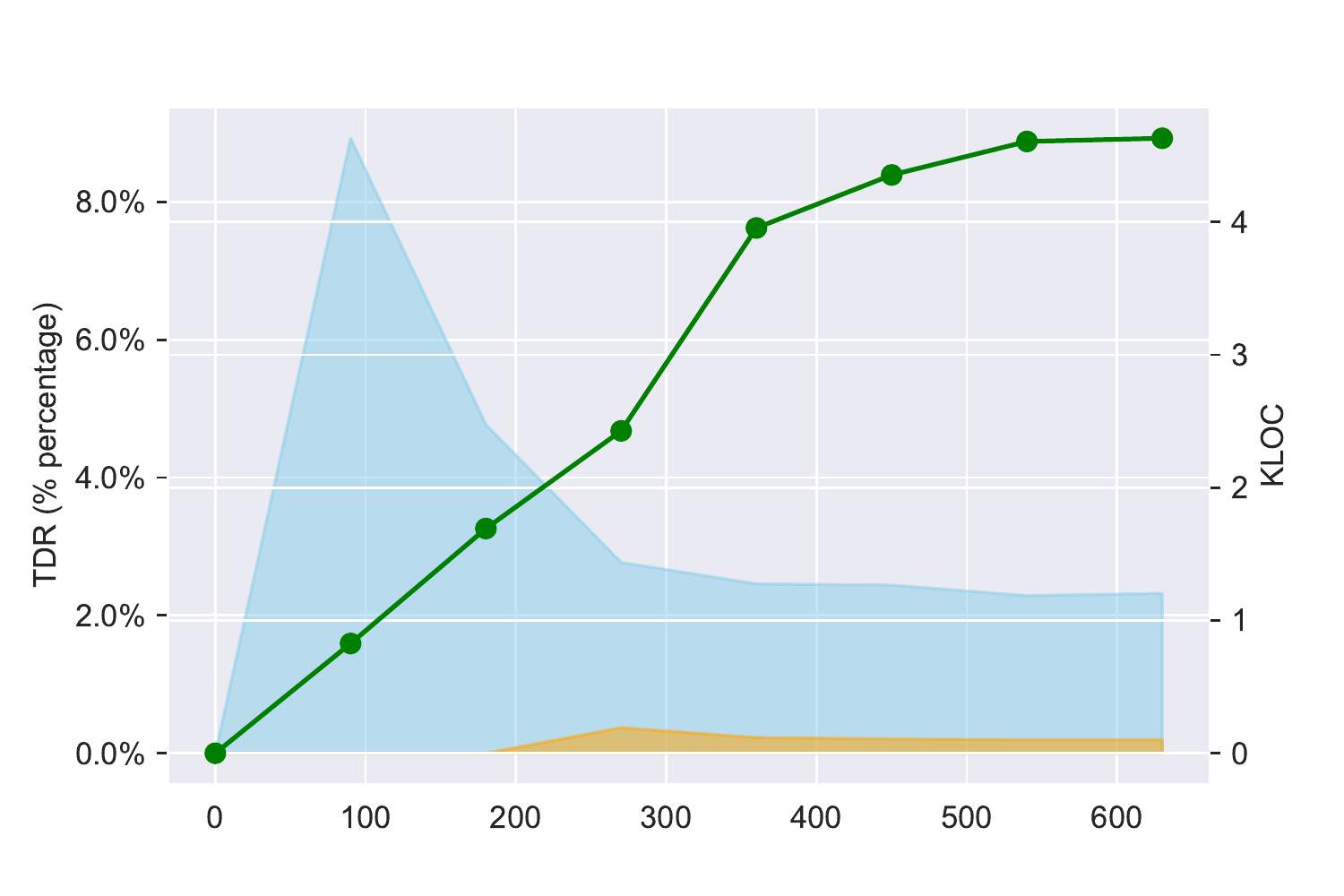}
                \caption{Quantum algorithm (grove)}
                \label{fig:sub-None-Quantum1}
        \end{subfigure}%
        \begin{subfigure}[b]{0.4\textwidth}
                \includegraphics[width=\linewidth]{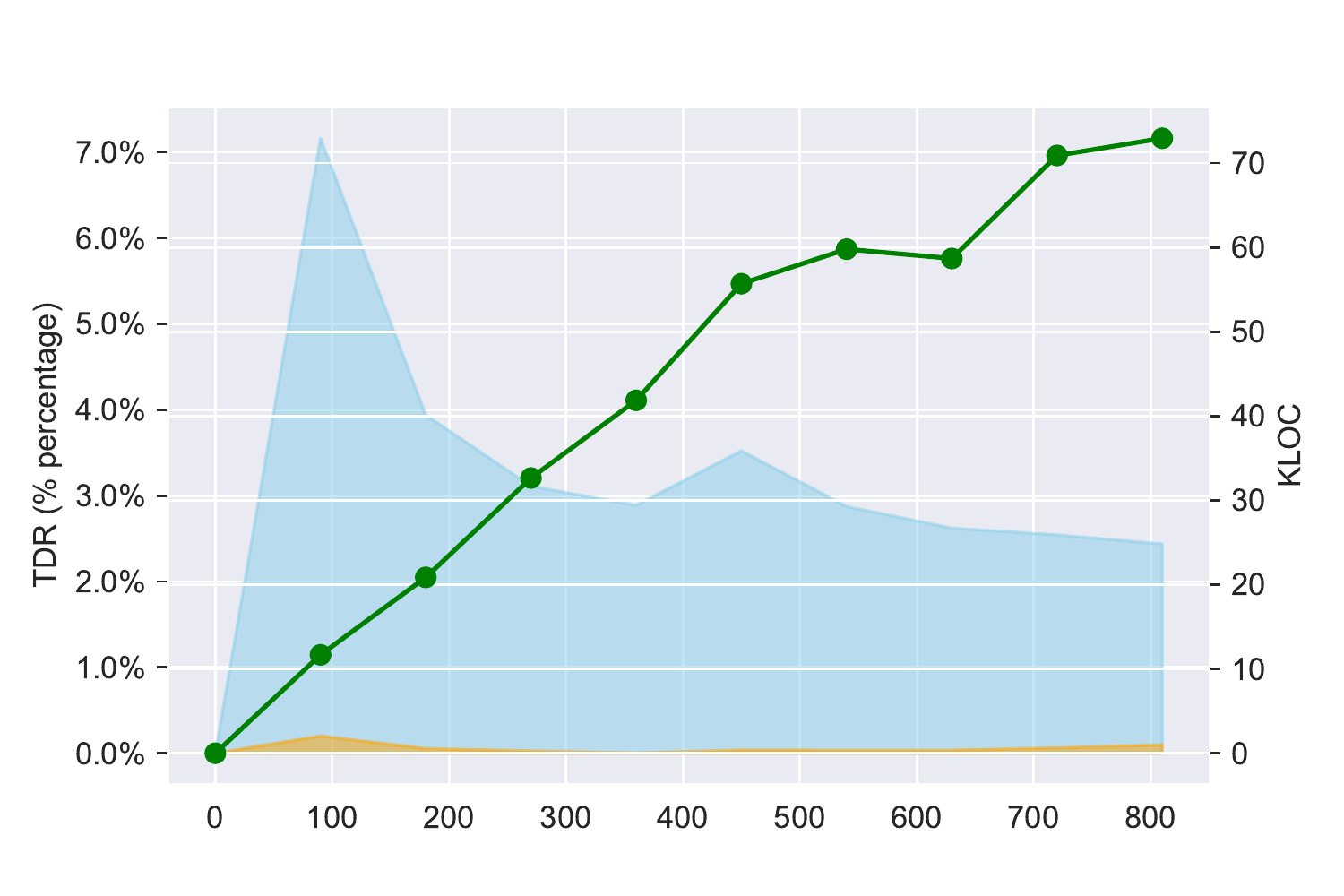}
                \caption{Quantum algorithm (qiskit-aqua)}
                \label{fig:sub-None-Quantum2}
        \end{subfigure}
        }

        %%% Cryptography and compiler
        \end{figure}
         \begin{figure}\ContinuedFloat
         \scalebox{1.0}{ 
         \begin{subfigure}[b]{0.4\textwidth}
                \includegraphics[width=\linewidth]{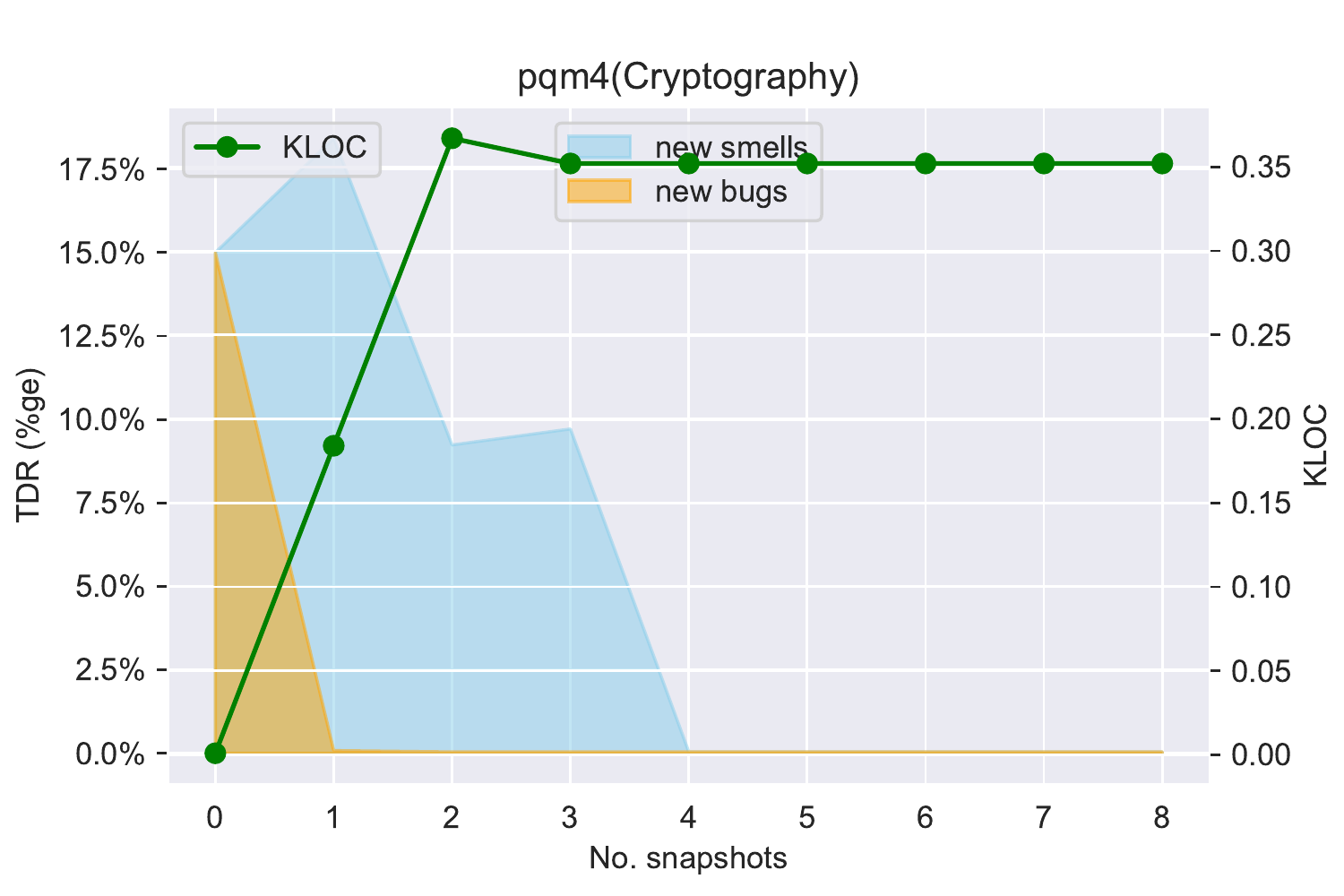}
                \caption{Cryptography (pqm4)}
                \label{fig:sub-simulator1}
        \end{subfigure}%
        \begin{subfigure}[b]{0.4\textwidth}
                \includegraphics[width=\linewidth]{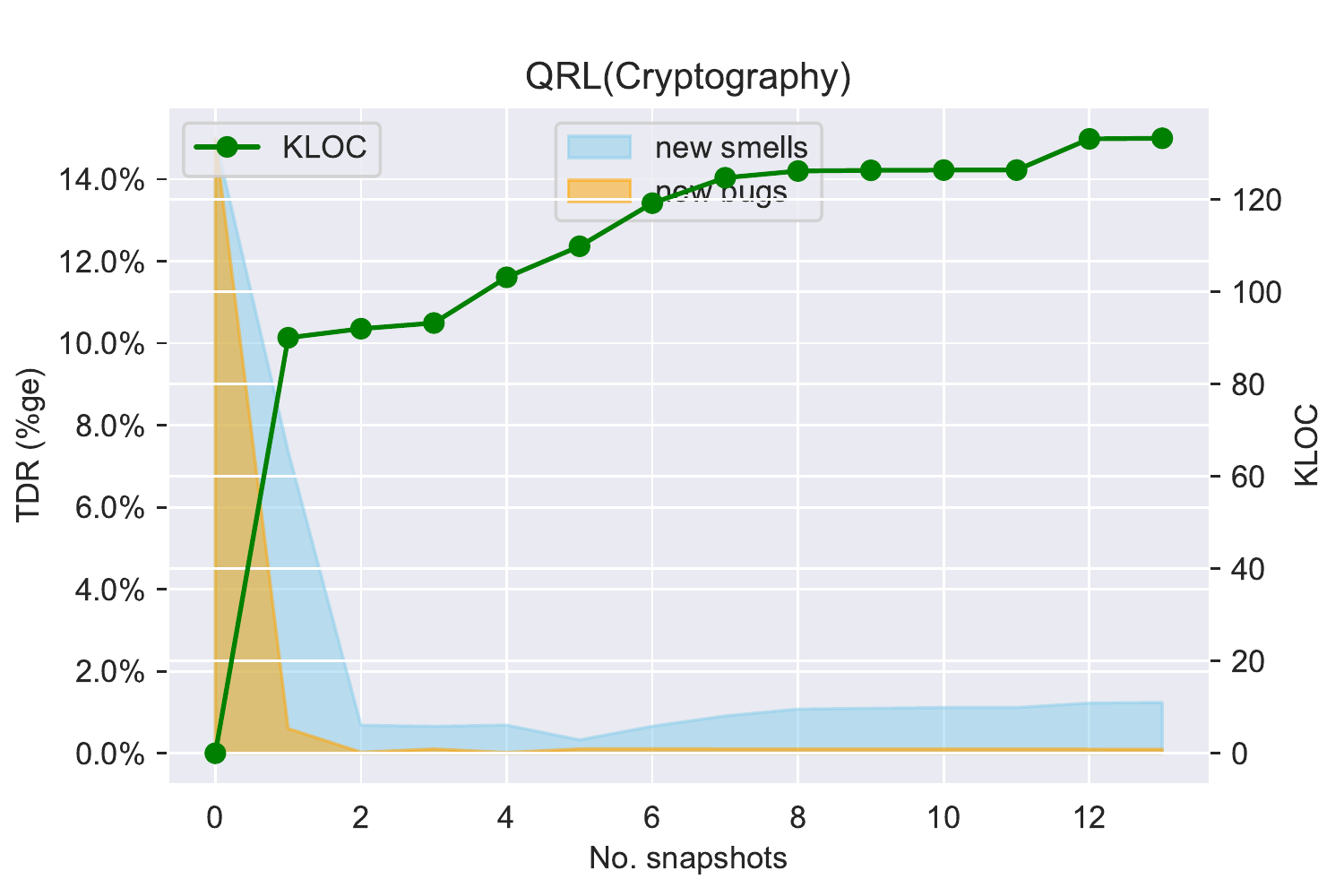}
                \caption{Cryptography (QRL)}
                \label{fig:sub-simulator2}
        \end{subfigure}%
        }
         \hfill
        \scalebox{1.0}{
        \begin{subfigure}[b]{0.4\textwidth}
                \includegraphics[width=\linewidth]{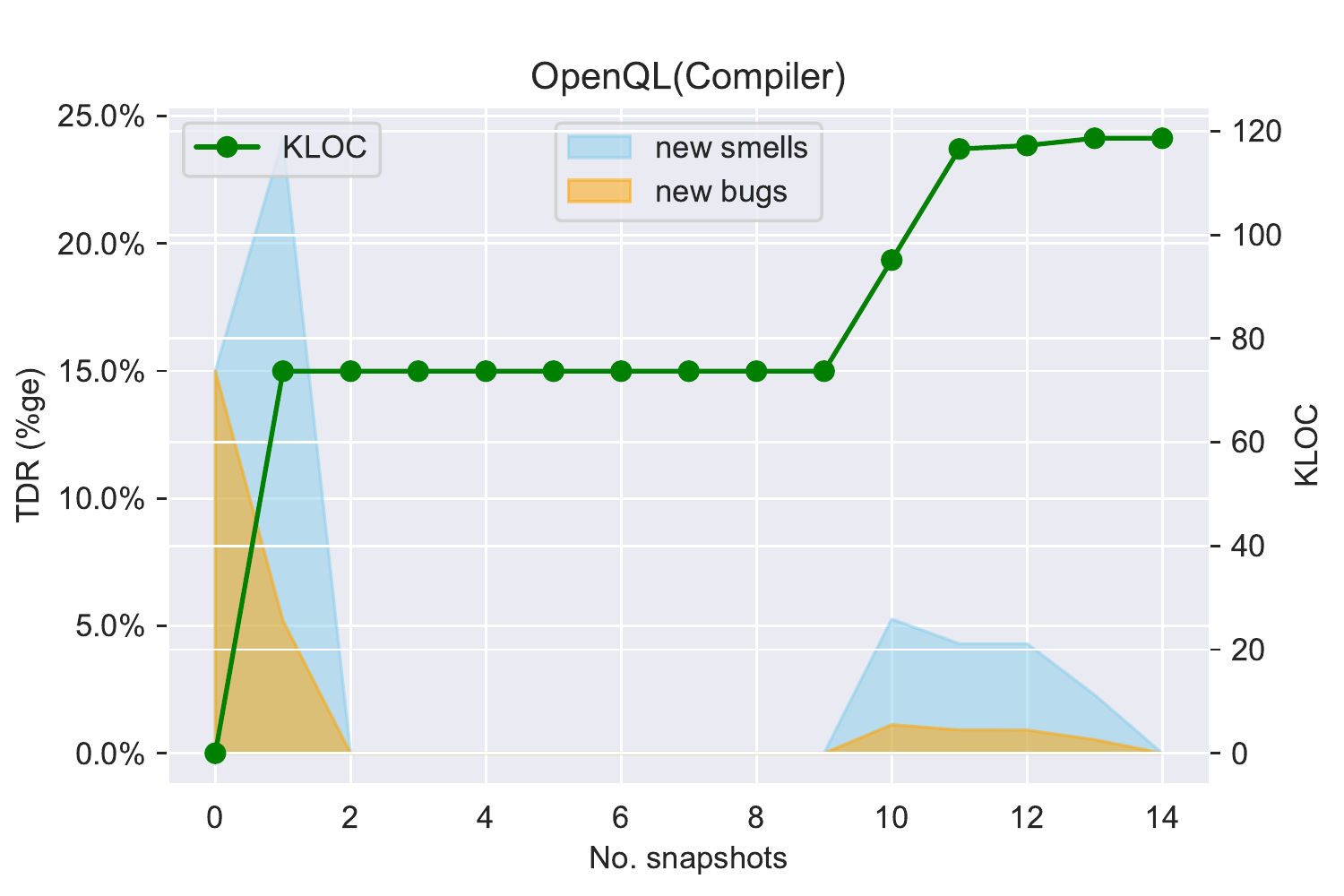}
                \caption{Compiler (OpenQL)}
                \label{fig:sub-None-Quantum1}
        \end{subfigure}%
        \begin{subfigure}[b]{0.4\textwidth}
                \includegraphics[width=\linewidth]{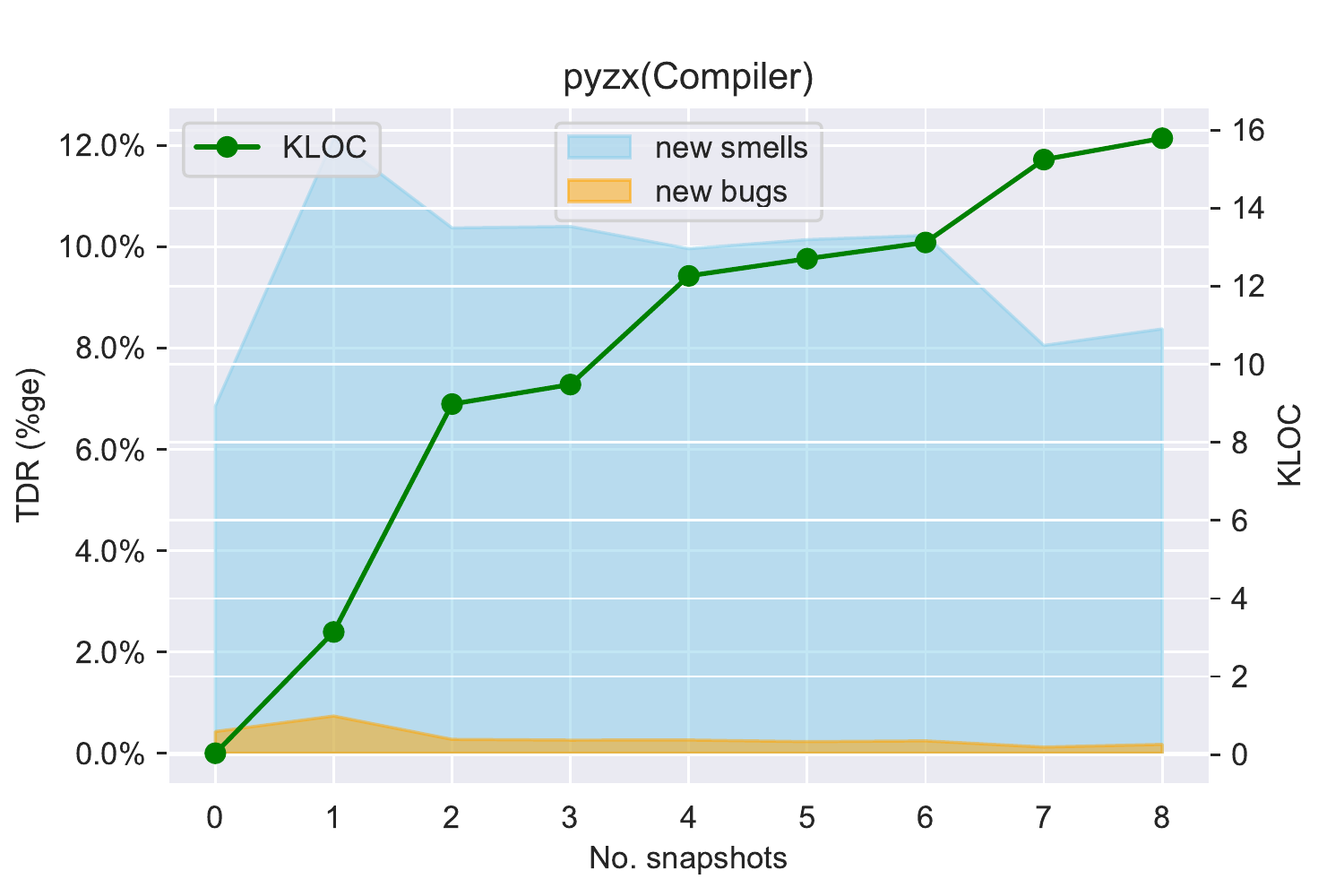}
                \caption{Compiler (pyzx)}
                \label{fig:sub-None-Quantum2}
        \end{subfigure}
        }
	\caption{
% 	\Mehdi{we need to explain the meaning of colored areas and green line} \MO{Not sure how the color keys disappeared! these were the keys;- light blue: code smells, yellow: the coding errors(bugs), green line: the KLOC;} \MO{Also the axis of the figures should be labelled, y-axis left is TDR, y-axis on the right is KLOC, x-axis are the snapshot num (from 0 to latest or we could label x-axis as 0, 90,180,270,.. based on 90days interval)}
	How the technical debts (coding errors \crule[light-yellow]{0.3cm}{0.3cm} and code smells \crule[light-blue]{0.3cm}{0.3cm}) evolve with development activities (the added and deleted lines of code (KLOC) \crule[strong-green]{0.5cm}{0.07cm})  for two randomly selected projects in each of the eight studied categories with high rate of technical debts. We show the technical debt ratio (TDR, described in Section \ref{subsec:background:sonar}) on the left side of y-axis and the KLOC on the right side (calculated from LOC returned by SonarQube) and the number of days based on 90 days interval (snapshots) on the x-axis.}\label{fig:evolution-top-6}
\end{figure}

\fig\ref{fig:evolution-top-6} (\ref{fig:sub-assembly1} to \ref{fig:sub-None-Quantum2}) illustrates how new technical debts were added to the projects over time across the studied snapshots. The technical debt in the first snapshot of each application is new; the divergence from the horizontal line indicates the supplementary debt that has been introduced or removed or both. Each sub-figure is scaled to indicate the technical debts ration (TDR) on the left and KLOC on the right. The KLOC allows us to  identify the snapshots where key development activities took place, such as new code added or deleted and how these activities are related to technical debts.  

In the figure, we observed that there is a general sharp raise of TDR at the initial phase of the software development (more than $10\%$ of corresponding to SQALE maintainability rating from $B$ and $C$), which later gradually decreased as the code base matured in most of the target projects. Also, according to \fig\ref{fig:evolution-top-6}, we found that the introduction of technical debts is highly related to the project size (KLOC). For the Assembly project \texttt{artiste-qb-net/qubiter}, we can see that most of the code was added within the first two snapshots (corresponding to the development time of up to 180 days), in which most new added debts are related to code smells. We observed fewer code changes in the later snapshots of the project. These changes have little relationship with technical debts. 
Similarly, we observed a sharp rise of technical debts in the \texttt{BBN-Q/pyqgl2} project.
%and most of the projects where the type code smells dominates across most of the selected project's snapshots except for projects \texttt{dwavesystems/qbsolv} and texttt{Blueqat/Blueqat}. 

For the case of the Quantum Annealing project \texttt{dwavesystems/qbsolv}, the majority of the codebase was added between the first $90$ and $180$ days. The added code contains more coding errors than code smells. The number of technical debts remains stable in the later snapshots of the project. Similar trends are observed in the Full-stack library project
% \texttt{Blueqat/Blueqat}\footnote{\url{https://github.com/Blueqat/Blueqat}} 
\href{https://github.com/Blueqat/Blueqat}{{\texttt{Blueqat/Blueqat}}}
where more coding errors were detected mainly in the initial snapshot of the projects compared to the later snapshots. 

Higher debts and the sharp rise of code lines at the initial snapshots may be due to the fact that most new files are added at the initial phase of the development. When a project becomes mature, fewer added and deleted lines are observed, which can be associated with minor or modification changes.
%KLOC may be associated with minor or modification changes, such as fewer lines within the existing files. 
This was also observed in some of the projects such as \texttt{artiste-qb-net/qubiter} where a single commit
\href{https://api.github.com/repos/artiste-qb-net/qubiter/commits/e3f55398d7dc145641ff232b4ac78b8a5024f284}{{\texttt{\#e3f5539}}} added 71 new files with over 8,800 new lines of code in the first snapshot. With a manual investigation on 30 of these files, we found both coding errors and code smells existing in the files. Some of the files were later refactored, which implies that developers realized the technical debts and removed them.
Tsoukalas et al.~\cite{tsoukalas2020technical} carried out research on 15 open-source projects and showed that complexity and LOC are two of the most significant indicators of technical debts. Siavvas et al.~\cite{siavvas2020technical} also studied 150 open-source software projects and reported that cyclomatic complexity of code is one of the key indicators of technical debts. 
% \texttt{`MultiplexorSEO\_writer.py'}\footnote{\url{https://raw.githubusercontent.com/artiste-qb-net/qubiter/e3f55398d7dc145641ff232b4ac78b8a5024f284/quantum_compiler/MultiplexorSEO_writer.py}}, 
%\href{https://raw.githubusercontent.com/artiste-qb-net/qubiter/e3f55398d7dc145641ff232b4ac78b8a5024f284/quantum_compiler/MultiplexorSEO_writer.py}{\texttt{`MultiplexorSEO\_writer.py'}} and
% \texttt{`MultiplexorExpander.py'}\footnote{\url{https://github.com/artiste-qb-net/qubiter/raw/e3f55398d7dc145641ff232b4ac78b8a5024f284/quantum_compiler/MultiplexorExpander.py}}, 
%\href{https://github.com/artiste-qb-net/qubiter/raw/e3f55398d7dc145641ff232b4ac78b8a5024f284/quantum_compiler/MultiplexorExpander.py}{\texttt{`MultiplexorExpander.py'}}, we found both coding errors and code smells 
\

%which reported technical debts of either code smells, SQ-bugs or both. Some of these files were refactored in the later commits
% \#7f60450\footnote{\url{https://github.com/artiste-qb-net/qubiter/commit/7f6045003d313158bcc50f03df9586f0b4356b5b}} 
%\href{https://github.com/artiste-qb-net/qubiter/commit/7f6045003d313158bcc50f03df9586f0b4356b5b}{\hashid{\texttt{\#7f60450}}}
%where over 1,982 lines of codes where removed. 

% \Le{We need to compare the above results with traditional software}

\begin{table}[t]
	\centering
 	\caption{The Spearman rank correlation coefficient that evaluates the correlations between the total technical debts in a file and the lines of code of the file. 
%between technical debts and source file LOC 
We calculated the correlation coefficient for two randomly selected quantum projects from each quantum categories in \fig\ref{fig:evolution-top-6}.\\ 
Statistical significance denoted as: ***$< 0.001$,  **$< 0.01$, *$< 0.05$}
 	\label{table:spearman-test}
 	\begin{tabular}{l|l r r}
        \toprule
        \textbf{Category}&\textbf{Target project}&\textbf{correlation}&\textbf{p-value}\\  \hline
        
        \multirow{2}{*}{Assembly} 
          &\texttt{qubiter}&0.622 &$***$\\
        \cline{2-4}
                     &\texttt{pyqgl2}       &0.895 &$***$\\
                     
        \hline       
        \multirow{2}{*}{Quantum-Annealing} 
          &\texttt{dwave-system}&0.893&$***$\\
        \cline{2-4}
        &\texttt{qbsolv}&0.879 &$***$\\
        \hline
        %\rowcolor{gray!10}
        \multirow{2}{*}{Experimental} 
          &\texttt{PyQLab}&0.940&$**$\\
        %\rowcolor{gray!10} 
        \cline{2-4}
        &\texttt{QGL} &0.936 &$**$\\
        \hline
        \multirow{2}{*}{Full-stack Library} 
          &\texttt{strawberryfields}&0.895&$***$\\
        \cline{2-4}
        &\texttt{Blueqat}  &0.902 &$***$\\
        \hline 
        \multirow{2}{*}{Quantum-Chemistry} 
          &\texttt{QCFractal}&0.903&$***$\\
        \cline{2-4}
                     &\texttt{CheMPS2}       &0.883&$***$\\
                     
        \hline
        \multirow{2}{*}{Toolkit} 
          &\texttt{qucat}&0.820&$***$\\
        \cline{2-4}
        &\texttt{bloomberg/quantum}       &0.949 &$***$\\
        
        \hline 
        \multirow{2}{*}{Simulator} 
          &\texttt{SimulaQron}&0.855&$***$\\
        \cline{2-4}
        &\texttt{Quirk} &0.584 &$***$\\
                     
        \hline
        \multirow{2}{*}{Quantum-Algorithms} 
          &\texttt{grove}&0.838&$***$\\
        \cline{2-4}
        &\texttt{qiskit-aqua} &0.860&$***$\\
                     
        \bottomrule
        \end{tabular}
\end{table}

\begin{table}[ht!]
		\centering
		\includegraphics[width=0.9\textwidth]{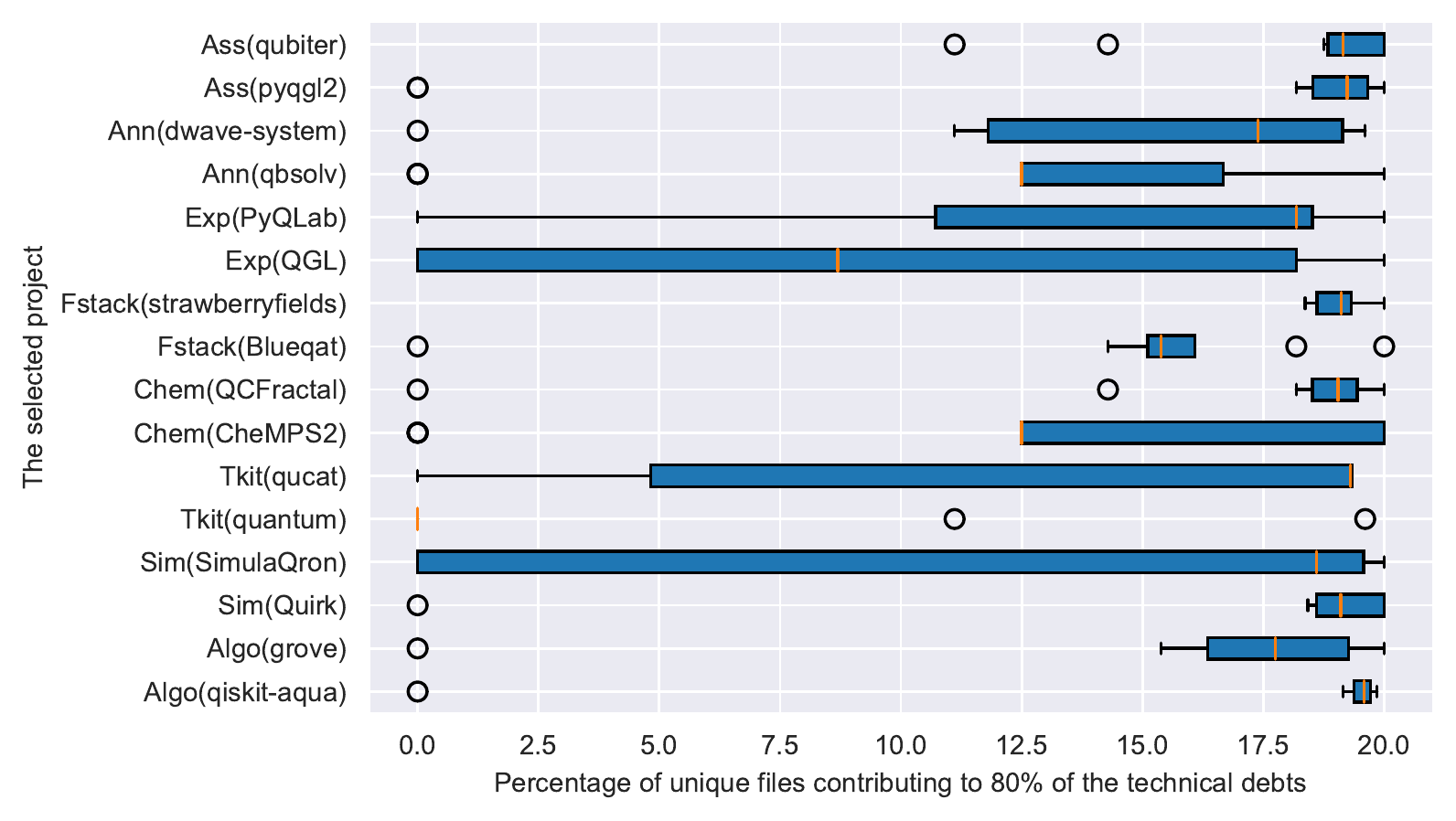}
		\captionof{figure}{Percentage of unique files contributing to 80\% of the overall technical debts across the snapshots of the selected projects. %\Le{Some text need to be revised}
		}
		\label{figu:box-unique-files}
\end{table}

%\Le{Let's consider to remove this}
We observed a similar relation for quantum projects. We observed a high positive correlation between file size and the amount of technical debts. Table \ref{table:spearman-test} presents the results of our correlation analysis (using the Spearman rank correlation test~\cite{zar2005:spearman}) between file size (LOC) and reported new technical debts across the snapshots of the randomly selected quantum projects from \fig\ref{fig:evolution-top-6}). 
%carried out the Spearman rank correlation test~\cite{zar2005:spearman} to further investigate the correlation between file size (LOC) and the  The results is shown in . We 
The low values of standard deviation indicate that the results are consistent across the project snapshots. We also found that at least half of the reported technical debts are contributed by the top 20\% of files (according to LOC) in all the studied snapshots, while the bottom 20\% of files only introduced a few technical debts ($\leq5\%$). \fig\ref{figu:box-unique-files} illustrates the percentage of unique files contributing to at least 80\% of the technical debts across the snapshots of our studied projects. Our results 
%are in line with a previous work~\cite{molnar:2020:long} that studied technical debts in traditional software, which 
confirms the Pareto's principle (80-20 rule)~\cite{dunford:2014:pareto} and is in line with
%Based on these observations, we summarize that most of the technical debts in quantum software are introduced by only a few amount of files with larger file size, which is inline with 
the previous studies~\cite{walkinshaw:201820,molnar:2020:long} on software defects and technical debts in traditional software. 
Walkinshaw et al.~\cite{walkinshaw:201820} studied 100 open-source projects  
and reported that 80.5\% of fixes are related to just 20\% of the files in all projects. Besides, they identified that near to 73\% of LOC of the top 20\% files are involved in the top 80\% of fixes. 

%\MO{Not sure about code complexity here, Medhi can you explain this?}
\begin{tcolorbox}
Technical debts tend to be added in the initial versions of a project in quantum computing software systems. Furthermore, similar to other types of software systems, we show that LOC %and code complexity 
is a key indicator of the occurrence of technical debts in quantum computing software systems. 
\end{tcolorbox}
% \Le{We need a summary box here}

\subsection{\textbf{RQ3: What Is the Relationship Between Technical Debts and Faults?}}\label{sec:result-correlation}

\begin{table}
	\caption{$p$-values for statistical significance and the coefficient for the Multiple Linear Regression (MLR) model based on all quantum projects (ALL-QT). The model uses the fault-inducing commits as dependent variable and the list of independent variables derived from technical debts types (as described in Section \ref{subsec:predict-step}). \\ Statistical significance denoted as: ***$< 0.001$,  **$< 0.01$, *$< 0.05$,\\ $\dag$ variable removed during VIF analysis,\\ $\ddag$ variable removed during stepwise selection criteria.
} 
	\label{tab:tbl_tags_corr_debt}
	\centering
    \begin{tabular}{l|c}\toprule
        \rowcolor{gray!20}
        \textbf{coef/p-value}  & \textbf{ALL-QT}\\ \midrule
        is\_smelly&\hlcyan{2.045e-06$^{***}$}\\ %hline
        is\_erroneous&\hlcyan{2.045e-06$^{***}$}\\ \midrule
        project&$\ddag$\\
        \midrule
        accessibility&$\ddag$\\
        brain-overload&\hlcyan{2.497e-05$^{***}$}\\
        design&\hlcyan{1.853e-09$^{**}$}\\
        unused&$\ddag$\\
        convention&\hlcyan{3.867e-06$^{*}$}\\
        cwe&$\ddag$\\
        redundant&\hlcyan{0.0003$^{***}$}\\
        confusing&$\ddag$\\
        error-handling&\hlcyan{1.687e-09$^{**}$}\\
        obsolete&$\ddag$\\
        user-experience&$\ddag$\\
        suspicious&$\ddag$\\ %hline
        \bottomrule
        Adj. $R^{2}$&0.074\\ %hline
        Prob (F-stat)&0.00\\ %hline
        AIC&-2.039e+06\\ %hline
        \bottomrule
    \end{tabular}
\end{table}

\begin{table}
    \caption{$p$-values for statistical significance and the coefficient for the MLR model based on individual quantum categories. The model uses the number of fault-inducing commits as dependent variable and the list of independent variables derived from technical debts types. \\ Statistical significance denoted as: $***< 0.001,  ** < 0.01, * < 0.05 $,\\ $\dag$ variable removed during VIF analysis,\\ $\ddag$ variable removed during stepwise criteria.
    } 
    \label{tab:tbl_tags_corr_debt_qt_category}
	\centering
	\begin{adjustbox}{width=\columnwidth,center}
	\small
        \begin{tabular}{l| c c c c c c c c c c}\toprule
        \rowcolor{gray!20}
        \textbf{coef/p-value}  & \textbf{Fstack}&  \textbf{Tkit}&   \textbf{Exp}&  \textbf{Sim}&  \textbf{Crypt}&   \textbf{Comp}& \textbf{Alg}& \textbf{Chem}&   \textbf{Asm}&\textbf{Ann}\\ \midrule
        is\_smelly&\hlcyan{0.1115$^{***}$}&\hlcyan{0.0012$^{***}$}&\hlcyan{0.0023$^{***}$}&\hlcyan{0.0938$^{***}$}&\hlcyan{0.0002$^{***}$}&\hlcyan{0.0410$^{***}$}&\hlcyan{0.0259$^{***}$}&\hlcyan{0.0632$^{***}$}&\hlcyan{0.0073$^{***}$}&\hlcyan{0.0088$^{**}$}\\

        %hline
        is\_erroneous&\hlcyan{0.2100$^{***}$}&\hlcyan{0.0013$^{***}$}&\hlcyan{0.0042$^{***}$}&$\ddag$&\hlcyan{0.0003$^{***}$}&\hlcyan{0.0230$^{**}$}&\hlcyan{0.0080$^{**}$}&\hlcyan{0.1433$^{***}$}&\hlcyan{0.0084$^{*}$}&\hlcyan{0.0136$^{**}$}\\ \midrule
        
        %% project
        \rowcolor{gray!10}
        project&$\ddag$&\hlcyan{0.0005}$^{***}$&\hlcyan{0.0108}$^{*}$&$\ddag$&\hlcyan{0.0120$^{**}$}&\hlcyan{0.9320}$^{***}$&$\dag$&\hlcyan{2.9523}$^{***}$&$\ddag$&$\ddag$\\%\hline
        \midrule
        \rowcolor{gray!10}
        accessibility&$\ddag$&$\ddag$&$\dag$&$\ddag$&$\dag$&$\dag$&$\dag$&$\ddag$&$\ddag$&$\dag$\\%\hline
        
        brain-overload&\hlcyan{0.7717$^{***}$}&$\dag$&\hlcyan{0.0813$^{***}$}&$\dag$&$\dag$&$\dag$&\hlcyan{ 0.1778$^{***}$}&$\dag$&\hlcyan{0.1323$^{**}$}&\hlcyan{0.2321$^{***}$}\\%\hline
        \rowcolor{gray!10}
        design&\hlcyan{0.1035$^{***}$}&\hlcyan{0.0183$^{***}$}&\hlcyan{0.0023$^{***}$}&\hlcyan{0.0976$^{***}$}&\hlcyan{1.011e-05$^{***}$}&$\dag$&\hlcyan{0.0735$^{**}$}&$\ddag$&$\ddag$&$\ddag$\\%\hline

        unused&$\ddag$&$\ddag$&\hlcyan{0.0169$^{***}$}&\hlcyan{0.0497$^{***}$}&$\dag$&$\dag$&$\ddag$&\hlcyan{0.2327$^{***}$}&\hlcyan{0.0087$^{***}$}&$\ddag$\\%\hline
        \rowcolor{gray!10}
        convention&\hlcyan{0.0790$^{***}$}&\hlcyan{0.0085$^{***}$}&$\ddag$&\hlcyan{0.0475$^{***}$}&$\dag$&$\ddag$&\hlcyan{0.0214$^{***}$}&$\ddag$&$\dag$&$\ddag$\\%\hline
        
        cwe&$\ddag$&\hlcyan{0.0053$^{***}$}&\hlcyan{0.0231$^{*}$}&\hlcyan{4.2227$^{***}$}&\hlcyan{0.0001$^{**}$}&$\dag$&$\ddag$&$\ddag$&$\dag$&$\ddag$\\%\hline
        
        \rowcolor{gray!10}
        redundant&\hlcyan{1.8002$^{***}$}&\hlcyan{0.0385$^{***}$}&$\ddag$&\hlcyan{0.7558$^{***}$}&$\dag$&\hlcyan{2.4618$^{*}$}&$\ddag$&\hlcyan{2.2455$^{*}$}&$\ddag$&$\ddag$\\%\hline

        confusing&$\dag$&\hlcyan{0.0262$^{***}$}&\hlcyan{0.0281$^{***}$}&\hlcyan{0.4411$^{**}$}&$\dag$&$\dag$&\hlcyan{0.2032$^{***}$}&$\ddag$&$\ddag$&\hlcyan{0.2616$^{*}$}\\%\hline
        \rowcolor{gray!10}
        error-handling&$\ddag$&$\ddag$&\hlcyan{0.0050$^{***}$}&\hlcyan{0.0385$^{*}$}&\hlcyan{0.0005$^{***}$}&$\ddag$&\hlcyan{0.0791$^{*}$}&\hlcyan{0.0260$^{**}$}&\hlcyan{0.0206$^{*}$}&$\ddag$\\%\hline
        obsolete&$\dag$&$\ddag$&$\dag$&$\dag$&$\dag$&$\dag$&$\dag$&$\dag$&$\dag$&$\dag$\\%\hline
        
        \rowcolor{gray!10}
        user-experience&$\ddag$&$\ddag$&$\dag$&$\ddag$&$\dag$&$\dag$&$\ddag$&$\ddag$&$\ddag$&$\dag$\\%\hline
        suspicious&\hlcyan{0.3689$^{***}$}&$\ddag$&\hlcyan{0.0023$^{**}$}&$\ddag$&\hlcyan{5.244e-05$^{***}$}&$\dag$&$\ddag$&$\ddag$&$\ddag$&$\ddag$\\ %hline
        \bottomrule
        Adj. $R^{2}$&\multicolumn{1}{r}{0.210}&\multicolumn{1}{r}{0.197}&\multicolumn{1}{r}{0.291}&\multicolumn{1}{r}{0.203}&\multicolumn{1}{r}{0.554}&\multicolumn{1}{r}{0.094}&\multicolumn{1}{r}{0.236}&\multicolumn{1}{r}{0.028}&\multicolumn{1}{r}{0.359}&\multicolumn{1}{r}{0.175}\\ %hline
        Prob (F-stat)&\multicolumn{1}{r}{0.00}&\multicolumn{1}{r}{0.00}&\multicolumn{1}{r}{9.40e-249}&\multicolumn{1}{r}{0.00}&\multicolumn{1}{r}{0.00}&\multicolumn{1}{r}{3.28e-58}&\multicolumn{1}{r}{1.69e-274}&\multicolumn{1}{r}{2.36e-284}&\multicolumn{1}{r}{1.69e-44}&\multicolumn{1}{r}{1.72e-34}\\ %hline
        AIC:&\multicolumn{1}{r}{-9139.}&\multicolumn{1}{r}{-9.767e+04}&\multicolumn{1}{r}{-2.273e+04}&\multicolumn{1}{r}{-8373.}&\multicolumn{1}{r}{-5.372e+04}&\multicolumn{1}{r}{-1.243e+04}&\multicolumn{1}{r}{-2.136e+04}&\multicolumn{1}{r}{-5.435e+04}&\multicolumn{1}{r}{-3141.}&\multicolumn{1}{r}{-5411.}\\ %hline
        \bottomrule
        \end{tabular}
    \end{adjustbox}
\end{table}
%%% End of latest updated table

In Section \ref{subsec:predict-step}, we described the steps used to examine the relationship between technical debts and faults introduced, as well as metrics that are related to the fault-inducing changes. This section discusses the results of the analysis of Section \ref{subsec:predict-step}. 

Tables \ref{tab:tbl_tags_corr_debt} and \ref{tab:tbl_tags_corr_debt_qt_category} presents the Multiple Linear Regression (MLR) model fitting results showing the relationship between technical debt as dependent variables and the fault-inducing commits as the independent variable. The dependent variables $is\_smelly$ and $is\_errorneous$ indicate if a given file contains at least one technical debt of type code smell and code error, respectively. The rest of the independent  %\Foutse{independent?} 
variables are the total counts of different technical debts by tags (tags presented in Table~\ref{table:minimun-tags}) in a given file across the project's snapshots.

Table \ref{tab:tbl_tags_corr_debt} reports the MLR results after fitting the model on all the quantum projects as one dataset (\emph{i.e.,} the debts reported for all the target projects as independent variables and the total number fault-inducing commits as dependent variable at file level). In Table \ref{tab:tbl_tags_corr_debt_qt_category}, we present the MLR fit result for the 10 individual quantum categories. We only reported the MLR model results for the $p$-value that is less and equal to 0.05 (i.e., $p$-value $\leq$ 0.05), the positive coefficient for statistically significant independent variables, and the variance inflation factor (VIF) that is less and equal to 2.5. The positive coefficient indicates the positive correlation between the independent variables (number of technical debts) and the dependent variable (number of fault-inducing commits). 

From Tables \ref{tab:tbl_tags_corr_debt} and \ref{tab:tbl_tags_corr_debt_qt_category}, we observed a statistically significant correlation between code smells and the number of fault-inducing changes, which is shown by the positive coefficient values of `is\_smelly' in all the categories of studied projects. The high positive coefficient of `is\_smelly' for the quantum category `Full-stack Library', `Simulator', `Quantum Chemistry', `Compiler',  and `Quantum Algorithm' also shows that there is a particularly high correlation between the code smells and the number of fault-inducing commits in the respective target quantum software. In addition, we found statistically significant correlation between the coding errors and the number of fault-inducing commits in most ($80\%$) of the target quantum categories. In particular, the coding Error shows a higher positive correlation on the fault introduced in the `Full-stack Library', and `Quantum Chemistry' categories.   %\Foutse{please check your dependent variable and make sure to clarify what it is!!!}

%\begin{tcolorbox}
%\textit{\textbf{Summary of findings (5)}}: The technical debts have a statistically significant positive correlation to the bug-inducing commits in the target quantum software systems.
%\end{tcolorbox}
%\Le{I removed this summary box, instead, we should compare the result with a study on traditional software.}

Moreover, Table \ref{tab:tbl_tags_corr_debt_qt_category} presents the correlation of technical debts based on the tag and the number of fault-inducing commits. %Table \ref{tab:tbl_tags_corr_debt_qt_category} further presents the correlation between different types of technical debts (shown in Table \ref{table:minimun-tags}) and the number of fault-inducing commits. 
Our results show that the tags `design', `redundant', `error-handling', and `brain-overload' have a statistically significant correlation with the number of fault-inducing commits in most ($\geq 50\%$) of the quantum categories. This result suggests that the tags `design', `redundant', `error-handling', and `brain-overload' would help predict the occurrences of fault-inducing commits in quantum software systems. %In particular, the type `brain-overload' has such a statistically significant correlation in about $90\%$ of the studied quantum categories. This result suggests that the type `brain-overload' would help predict the occurrences of fault-inducing commits in quantum software systems. 
The design technical debt has higher positive correlation to fault-inducing in Full-stack libraries, quantum simulators and Toolkit categories, `brain-overload' shows a higher positive correlation with the number of fault-inducing commits in the quantum categories in the order `Full-stack Library', `Quantum-Annealing', `Quantum-algorithms' and `Assembly'.

Comparing the technical debts  tags across the project categories, `brain-overload' shows a higher positive correlation with the number of fault-inducing commits in the quantum categories  in the order `Full-stack Library', `Quantum-Annealing', `Quantum-algorithms' and `Assembly'. Similarly, the design related technical debt has higher positive correlation to fault-inducing in `Full-stack Library'. Moreover, the technical debt of tag `redundant' introduces more faults in the categories `Compiler', `Quantum-Chemistry'. %Moreover, we observed that `user-experience' also has a higher correlation with the number of fault-inducing commits in the categories of `Experimental', `Compiler', `Assembly', and `Quantum-Annealing'. 
Finally, in Table~\ref{tab:tbl_tags_corr_debt_qt_category} we observed that most of the tags are statistically significantly correlated with the number of fault-inducing commits in the quantum categories of `Experimentation', `Simulator' and `Quantum Algorithms', whereas only fewer ($\leq 20\%$) of the tags are correlated with the fault-inducing commits in the categories `Compiler' and `Quantum Annealing'. This implies that these technical debt tags can be included to predict the number of fault-inducing commits in the respective quantum categories.

%\begin{tcolorbox}
%\textit{\textbf{Summary of findings (6)}}: Technical debts with the tags `brain-overload', `error-handling' and `suspicious' have a statistically significant correlation with bug-inducing commits in most of the quantum software. In particular, the debts `brain-overload, shows statistically significant correlation with bug-inducing in most ($90\%$) of the quantum categories compared to other tags.
%\end{tcolorbox}
Previous studies also reviewed the relationship between technical debts and types of faults in different types of software systems. Digkas et al.~\cite{digkas:2017:evolution} carried out a research on 66 open-source java based software projects developed by the Apache software foundation. They showed that 'literal duplicate' is the most frequent type of technical debts. They also observed that this type of technical debts are equally distributed among their analyzed systems.
%with a Gini index of 0.31. Gini index is a statistical measure of the degree of inequality in a set of values. The Gini index value is between 0\% and 100\%. 0 represents complete equality and 100\% represents complete inequality.
%which means that it's very equally distributed among their analyzed systems.
%was equally distributed amongst their studied software systems. 
Moreover, they revealed that `using diamond operation' and `code comment-out' 
%with the highest Gini index 
are the least equally distributed technical debts among their studied software systems. %\Foutse{what does this Gini index score means concretely? can you guys better explain these reported results?}.
Tan et al.~\cite{tan:evolution} studied 44 open-source software systems based on Python belonging to the Apache software foundation. They indicated that `defining docstring', `too long lines' and `undefined variables' are the top three types of technical debts identified in those projects. Moreover, they mentioned that `empty nested block' and `too many line of code in a file' are the least frequent types of technical debt.

% \Le{We need a better summary box, which should mention the comparison with traditional software.}

\begin{tcolorbox}
As expected, the correlation between technical debts and the number of faults introduced are different based on the nature of software systems. %On the other hand, used programming language can play a key role in the  existence of correlation between technical debts and faults. 
On the other hand, the technical debts of types `design', `redundant', `error-handling' and `brain-overload' can be a good indicator to predict the number of fault-inducing commits in quantum software since they have the highest significance in most of our studied quantum projects. 
\end{tcolorbox}

\section{Discussion and Implication}\label{sec:discussion}

In this section, we further elaborate on the results shown in Section \ref{sec:results} and highlight the implications of the results to the researchers and the developers of quantum software.

As the initial step, we investigated the representation of maintainability and reliability in the overall technical debts in quantum software systems as shown in Figure~\ref{fig:debt-type}. We then broke down the technical debt based on the severity (in Figure~\ref{fig:debt-severity}) and examine the rules distributions across the different categories of technical debt. %\Foutse{across the different categories of technical debt?} the overall technical debts.
We found that about 80\% of the technical debts are related to code smells and most of these debts are related to critical and major severity types. We also highlighted some of the examples for critical debts as reported in Table \ref{tab:tags-rules}. We believe that this information can help practitioners prioritize resource allocation to refactor their code, specifically on the code with critical technical debts.

In our manual investigation of the quantum project of compiler category, we observed that technical debts are found most in source code for handling and generating graphs, benchmarks, unreachable code %\Foutse{unreachable code is a type of technical debt!}
(due to inappropriate exit points, control flow or jump statements), and unexpected expressions. %\Foutse{what is this exactly? this is debt too...}. 
We also observed code smells due to multiple empty blocks of codes detected as `suspicious' in the source code generating Python list to store gates which is of Hermitian conjugate, in the Assembly project. The related source code implementing the in-memory storage of circuits were also associated with other technical debts such as design, unused code, brain-overload (related to cognitive complexity). Implementing in-memory storage of circuits can be helpful in quantum Assembly for easy randomly access any gate of the circuit, for instance, during circuit optimizations. These features are particular in quantum software. Usually, a typical quantum program consists of blocks of code, each of which contains both classical and quantum components to execute classical instructions and quantum instruction, respectively. Classical instructions operate on the state of classical bits and apply conditional statements. They are also used for post processing the outcome of measurements on qubits, %\Foutse{classical instructions are also used for post processing the outcome of measurements on qbits!},
while quantum instructions operate on the state of qubits and measure the qubit values~\cite{cross2017open}. Studies on traditional software by Marcilio et al.~\cite{marcilio:2019:static} revealed that the problems regarding packages and exception handling contain the biggest proportion of technical debts in their studied traditional projects. Digkas et al.~\cite{digkas:2018:developers} also reported that the most frequent technical debts are related to resource management, null pointer, and exception handling problems. 

In Table \ref{table:minimun-tags} we showed that a few types of technical debts
contribute to about 80\% and more of the overall technical debts. These dominating types include ‘code convention’, ‘unused’, `design issues' (the design of the code is questionable, \emph{e.g.,} duplicate string literals), `brain-overload', `obsolete', `accessibility,' `cwe,' `confusing' and `error-handling'.  %, impacting highly quantum applications based on the category. 
Our finding is in line with the Pareto principle hence its application. We have also shown that the type `code convention' dominates in more than 50\% of the studied quantum categories. Therefore, we recommend that quantum developers should follow the coding convention to avoid confusion and allow efficient team collaboration.

Lenarduzzi et al.~\cite{Lenarduzzi:2019:Empirical-TechnicalDebt} have shown that constraints such as time and budgets are the root causes of technical debts. Also, factors such as refactoring efforts and architectural changes have been mentioned to influence technical debt in the software projects as compared to the changes in the line of code~\cite{martini2015investigating, Arthur:Discovering:2017}. We investigated how and when the technical debts are introduced into the code base of quantum software systems to improve our understanding and verify whether similar trends can be identified for technical debt in a quantum software system. \fig{\ref{fig:evolution-top-6}} shows how the code changes are related to technical debts and how these trends evolve over time. We found that most of the technical debts are associated with the initial versions, where most new code or source files are added. Our result indicates that less than 20\% of the studied files with large file size introduced most of the technical debts across the studied snapshots of the quantum projects. This result is inline with the studies on traditional software~\cite{Arthur:Discovering:2017,Arthur-Jozsef:Longitudinal:2019}. Our finding suggests that quantum software developers should pay attention to the code quality and code size, especially when new files are added. They should more carefully review and test the commits with a large number of changes.
%They should consider to have more effort in code review, use static analysis tools and IDE to avoid potential errors and technical debts in their source code.
%adopting some traditional approaches such as code review, following the coding convention, and coding with best practices. 
Quantum developers should also consider using the traditional static code analysis tools to detect and monitor their source code at the early stage of development to reduce future maintenance costs. %We explored the evolution of technical debts and provided an overview of how technical debts were introduced in quantum software applications based on the types. 
In the future, we plan to examine the lifespan of quantum software faults and how developers prioritize resolving these faults based on the types and severity.  % \Foutse{this is fine, but a bit generic...i wonder if we did observed some types of issues that are specific to the probabilistic nature of quantum programs for examples? some issues related to quantum circuits, their conversions etc ? showing such examples would make your discussions more interesting!}

%\Mona{Please check the code complexity part as requested by reviewer 1}
In Table \ref{tab:tbl_tags_corr_debt}, we found that there is a statistically significant correlation between the technical debts and the number of fault-inducing commits in the studied quantum projects. We observed that the debt types related to cognitive complexity (brain-overload), code design, redundant code and error-handling show a significant correlation with fault-inducing commit occurrences in about 60\% of the quantum categories. 
%indicate a statistically significant correlated with fault-inducing commits in about $90\%$ of the studied quantum categories. 
We recommend that developers should use these metrics as measures to predict faults in quantum software systems especially when the developers are working on projects related to `Full-stack Library', `Quantum-annealing', `Quantum-chemistry', `Quantum-algorithms', and `Compiler'. Also, as we observed the statistical significant correlation between design-related technical and faults introduced  in most (60\%) of the studied quantum categories, % as a result of  \Foutse{this is claiming causation! do you have evidence that these faults are introduced by design-related technical debts?} design-related technical debts, 
we recommend that practitioners should implement more systematic testing on quantum software systems that reflect the structural nature of the quantum software.  
We suggest that the quality assurance team for quantum software systems should effectively allocate more resources and time to thoroughly validate and test the complex components and the components related to error handling before integrating new commits. %\Foutse{it would be nice if we could also comment on some errors that are specific to the nature of quantum programs!}% to the main branch.%\Le{we may say before merge new commits to the master branch. I think there's often one repo for a project but there can be many branches}.

\begin{comment}
From the messages of the changes that introduced faults, we highlighted 22 key topics (in Table \ref{tab:topics}) \Mona{this reference is broken} related to optimization procedures, release engineering process, environment, file management, and quantum/molecular mechanics. 
%These topics are generated from the faulty commits during quantum software development activities as a result of that calls for further investigation by both practitioners and the researchers.
Although all these topics are important in their respective area, we further drilled down to the most dominating topics with fault-inducing commits and those requiring a longer time to get fixed. The top dominating topics include `Configuration files', `Dependency Management', `Compiler Optimization', `Code Refactoring', `Calibration \& Visualization' and `Program Optimization'. And the topics that require longer time to fix include `Chemical Calculation', `Configuration Interaction', `Path Integral', `Message Passing', `Parameter checks and Selection', `Arrays and Matrix Operations' and Memory/ Performance Optimization'. 
%These topics may suggest quantum developers may find them challenging.
These topics imply the challenges that quantum developers faced. 
%Tools should be implemented to assist developers in the respective area. Also, identifying the actual problems related to each of these topics is out of this study's scope, and we call for more investigations by the research community.
%We call for future investigations by the research community.
\end{comment}
We suggest that developers should use existing static analysis tools to detect potential problems that also happen in traditional software. We also appeal the quantum software community to introduce new tools to detect quantum-specific errors or technical debts.

%\subsection{Verifying quantum software programs} The unique features of quantum programs such as superposition, entanglement, and no-cloning make it difficult to debug and test than traditional software programs.

\section{Threats to Validity} \label{sec:threats}

%\Le{I don't think this first paragraph is needed, since we need to reveal the threats in this section. Please consider to %remove this paragraph.}
%In this work, we followed a set of research methodology to achieve. Initially, we derived a set of research objectives and the research questions. Given the research objectives, we selected the target projects consisting of both quantum and traditional software repositories, which were then preprocessed and analyzed to answer our proposed research question. 
There are several threats that can potentially affect the validity of our study. In this section, we discuss the threats to validity of our study by following the guidelines for case study research. We followed the best practice defined to evaluate our study as presented by Runeson et al.~\cite{Runeson:2009}. %, Per, and Martin Höst~\cite{Runeson:2009}.

\paragraph{Threats to construct validity} are concerned with the relationship between theory and observation. We used the SZZ algorithm to identify the changes that introduced faults. This algorithm assumes that faults are introduced by the lines that are later fixed by fault-fixing commits. However, in a fault-fixing commit, not all of the changed lines are used to fix faults. To mitigate this threat, we used the PyDriller which eliminate the candidates of fault-inducing commits that only changed white spaces. Muse et al. \cite{muse2020prevalence} manually validated the precision of PyDriller and showed that the tool only yielded 6\% of false positives. 
%\BLUE{Abidi et al.~\cite{abidi2020multi} also manually examined the performance of PyDriller on multi-language projects. They found that the tool can precisely identify more than 80\% of the fault-inducing commits in the projects.}
%the precision of fault-inducing detection is higher than 80\% in the projects.}
%results indicate the precision of at least 80\%.}  % when detecting faults in SQL code.
%Previous work \cite{muse2020prevalence} showed that Pydriller only yielded 6\% of false positives when detecting faults in SQL code.
%\Foutse{Moses, you can also cite your recent TOSEM paper with Mouna where you guys analyzed a sample of fault-inducing commits to assess the precision of SZZ and report the precision that was found there!} 
In addition, we used the topic modeling technique to automatically extract the characteristics of the fault-inducing changes. Some characteristics may be missed during this automatic process. In future work, we plan to select a sample of the fault-inducing commits, and manually analyze the root causes and the symptoms of the faults.

\paragraph{Threats to internal validity} are concerned with the factors that may affect a dependent variable and were not considered in the study.
We manually selected quantum related projects and grouped the projects into different categories. To minimize this threat, the project selection and categorization were performed by two of the authors independently. They discussed each of the discrepancies until a mutual agreement was reached.
%Another threat may be introduced during the manual labeling of the fault-inducing commits and determining the optimal number of $K$ and iterations value $I$ for the fault-inducing topics. We followed a well-defined technique adapted from~\cite{Linstead:GNOME-bug-report,Hindle:Automated-Topic:2011,Liao:Issue-Related:2018,Mehdi:GoingBig:2019,Christoffer:Mobile-Developers:2016,Kartik:Web-Developers:2014,Syed:Concurrency:2018} together with multiple experiments to ensure that we identified the best $K$ and $I$ values. We referred to multiple resources during the manual labeling process, including quantum publications and official documentation of target projects, to ensure the correct labels were assigned to the topics. In addition, each label was only assigned after complete agreements among the authors. 
We shared our dataset online \footnote{\url{https://github.com/openjamoses/JSS-Replication}}.
% \Mehdi{We need to put the link of repository here}\Foutse{please add the replication package link}. 
Future studies can replicate and validate our results.

\paragraph{Threats to conclusion validity} are concerned with the relationship between the treatment and the outcome.
%Another major threat is that 
We detect the technical debts from quantum software by using a static analysis tool (SonarQube), which is designed for traditional software. However, to mitigate this threat we limit our studied projects to be those that are written in traditional programming languages and can be analyzed by traditional static code analysis tools. Validating our results on the software running specifically on quantum computers will be our future work.
%\nd Another major threat is that this study relied on the existing code analysis tool to detect the TDs and given the difference between quantum systems and  traditional software systems. A similar reason why our study is limited to only quantum projects that use the traditional programming languages supported by the existing static code analysis tool to help us replicate the study. This rule may not be easily generalized to quantum software systems' significant issues.  However, since this software category are developed to run on classical computers and a similar tool has been applied to many software systems of different domains sharing the related programming language. Hence, similar rules may apply to quantum systems running on classical computers. 
%\Le{We need to carefully discuss the threats from the SZZ analysis}
%\nd Finally, to validate our results and any limitations that may exist with SonarQube as well as the limitations during automated steps to detect fault-inducing commits, 
%Finally, 
To validate the detection accuracy of SonarQube on technical debts, we manually examined 500 technical debts identified by the tool and observed that 98\% of the technical debts are correct. Throughout our discussion in Section \ref{sec:results}, we highlighted the technical debts that are validated in our manual validation.
In addition, we relied on a set of rules defined by SonarQube to identify different types of technical debts. These rules may not capture all possible technical debts. In the future, we plan to use other equivalent tools to verify our results and perform manual analyses to find out which kinds of debts cannot be identified by the state-of-the-art tools and will suggest the software engineering community to improve the static analysis tools on this aspect. 

\paragraph{Threats to external validity} are concerned with the generalizability of our results. 
To study the characteristics of technical debts and faults in quantum software, we analyzed 118 open-source quantum projects from GitHub. Our selected projects are related to different domains of quantum computing. Our results can be considered as a reference for quantum developers and researchers to improve their software quality in terms of maintainability and reliability. However, our results may not generalize to all quantum software. Future studies are welcome to replicate and validate our work in other quantum projects.

%of diverse categories, we relied entirely on open-source projects hosted on GitHub, used the GitHub rest API, and searched for repositories based on the keyword. This step may have missed some relevant %and sizeable 
%quantum repositories 
%from other repositories or GitHub 
%that could improve our results and the generalization. Also, to study the correlation between the technical debts and fault introduced during quantum software development, we relied on a set of rules and tags from SonarQube analysis. However, these SQ-issues may not be considered exhaustive enough and do not represent the complete set of TD related issues that manifests in quantum source code. Some of these SQ-issues are specific to the programming language, which limits us from grouping the tags across multiple languages in some cases. Our steps to partially mitigate this threat is to use the latest SonarQube version 8.5 that has updated rules and tags for more languages. 

%\paragraph{Construct Threats:}
%\Le{I revised this paragraph based on my understanding. The original text is not easy to understand. Please check if this is what you meant.}

%\Le{We also need to carefully discuss why bugs about tests and documentation are included in this study.}

\section{Conclusion} \label{sec:conclusion}

As for traditional software, the maintenance of quantum software systems is equally important because the maintenance practices can affect the quality of a whole quantum computing system. In this study, we selected 118 open-source quantum computing related projects from GitHub. We empirically studied the distribution and evolution of technical debts in these projects and the relationship between technical debts and fault occurrences.
We observed that a few types of technical debts (related to code convention violation, error-handling, and code complexity) dominate the total number of technical debts detected from our studied projects. There is a strong correlation between technical debts and post-release faults. Particularly, files with high code complexity are more likely to lead to faults. We also found that code changes related to configuration files and dependency management tend to be highly fault-prone, which need more attention when developers perform code review and testing. Based on the findings reported in this paper, we formulate the following recommendations:
%summarize the implications of our results and the recommendations as follows: %We also suggest the quantum community to introduce tools for detecting potential errors that are specific to quantum computing software.

\begin{itemize}
    \item  For the technical debt detection, we recommend that quantum software developers use the existing static analysis tools to examine their code.
    
    \item We also recommend that practitioners prioritize resource allocation to refactor their source code, especially the code with critical technical debts. Also, quantum developers should follow the coding convention to avoid confusion, especially to support team collaboration.
    
    \item Quantum developers should pay attention to the code quality and code size, especially when new files are added. They should review and test carefully the commits with a large number of changes.
    
    \item Code reviewers and quantum quality assurance team should use metrics like code convention, code redundancy, error-handling, and the cognitive complexity of the code to predict faulty commits.
    
    \item New tools should be introduced to support identifying quantum-specific problems, such as the technical debts and faults that only occur in a quantum software system.
    
    \item Future works are appealed to study other aspects of quantum software in terms of maintenance and reliability, such as code review, verification methods to ensure the correctness of a quantum program, and practical fault detection techniques for supporting quantum systematic testing and debugging.
\end{itemize}

\section*{Acknowledgement}
We thank the Natural Sciences and Engineering Research Council of Canada (NSERC) for funding this project. 


%% References

%\bibliographystyle{ieeetr}
\bibliographystyle{elsarticle-num}
%\bibliographystyle{unsrt}
%\bibliographystyle{ieeetr}





%addbibresource{refs}
\bibliography{refs}
%\addbibresource{refs}
 %\end{thebibliography}

\begin{appendices}
%\newpage

\section{Sample quantum projects in each categorie}
\label{tab:example_repo}
The GitHub repositories are shown as: \texttt{<owner\_name>/<repository\_name>} 
\begin{table}[!h]
    %\scriptsize
    \centering
    \begin{tabular}{c| m{4.0cm} | m{10.5cm}| c }
\toprule
\rowcolor{gray!10}
\textbf{}&\textbf{Category} &\textbf{The studied repositories (separated by comma) %\Foutse{why don't you provide the full list? this is an appendix, so people expect details...otherwise why isn't this in the paper?}
}&Total\\ \midrule

1&Full-stack Library or Framework (Fstack)&Qiskit/qiskit-terra, microsoft/Quantum, microsoft/QuantumLibraries, microsoft/Quantum-NC, rigetti/pyquil, Qiskit/qiskit, Qiskit/qiskit-ibmq-provider, qiskit-community/qiskit-js, qiskit-community/qiskit-vscode, rigetti/rpcq, ProjectQ-Framework/ProjectQ, QuantumPackage/qp2, XanaduAI/strawberryfields, Blueqat/Blueqat, quantumlib/Cirq, softwareQinc/staq&16\\ %hline
\rowcolor{gray!10}
2&Experimentation (Exp)&m-labs/artiq, sedabull/quantum-shell, qutech/qupulse, iitis/QuantumInformation.jl, lneuhaus/pyrpl, BBN-Q/Qlab, BBN-Q/PyQLab&8\\ %hline

3&Simulator (Sim)&softwareQinc/qpp, QuantumBFS/YaoBlocks.jl, issp-center-dev/HPhi, SoftwareQuTech/SimulaQron, QuantumBFS/Yao.jl, Strilanc/Quirk, Qiskit/qiskit-aer, qutip/qutip, QuEST-Kit/QuEST, vm6502q/qrack, qulacs/qulacs, quantumlib/OpenFermion-Cirq, microsoft/qmt, Approximates/dotBloch, Qiskit/qiskit-jku-provider, ngnrsaa/qflex, aparent/QCViewer, rigetti/qvm, marvel-nccr/quantum-mobile&20\\ %hline
\rowcolor{gray!10}
4&Cryptography (Crypt)&theQRL/QRL, exaexa/codecrypt, BBN-Q/QGL, mupq/pqm4&4\\ %hline

5&Compiler (Comp)&microsoft/qsharp-compiler, QE-Lab/OpenQL, Quantomatic/pyzx&3\\ %hline
\rowcolor{gray!10}
6&ToolKit (Tkit)&evaleev/libint, qojulia/QuantumOptics.jl, jcmgray/quimb, XanaduAI/pennylane, OriginQ/QPanda-2, CQCL/pytket, bloomberg/quantum, rigetti/forest-benchmarking, QInfer/python-qinfer, zoran-cuckovic/QGIS-visibility-analysis, qubekit/QUBEKit, tensorflow/quantum, redhat-cip/openstack-quantum-puppet, boschmitt/tweedledum, TRIQS/triqs, QuTech-Delft/qtt, lmacken/quantumrandom, qucat/qucat, Qiskit/qiskit-ignis, orbkit/orbkit,deepchem/deepchem, aiidateam/aiida-quantumespresso&22\\ %hline

7&Algorithms (Algo)&qrefine/qrefine, mabuchilab/QNET, Qiskit/qiskit-aqua, quantumlib/OpenFermion, rigetti/grove, netket/netket, XanaduAI/thewalrus, qucontrol/krotov, aeantipov/pomerol, Q-solvers/EDLib, dwave-examples/factoring, PanPalitta/phase\_estimation, ProjectQ-Framework/FermiLib, JoshuaSBrown/QC\_Tools&14\\ %hline
\rowcolor{gray!10}
8&Annealing (Ann)&shinmorino/sqaod, dwavesystems/qbsolv, dwavesystems/dimod, dwavesystems/dwave-system, dwavesystems/dwavebinarycsp, dwavesystems/penaltymodel&6\\ %hline

% 10&Web-Extension&foxyproxy/firefox-extension, spikespaz/firefox-nativedark, quantacms/quanta, dauphine-dev/drop-feeds\\ %hline
% \rowcolor{gray!10}
9&Chemistry (Chem)&ValeevGroup/mpqc, pyscf/pyscf, MolSSI-BSE/basis\_set\_exchange, MolSSI/QCElemental, MolSSI/QCEngine, MolSSI/QCFractal, SebWouters/CheMPS2, Quantum-Dynamics-Hub/libra-code, QMCPACK/qmcpack, cp2k/cp2k, votca/xtp, hande-qmc/hande, tmancal74/quantarhei, LCPQ/quantum\_package, vonDonnerstein/QuantumLab.jl, ericchansen/q2mm, aoterodelaroza/critic2, GQCG/GQCP, qcdb/qcdb&19\\ %hline
\rowcolor{gray!10}
10&Assembly (Ass)&valeevGroup/mpqc, MolSSI/QCElemental, MolSSI/QCEngine, QMCPACK/qmcpack, SebWouters/CheMPS2&5\\ %hline

\bottomrule

    \end{tabular} 

    %\caption{Sample quantum projects related to each categories}
    \label{tbl:quantum_projects_categories}
\end{table}

\newpage

\end{appendices}
\end{document}